\documentclass[longauth]{aa}

\usepackage{graphicx}
\usepackage{scalerel}
\usepackage{multirow}
\usepackage{lastpage}
\usepackage{csquotes}
\usepackage{dcolumn} 
\usepackage{tablefootnote}
\usepackage[table]{xcolor}

\usepackage{txfonts}
\usepackage[pdfencoding=auto,psdextra]{hyperref}
\hypersetup{
    colorlinks=true,
    linkcolor=blue,
    filecolor=magenta,      
    urlcolor=blue,
    citecolor=blue
}
\usepackage[nameinlink,capitalise]{cleveref}
\crefname{section}{Sect.}{Sects.}
\Crefname{section}{Section}{Sections}
\crefname{figure}{Fig.}{Figs.}
\Crefname{figure}{Figure}{Figures}
\crefname{equation}{Eq.}{Eqs.}
\Crefname{equation}{Equation}{Equations}
\crefname{table}{Table}{Tables}
\crefname{appendix}{Appendix}{Appendices}
\creflabelformat{equation}{#2\textup{#1}#3}

\makeatletter
\renewcommand*\aa@pageof{, page \thepage{} of \pageref*{LastPage}}
\makeatother

\usepackage[utf8]{inputenc}

\usepackage[switch, modulo]{lineno}

\usepackage{euclid}

\defcitealias{Bisigello_2024}{B24}
\defcitealias{Q1-SP003}{R25}
\defcitealias{Q1-SP027}{M25}

\begin{document}

\title{Euclid Quick Data Release (Q1)}
\subtitle{The impact of AGN emission on SED-derived physical properties}

\newcommand{\orcid}[1]{} 
\author{Euclid Collaboration: B.~Laloux\orcid{0000-0001-9996-9732}\thanks{\email{brivael.laloux@inaf.it}}\inst{\ref{aff1},\ref{aff2}}
\and A.~Bongiorno\orcid{0000-0002-0101-6624}\inst{\ref{aff3}}
\and M.~Salvato\orcid{0000-0001-7116-9303}\inst{\ref{aff2}}
\and V.~Allevato\orcid{0000-0001-7232-5152}\inst{\ref{aff1}}
\and M.~Mezcua\orcid{0000-0003-4440-259X}\inst{\ref{aff4},\ref{aff5}}
\and W.~Roster\orcid{0000-0002-9149-6528}\inst{\ref{aff2}}
\and T.~Matamoro~Zatarain\orcid{0009-0007-2976-293X}\inst{\ref{aff6}}
\and S.~Paltani\orcid{0000-0002-8108-9179}\inst{\ref{aff7}}
\and R.~Shirley\orcid{0000-0002-1114-0135}\inst{\ref{aff2}}
\and F.~Tarsitano\orcid{0000-0002-5919-0238}\inst{\ref{aff8},\ref{aff7}}
\and C.~Saulder\orcid{0000-0002-0408-5633}\inst{\ref{aff2},\ref{aff9}}
\and S.~Fotopoulou\orcid{0000-0002-9686-254X}\inst{\ref{aff6}}
\and C.~Andonie\orcid{0000-0002-5580-4298}\inst{\ref{aff2}}
\and J.~Buchner\orcid{0000-0003-0426-6634}\inst{\ref{aff2}}
\and F.~La~Franca\orcid{0000-0002-1239-2721}\inst{\ref{aff10},\ref{aff3}}
\and V.~Le~Brun\orcid{0000-0002-5027-1939}\inst{\ref{aff11}}
\and F.~Ricci\orcid{0000-0001-5742-5980}\inst{\ref{aff10},\ref{aff3}}
\and D.~Scott\orcid{0000-0002-6878-9840}\inst{\ref{aff12}}
\and F.~Shankar\orcid{0000-0001-8973-5051}\inst{\ref{aff13}}
\and M.~Siudek\orcid{0000-0002-2949-2155}\inst{\ref{aff4},\ref{aff14},\ref{aff15}}
\and J.~G.~Sorce\orcid{0000-0002-2307-2432}\inst{\ref{aff16},\ref{aff17}}
\and L.~Spinoglio\orcid{0000-0001-8840-1551}\inst{\ref{aff18}}
\and Y.~Toba\orcid{0000-0002-3531-7863}\inst{\ref{aff19},\ref{aff20}}
\and A.~Viitanen\orcid{0000-0001-9383-786X}\inst{\ref{aff21},\ref{aff7},\ref{aff3}}
\and L.~Wang\orcid{0000-0002-6736-9158}\inst{\ref{aff22},\ref{aff23}}
\and G.~Zamorani\orcid{0000-0002-2318-301X}\inst{\ref{aff24}}
\and S.~Andreon\orcid{0000-0002-2041-8784}\inst{\ref{aff25}}
\and N.~Auricchio\orcid{0000-0003-4444-8651}\inst{\ref{aff24}}
\and C.~Baccigalupi\orcid{0000-0002-8211-1630}\inst{\ref{aff26},\ref{aff27},\ref{aff28},\ref{aff29}}
\and M.~Baldi\orcid{0000-0003-4145-1943}\inst{\ref{aff30},\ref{aff24},\ref{aff31}}
\and A.~Balestra\orcid{0000-0002-6967-261X}\inst{\ref{aff32}}
\and S.~Bardelli\orcid{0000-0002-8900-0298}\inst{\ref{aff24}}
\and P.~Battaglia\orcid{0000-0002-7337-5909}\inst{\ref{aff24}}
\and A.~Biviano\orcid{0000-0002-0857-0732}\inst{\ref{aff27},\ref{aff26}}
\and E.~Branchini\orcid{0000-0002-0808-6908}\inst{\ref{aff33},\ref{aff34},\ref{aff25}}
\and M.~Brescia\orcid{0000-0001-9506-5680}\inst{\ref{aff35},\ref{aff1}}
\and S.~Camera\orcid{0000-0003-3399-3574}\inst{\ref{aff36},\ref{aff37},\ref{aff38}}
\and G.~Ca\~nas-Herrera\orcid{0000-0003-2796-2149}\inst{\ref{aff39},\ref{aff40}}
\and V.~Capobianco\orcid{0000-0002-3309-7692}\inst{\ref{aff38}}
\and C.~Carbone\orcid{0000-0003-0125-3563}\inst{\ref{aff41}}
\and J.~Carretero\orcid{0000-0002-3130-0204}\inst{\ref{aff42},\ref{aff43}}
\and S.~Casas\orcid{0000-0002-4751-5138}\inst{\ref{aff44},\ref{aff45}}
\and M.~Castellano\orcid{0000-0001-9875-8263}\inst{\ref{aff3}}
\and G.~Castignani\orcid{0000-0001-6831-0687}\inst{\ref{aff24}}
\and S.~Cavuoti\orcid{0000-0002-3787-4196}\inst{\ref{aff1},\ref{aff46}}
\and K.~C.~Chambers\orcid{0000-0001-6965-7789}\inst{\ref{aff47}}
\and A.~Cimatti\inst{\ref{aff48}}
\and C.~Colodro-Conde\inst{\ref{aff14}}
\and G.~Congedo\orcid{0000-0003-2508-0046}\inst{\ref{aff39}}
\and C.~J.~Conselice\orcid{0000-0003-1949-7638}\inst{\ref{aff49}}
\and L.~Conversi\orcid{0000-0002-6710-8476}\inst{\ref{aff50},\ref{aff51}}
\and Y.~Copin\orcid{0000-0002-5317-7518}\inst{\ref{aff52}}
\and A.~Costille\inst{\ref{aff11}}
\and F.~Courbin\orcid{0000-0003-0758-6510}\inst{\ref{aff53},\ref{aff54},\ref{aff55}}
\and H.~M.~Courtois\orcid{0000-0003-0509-1776}\inst{\ref{aff56}}
\and M.~Cropper\orcid{0000-0003-4571-9468}\inst{\ref{aff57}}
\and A.~Da~Silva\orcid{0000-0002-6385-1609}\inst{\ref{aff58},\ref{aff59}}
\and H.~Degaudenzi\orcid{0000-0002-5887-6799}\inst{\ref{aff7}}
\and G.~De~Lucia\orcid{0000-0002-6220-9104}\inst{\ref{aff27}}
\and H.~Dole\orcid{0000-0002-9767-3839}\inst{\ref{aff17}}
\and F.~Dubath\orcid{0000-0002-6533-2810}\inst{\ref{aff7}}
\and C.~A.~J.~Duncan\orcid{0009-0003-3573-0791}\inst{\ref{aff39}}
\and X.~Dupac\inst{\ref{aff51}}
\and S.~Escoffier\orcid{0000-0002-2847-7498}\inst{\ref{aff60}}
\and M.~Fabricius\orcid{0000-0002-7025-6058}\inst{\ref{aff2},\ref{aff9}}
\and M.~Farina\orcid{0000-0002-3089-7846}\inst{\ref{aff18}}
\and R.~Farinelli\inst{\ref{aff24}}
\and S.~Ferriol\inst{\ref{aff52}}
\and F.~Finelli\orcid{0000-0002-6694-3269}\inst{\ref{aff24},\ref{aff61}}
\and P.~Fosalba\orcid{0000-0002-1510-5214}\inst{\ref{aff5},\ref{aff4}}
\and N.~Fourmanoit\orcid{0009-0005-6816-6925}\inst{\ref{aff60}}
\and M.~Frailis\orcid{0000-0002-7400-2135}\inst{\ref{aff27}}
\and E.~Franceschi\orcid{0000-0002-0585-6591}\inst{\ref{aff24}}
\and M.~Fumana\orcid{0000-0001-6787-5950}\inst{\ref{aff41}}
\and S.~Galeotta\orcid{0000-0002-3748-5115}\inst{\ref{aff27}}
\and K.~George\orcid{0000-0002-1734-8455}\inst{\ref{aff62}}
\and B.~Gillis\orcid{0000-0002-4478-1270}\inst{\ref{aff39}}
\and C.~Giocoli\orcid{0000-0002-9590-7961}\inst{\ref{aff24},\ref{aff31}}
\and J.~Gracia-Carpio\inst{\ref{aff2}}
\and A.~Grazian\orcid{0000-0002-5688-0663}\inst{\ref{aff32}}
\and F.~Grupp\inst{\ref{aff2},\ref{aff9}}
\and S.~Gwyn\orcid{0000-0001-8221-8406}\inst{\ref{aff63}}
\and S.~V.~H.~Haugan\orcid{0000-0001-9648-7260}\inst{\ref{aff64}}
\and H.~Hoekstra\orcid{0000-0002-0641-3231}\inst{\ref{aff40}}
\and W.~Holmes\inst{\ref{aff65}}
\and F.~Hormuth\inst{\ref{aff66}}
\and A.~Hornstrup\orcid{0000-0002-3363-0936}\inst{\ref{aff67},\ref{aff68}}
\and K.~Jahnke\orcid{0000-0003-3804-2137}\inst{\ref{aff69}}
\and M.~Jhabvala\inst{\ref{aff70}}
\and B.~Joachimi\orcid{0000-0001-7494-1303}\inst{\ref{aff71}}
\and S.~Kermiche\orcid{0000-0002-0302-5735}\inst{\ref{aff60}}
\and A.~Kiessling\orcid{0000-0002-2590-1273}\inst{\ref{aff65}}
\and B.~Kubik\orcid{0009-0006-5823-4880}\inst{\ref{aff52}}
\and M.~K\"ummel\orcid{0000-0003-2791-2117}\inst{\ref{aff9}}
\and M.~Kunz\orcid{0000-0002-3052-7394}\inst{\ref{aff72}}
\and H.~Kurki-Suonio\orcid{0000-0002-4618-3063}\inst{\ref{aff73},\ref{aff74}}
\and A.~M.~C.~Le~Brun\orcid{0000-0002-0936-4594}\inst{\ref{aff75}}
\and S.~Ligori\orcid{0000-0003-4172-4606}\inst{\ref{aff38}}
\and P.~B.~Lilje\orcid{0000-0003-4324-7794}\inst{\ref{aff64}}
\and V.~Lindholm\orcid{0000-0003-2317-5471}\inst{\ref{aff73},\ref{aff74}}
\and I.~Lloro\orcid{0000-0001-5966-1434}\inst{\ref{aff76}}
\and G.~Mainetti\orcid{0000-0003-2384-2377}\inst{\ref{aff77}}
\and D.~Maino\inst{\ref{aff78},\ref{aff41},\ref{aff79}}
\and E.~Maiorano\orcid{0000-0003-2593-4355}\inst{\ref{aff24}}
\and O.~Mansutti\orcid{0000-0001-5758-4658}\inst{\ref{aff27}}
\and S.~Marcin\inst{\ref{aff80}}
\and O.~Marggraf\orcid{0000-0001-7242-3852}\inst{\ref{aff81}}
\and M.~Martinelli\orcid{0000-0002-6943-7732}\inst{\ref{aff3},\ref{aff82}}
\and N.~Martinet\orcid{0000-0003-2786-7790}\inst{\ref{aff11}}
\and F.~Marulli\orcid{0000-0002-8850-0303}\inst{\ref{aff83},\ref{aff24},\ref{aff31}}
\and R.~J.~Massey\orcid{0000-0002-6085-3780}\inst{\ref{aff84}}
\and E.~Medinaceli\orcid{0000-0002-4040-7783}\inst{\ref{aff24}}
\and S.~Mei\orcid{0000-0002-2849-559X}\inst{\ref{aff85},\ref{aff86}}
\and Y.~Mellier\thanks{Deceased}\inst{\ref{aff87},\ref{aff88}}
\and M.~Meneghetti\orcid{0000-0003-1225-7084}\inst{\ref{aff24},\ref{aff31}}
\and E.~Merlin\orcid{0000-0001-6870-8900}\inst{\ref{aff3}}
\and G.~Meylan\inst{\ref{aff89}}
\and A.~Mora\orcid{0000-0002-1922-8529}\inst{\ref{aff90}}
\and M.~Moresco\orcid{0000-0002-7616-7136}\inst{\ref{aff83},\ref{aff24}}
\and L.~Moscardini\orcid{0000-0002-3473-6716}\inst{\ref{aff83},\ref{aff24},\ref{aff31}}
\and R.~Nakajima\orcid{0009-0009-1213-7040}\inst{\ref{aff81}}
\and C.~Neissner\orcid{0000-0001-8524-4968}\inst{\ref{aff91},\ref{aff43}}
\and R.~C.~Nichol\orcid{0000-0003-0939-6518}\inst{\ref{aff92}}
\and S.-M.~Niemi\orcid{0009-0005-0247-0086}\inst{\ref{aff93}}
\and C.~Padilla\orcid{0000-0001-7951-0166}\inst{\ref{aff91}}
\and F.~Pasian\orcid{0000-0002-4869-3227}\inst{\ref{aff27}}
\and K.~Pedersen\inst{\ref{aff94}}
\and W.~J.~Percival\orcid{0000-0002-0644-5727}\inst{\ref{aff95},\ref{aff96},\ref{aff97}}
\and V.~Pettorino\orcid{0000-0002-4203-9320}\inst{\ref{aff93}}
\and S.~Pires\orcid{0000-0002-0249-2104}\inst{\ref{aff98}}
\and G.~Polenta\orcid{0000-0003-4067-9196}\inst{\ref{aff99}}
\and M.~Poncet\inst{\ref{aff100}}
\and L.~A.~Popa\inst{\ref{aff101}}
\and L.~Pozzetti\orcid{0000-0001-7085-0412}\inst{\ref{aff24}}
\and G.~D.~Racca\orcid{0000-0002-9883-8981}\inst{\ref{aff40},\ref{aff93}}
\and F.~Raison\orcid{0000-0002-7819-6918}\inst{\ref{aff2}}
\and A.~Renzi\orcid{0000-0001-9856-1970}\inst{\ref{aff102},\ref{aff103}}
\and J.~Rhodes\orcid{0000-0002-4485-8549}\inst{\ref{aff65}}
\and G.~Riccio\inst{\ref{aff1}}
\and E.~Romelli\orcid{0000-0003-3069-9222}\inst{\ref{aff27}}
\and M.~Roncarelli\orcid{0000-0001-9587-7822}\inst{\ref{aff24}}
\and R.~Saglia\orcid{0000-0003-0378-7032}\inst{\ref{aff9},\ref{aff2}}
\and Z.~Sakr\orcid{0000-0002-4823-3757}\inst{\ref{aff104},\ref{aff105},\ref{aff106}}
\and D.~Sapone\orcid{0000-0001-7089-4503}\inst{\ref{aff107}}
\and B.~Sartoris\orcid{0000-0003-1337-5269}\inst{\ref{aff9},\ref{aff27}}
\and P.~Schneider\orcid{0000-0001-8561-2679}\inst{\ref{aff81}}
\and T.~Schrabback\orcid{0000-0002-6987-7834}\inst{\ref{aff108}}
\and M.~Scodeggio\inst{\ref{aff41}}
\and A.~Secroun\orcid{0000-0003-0505-3710}\inst{\ref{aff60}}
\and G.~Seidel\orcid{0000-0003-2907-353X}\inst{\ref{aff69}}
\and S.~Serrano\orcid{0000-0002-0211-2861}\inst{\ref{aff5},\ref{aff109},\ref{aff4}}
\and E.~Sihvola\orcid{0000-0003-1804-7715}\inst{\ref{aff21}}
\and P.~Simon\inst{\ref{aff81}}
\and C.~Sirignano\orcid{0000-0002-0995-7146}\inst{\ref{aff102},\ref{aff103}}
\and G.~Sirri\orcid{0000-0003-2626-2853}\inst{\ref{aff31}}
\and J.~Steinwagner\orcid{0000-0001-7443-1047}\inst{\ref{aff2}}
\and P.~Tallada-Cresp\'{i}\orcid{0000-0002-1336-8328}\inst{\ref{aff42},\ref{aff43}}
\and A.~N.~Taylor\inst{\ref{aff39}}
\and I.~Tereno\orcid{0000-0002-4537-6218}\inst{\ref{aff58},\ref{aff110}}
\and N.~Tessore\orcid{0000-0002-9696-7931}\inst{\ref{aff57}}
\and S.~Toft\orcid{0000-0003-3631-7176}\inst{\ref{aff111},\ref{aff112}}
\and R.~Toledo-Moreo\orcid{0000-0002-2997-4859}\inst{\ref{aff113}}
\and F.~Torradeflot\orcid{0000-0003-1160-1517}\inst{\ref{aff43},\ref{aff42}}
\and I.~Tutusaus\orcid{0000-0002-3199-0399}\inst{\ref{aff4},\ref{aff5},\ref{aff105}}
\and L.~Valenziano\orcid{0000-0002-1170-0104}\inst{\ref{aff24},\ref{aff61}}
\and J.~Valiviita\orcid{0000-0001-6225-3693}\inst{\ref{aff73},\ref{aff74}}
\and T.~Vassallo\orcid{0000-0001-6512-6358}\inst{\ref{aff27}}
\and A.~Veropalumbo\orcid{0000-0003-2387-1194}\inst{\ref{aff25},\ref{aff34},\ref{aff33}}
\and Y.~Wang\orcid{0000-0002-4749-2984}\inst{\ref{aff114}}
\and J.~Weller\orcid{0000-0002-8282-2010}\inst{\ref{aff9},\ref{aff2}}
\and A.~Zacchei\orcid{0000-0003-0396-1192}\inst{\ref{aff27},\ref{aff26}}
\and E.~Zucca\orcid{0000-0002-5845-8132}\inst{\ref{aff24}}
\and M.~Ballardini\orcid{0000-0003-4481-3559}\inst{\ref{aff115},\ref{aff116},\ref{aff24}}
\and M.~Bolzonella\orcid{0000-0003-3278-4607}\inst{\ref{aff24}}
\and E.~Bozzo\orcid{0000-0002-8201-1525}\inst{\ref{aff7}}
\and C.~Burigana\orcid{0000-0002-3005-5796}\inst{\ref{aff117},\ref{aff61}}
\and R.~Cabanac\orcid{0000-0001-6679-2600}\inst{\ref{aff105}}
\and M.~Calabrese\orcid{0000-0002-2637-2422}\inst{\ref{aff118},\ref{aff41}}
\and A.~Cappi\inst{\ref{aff119},\ref{aff24}}
\and T.~Castro\orcid{0000-0002-6292-3228}\inst{\ref{aff27},\ref{aff28},\ref{aff26},\ref{aff120}}
\and J.~A.~Escartin~Vigo\inst{\ref{aff2}}
\and L.~Gabarra\orcid{0000-0002-8486-8856}\inst{\ref{aff121}}
\and J.~Garc\'ia-Bellido\orcid{0000-0002-9370-8360}\inst{\ref{aff122}}
\and R.~Maoli\orcid{0000-0002-6065-3025}\inst{\ref{aff123},\ref{aff3}}
\and J.~Mart\'{i}n-Fleitas\orcid{0000-0002-8594-569X}\inst{\ref{aff124}}
\and M.~Maturi\orcid{0000-0002-3517-2422}\inst{\ref{aff104},\ref{aff125}}
\and N.~Mauri\orcid{0000-0001-8196-1548}\inst{\ref{aff48},\ref{aff31}}
\and R.~B.~Metcalf\orcid{0000-0003-3167-2574}\inst{\ref{aff83},\ref{aff24}}
\and A.~Pezzotta\orcid{0000-0003-0726-2268}\inst{\ref{aff25}}
\and M.~P\"ontinen\orcid{0000-0001-5442-2530}\inst{\ref{aff73}}
\and C.~Porciani\orcid{0000-0002-7797-2508}\inst{\ref{aff81}}
\and I.~Risso\orcid{0000-0003-2525-7761}\inst{\ref{aff25},\ref{aff34}}
\and V.~Scottez\orcid{0009-0008-3864-940X}\inst{\ref{aff87},\ref{aff126}}
\and M.~Sereno\orcid{0000-0003-0302-0325}\inst{\ref{aff24},\ref{aff31}}
\and M.~Tenti\orcid{0000-0002-4254-5901}\inst{\ref{aff31}}
\and M.~Viel\orcid{0000-0002-2642-5707}\inst{\ref{aff26},\ref{aff27},\ref{aff29},\ref{aff28},\ref{aff120}}
\and M.~Wiesmann\orcid{0009-0000-8199-5860}\inst{\ref{aff64}}
\and Y.~Akrami\orcid{0000-0002-2407-7956}\inst{\ref{aff122},\ref{aff127}}
\and I.~T.~Andika\orcid{0000-0001-6102-9526}\inst{\ref{aff62},\ref{aff128}}
\and G.~Angora\orcid{0000-0002-0316-6562}\inst{\ref{aff1},\ref{aff115}}
\and S.~Anselmi\orcid{0000-0002-3579-9583}\inst{\ref{aff103},\ref{aff102},\ref{aff129}}
\and M.~Archidiacono\orcid{0000-0003-4952-9012}\inst{\ref{aff78},\ref{aff79}}
\and F.~Atrio-Barandela\orcid{0000-0002-2130-2513}\inst{\ref{aff130}}
\and L.~Bazzanini\orcid{0000-0003-0727-0137}\inst{\ref{aff115},\ref{aff24}}
\and D.~Bertacca\orcid{0000-0002-2490-7139}\inst{\ref{aff102},\ref{aff32},\ref{aff103}}
\and M.~Bethermin\orcid{0000-0002-3915-2015}\inst{\ref{aff131}}
\and L.~Bisigello\orcid{0000-0003-0492-4924}\inst{\ref{aff32}}
\and A.~Blanchard\orcid{0000-0001-8555-9003}\inst{\ref{aff105}}
\and L.~Blot\orcid{0000-0002-9622-7167}\inst{\ref{aff132},\ref{aff75}}
\and M.~Bonici\orcid{0000-0002-8430-126X}\inst{\ref{aff95},\ref{aff41}}
\and M.~L.~Brown\orcid{0000-0002-0370-8077}\inst{\ref{aff49}}
\and S.~Bruton\orcid{0000-0002-6503-5218}\inst{\ref{aff133}}
\and A.~Calabro\orcid{0000-0003-2536-1614}\inst{\ref{aff3}}
\and B.~Camacho~Quevedo\orcid{0000-0002-8789-4232}\inst{\ref{aff26},\ref{aff29},\ref{aff27}}
\and F.~Caro\inst{\ref{aff3}}
\and C.~S.~Carvalho\inst{\ref{aff110}}
\and F.~Cogato\orcid{0000-0003-4632-6113}\inst{\ref{aff83},\ref{aff24}}
\and S.~Conseil\orcid{0000-0002-3657-4191}\inst{\ref{aff52}}
\and A.~R.~Cooray\orcid{0000-0002-3892-0190}\inst{\ref{aff134}}
\and S.~Davini\orcid{0000-0003-3269-1718}\inst{\ref{aff34}}
\and F.~De~Paolis\orcid{0000-0001-6460-7563}\inst{\ref{aff135},\ref{aff136},\ref{aff137}}
\and G.~Desprez\orcid{0000-0001-8325-1742}\inst{\ref{aff23}}
\and A.~D\'iaz-S\'anchez\orcid{0000-0003-0748-4768}\inst{\ref{aff138}}
\and S.~Di~Domizio\orcid{0000-0003-2863-5895}\inst{\ref{aff33},\ref{aff34}}
\and J.~M.~Diego\orcid{0000-0001-9065-3926}\inst{\ref{aff139}}
\and M.~Y.~Elkhashab\orcid{0000-0001-9306-2603}\inst{\ref{aff27},\ref{aff28},\ref{aff140},\ref{aff26}}
\and A.~Enia\orcid{0000-0002-0200-2857}\inst{\ref{aff24}}
\and Y.~Fang\orcid{0000-0002-0334-6950}\inst{\ref{aff9}}
\and A.~Finoguenov\orcid{0000-0002-4606-5403}\inst{\ref{aff73}}
\and F.~Fontanot\orcid{0000-0003-4744-0188}\inst{\ref{aff27},\ref{aff26}}
\and A.~Franco\orcid{0000-0002-4761-366X}\inst{\ref{aff136},\ref{aff135},\ref{aff137}}
\and T.~Gasparetto\orcid{0000-0002-7913-4866}\inst{\ref{aff3}}
\and E.~Gaztanaga\orcid{0000-0001-9632-0815}\inst{\ref{aff4},\ref{aff5},\ref{aff141}}
\and F.~Giacomini\orcid{0000-0002-3129-2814}\inst{\ref{aff31}}
\and F.~Gianotti\orcid{0000-0003-4666-119X}\inst{\ref{aff24}}
\and G.~Gozaliasl\orcid{0000-0002-0236-919X}\inst{\ref{aff142},\ref{aff73}}
\and M.~Guidi\orcid{0000-0001-9408-1101}\inst{\ref{aff30},\ref{aff24}}
\and C.~M.~Gutierrez\orcid{0000-0001-7854-783X}\inst{\ref{aff143}}
\and A.~Hall\orcid{0000-0002-3139-8651}\inst{\ref{aff39}}
\and C.~Hern\'andez-Monteagudo\orcid{0000-0001-5471-9166}\inst{\ref{aff15},\ref{aff14}}
\and H.~Hildebrandt\orcid{0000-0002-9814-3338}\inst{\ref{aff144}}
\and J.~Hjorth\orcid{0000-0002-4571-2306}\inst{\ref{aff94}}
\and L.~K.~Hunt\orcid{0000-0001-9162-2371}\inst{\ref{aff145}}
\and J.~J.~E.~Kajava\orcid{0000-0002-3010-8333}\inst{\ref{aff146},\ref{aff147},\ref{aff148}}
\and Y.~Kang\orcid{0009-0000-8588-7250}\inst{\ref{aff7}}
\and V.~Kansal\orcid{0000-0002-4008-6078}\inst{\ref{aff149},\ref{aff150}}
\and D.~Karagiannis\orcid{0000-0002-4927-0816}\inst{\ref{aff115},\ref{aff151}}
\and K.~Kiiveri\inst{\ref{aff21}}
\and J.~Kim\orcid{0000-0003-2776-2761}\inst{\ref{aff121}}
\and C.~C.~Kirkpatrick\inst{\ref{aff21}}
\and S.~Kruk\orcid{0000-0001-8010-8879}\inst{\ref{aff51}}
\and M.~Lattanzi\orcid{0000-0003-1059-2532}\inst{\ref{aff116}}
\and L.~Legrand\orcid{0000-0003-0610-5252}\inst{\ref{aff152},\ref{aff153}}
\and F.~Lepori\orcid{0009-0000-5061-7138}\inst{\ref{aff154}}
\and G.~Leroy\orcid{0009-0004-2523-4425}\inst{\ref{aff155},\ref{aff84}}
\and G.~F.~Lesci\orcid{0000-0002-4607-2830}\inst{\ref{aff83},\ref{aff24}}
\and J.~Lesgourgues\orcid{0000-0001-7627-353X}\inst{\ref{aff44}}
\and L.~Leuzzi\orcid{0009-0006-4479-7017}\inst{\ref{aff24}}
\and T.~I.~Liaudat\orcid{0000-0002-9104-314X}\inst{\ref{aff156}}
\and S.~J.~Liu\orcid{0000-0001-7680-2139}\inst{\ref{aff18}}
\and X.~Lopez~Lopez\orcid{0009-0008-5194-5908}\inst{\ref{aff24}}
\and J.~Macias-Perez\orcid{0000-0002-5385-2763}\inst{\ref{aff157}}
\and M.~Magliocchetti\orcid{0000-0001-9158-4838}\inst{\ref{aff18}}
\and E.~A.~Magnier\orcid{0000-0002-7965-2815}\inst{\ref{aff47}}
\and C.~Mancini\orcid{0000-0002-4297-0561}\inst{\ref{aff41}}
\and A.~Manj\'on-Garc\'ia\orcid{0000-0002-7413-8825}\inst{\ref{aff138}}
\and F.~Mannucci\orcid{0000-0002-4803-2381}\inst{\ref{aff145}}
\and C.~J.~A.~P.~Martins\orcid{0000-0002-4886-9261}\inst{\ref{aff158},\ref{aff159}}
\and L.~Maurin\orcid{0000-0002-8406-0857}\inst{\ref{aff17}}
\and M.~Miluzio\inst{\ref{aff51},\ref{aff160}}
\and P.~Monaco\orcid{0000-0003-2083-7564}\inst{\ref{aff140},\ref{aff27},\ref{aff28},\ref{aff26}}
\and A.~Montoro\orcid{0000-0003-4730-8590}\inst{\ref{aff4},\ref{aff5}}
\and C.~Moretti\orcid{0000-0003-3314-8936}\inst{\ref{aff27},\ref{aff26},\ref{aff28}}
\and G.~Morgante\inst{\ref{aff24}}
\and C.~Murray\inst{\ref{aff85}}
\and S.~Nadathur\orcid{0000-0001-9070-3102}\inst{\ref{aff141}}
\and K.~Naidoo\orcid{0000-0002-9182-1802}\inst{\ref{aff141},\ref{aff69}}
\and P.~Natoli\orcid{0000-0003-0126-9100}\inst{\ref{aff115},\ref{aff116}}
\and A.~Navarro-Alsina\orcid{0000-0002-3173-2592}\inst{\ref{aff81}}
\and S.~Nesseris\orcid{0000-0002-0567-0324}\inst{\ref{aff122}}
\and D.~Paoletti\orcid{0000-0003-4761-6147}\inst{\ref{aff24},\ref{aff61}}
\and F.~Passalacqua\orcid{0000-0002-8606-4093}\inst{\ref{aff102},\ref{aff103}}
\and K.~Paterson\orcid{0000-0001-8340-3486}\inst{\ref{aff69}}
\and A.~Pisani\orcid{0000-0002-6146-4437}\inst{\ref{aff60}}
\and D.~Potter\orcid{0000-0002-0757-5195}\inst{\ref{aff154}}
\and G.~W.~Pratt\inst{\ref{aff98}}
\and S.~Quai\orcid{0000-0002-0449-8163}\inst{\ref{aff83},\ref{aff24}}
\and M.~Radovich\orcid{0000-0002-3585-866X}\inst{\ref{aff32}}
\and G.~Rodighiero\orcid{0000-0002-9415-2296}\inst{\ref{aff102},\ref{aff32}}
\and K.~Rojas\orcid{0000-0003-1391-6854}\inst{\ref{aff80}}
\and S.~Sacquegna\orcid{0000-0002-8433-6630}\inst{\ref{aff161}}
\and M.~Sahl\'en\orcid{0000-0003-0973-4804}\inst{\ref{aff162}}
\and D.~B.~Sanders\orcid{0000-0002-1233-9998}\inst{\ref{aff47}}
\and E.~Sarpa\orcid{0000-0002-1256-655X}\inst{\ref{aff29},\ref{aff120},\ref{aff28}}
\and A.~Schneider\orcid{0000-0001-7055-8104}\inst{\ref{aff154}}
\and D.~Sciotti\orcid{0009-0008-4519-2620}\inst{\ref{aff3},\ref{aff82}}
\and E.~Sellentin\inst{\ref{aff163},\ref{aff40}}
\and L.~C.~Smith\orcid{0000-0002-3259-2771}\inst{\ref{aff164}}
\and K.~Tanidis\orcid{0000-0001-9843-5130}\inst{\ref{aff121}}
\and C.~Tao\orcid{0000-0001-7961-8177}\inst{\ref{aff60}}
\and G.~Testera\inst{\ref{aff34}}
\and R.~Teyssier\orcid{0000-0001-7689-0933}\inst{\ref{aff165}}
\and S.~Tosi\orcid{0000-0002-7275-9193}\inst{\ref{aff33},\ref{aff34},\ref{aff25}}
\and A.~Troja\orcid{0000-0003-0239-4595}\inst{\ref{aff102},\ref{aff103}}
\and M.~Tucci\inst{\ref{aff7}}
\and A.~Venhola\orcid{0000-0001-6071-4564}\inst{\ref{aff166}}
\and D.~Vergani\orcid{0000-0003-0898-2216}\inst{\ref{aff24}}
\and G.~Verza\orcid{0000-0002-1886-8348}\inst{\ref{aff167},\ref{aff168}}
\and P.~Vielzeuf\orcid{0000-0003-2035-9339}\inst{\ref{aff60}}
\and S.~Vinciguerra\orcid{0009-0005-4018-3184}\inst{\ref{aff11}}
\and N.~A.~Walton\orcid{0000-0003-3983-8778}\inst{\ref{aff164}}}
										   
%%%% please do not edit the affiliation list -- contact ECEB Bureau for changes
\institute{INAF-Osservatorio Astronomico di Capodimonte, Via Moiariello 16, 80131 Napoli, Italy\label{aff1}
\and
Max Planck Institute for Extraterrestrial Physics, Giessenbachstr. 1, 85748 Garching, Germany\label{aff2}
\and
INAF-Osservatorio Astronomico di Roma, Via Frascati 33, 00078 Monteporzio Catone, Italy\label{aff3}
\and
Institute of Space Sciences (ICE, CSIC), Campus UAB, Carrer de Can Magrans, s/n, 08193 Barcelona, Spain\label{aff4}
\and
Institut d'Estudis Espacials de Catalunya (IEEC),  Edifici RDIT, Campus UPC, 08860 Castelldefels, Barcelona, Spain\label{aff5}
\and
School of Physics, HH Wills Physics Laboratory, University of Bristol, Tyndall Avenue, Bristol, BS8 1TL, UK\label{aff6}
\and
Department of Astronomy, University of Geneva, ch. d'Ecogia 16, 1290 Versoix, Switzerland\label{aff7}
\and
Institute for Particle Physics and Astrophysics, Dept. of Physics, ETH Zurich, Wolfgang-Pauli-Strasse 27, 8093 Zurich, Switzerland\label{aff8}
\and
Universit\"ats-Sternwarte M\"unchen, Fakult\"at f\"ur Physik, Ludwig-Maximilians-Universit\"at M\"unchen, Scheinerstr.~1, 81679 M\"unchen, Germany\label{aff9}
\and
Department of Mathematics and Physics, Roma Tre University, Via della Vasca Navale 84, 00146 Rome, Italy\label{aff10}
\and
Aix-Marseille Universit\'e, CNRS, CNES, LAM, Marseille, France\label{aff11}
\and
Department of Physics and Astronomy, University of British Columbia, Vancouver, BC V6T 1Z1, Canada\label{aff12}
\and
School of Physics \& Astronomy, University of Southampton, Highfield Campus, Southampton SO17 1BJ, UK\label{aff13}
\and
Instituto de Astrof\'{\i}sica de Canarias, E-38205 La Laguna, Tenerife, Spain\label{aff14}
\and
Universidad de La Laguna, Dpto. Astrof\'\i sica, E-38206 La Laguna, Tenerife, Spain\label{aff15}
\and
Univ. Lille, CNRS, Centrale Lille, UMR 9189 CRIStAL, 59000 Lille, France\label{aff16}
\and
Universit\'e Paris-Saclay, CNRS, Institut d'astrophysique spatiale, 91405, Orsay, France\label{aff17}
\and
INAF-Istituto di Astrofisica e Planetologia Spaziali, via del Fosso del Cavaliere, 100, 00100 Roma, Italy\label{aff18}
\and
Department of Physical Sciences, Ritsumeikan University, Kusatsu, Shiga 525-8577, Japan\label{aff19}
\and
Academia Sinica Institute of Astronomy and Astrophysics (ASIAA), 11F of ASMAB, No.~1, Section 4, Roosevelt Road, Taipei 10617, Taiwan\label{aff20}
\and
Department of Physics and Helsinki Institute of Physics, Gustaf H\"allstr\"omin katu 2, University of Helsinki, 00014 Helsinki, Finland\label{aff21}
\and
SRON Netherlands Institute for Space Research, Landleven 12, 9747 AD, Groningen, The Netherlands\label{aff22}
\and
Kapteyn Astronomical Institute, University of Groningen, PO Box 800, 9700 AV Groningen, The Netherlands\label{aff23}
\and
INAF-Osservatorio di Astrofisica e Scienza dello Spazio di Bologna, Via Piero Gobetti 93/3, 40129 Bologna, Italy\label{aff24}
\and
INAF-Osservatorio Astronomico di Brera, Via Brera 28, 20122 Milano, Italy\label{aff25}
\and
IFPU, Institute for Fundamental Physics of the Universe, via Beirut 2, 34151 Trieste, Italy\label{aff26}
\and
INAF-Osservatorio Astronomico di Trieste, Via G. B. Tiepolo 11, 34143 Trieste, Italy\label{aff27}
\and
INFN, Sezione di Trieste, Via Valerio 2, 34127 Trieste TS, Italy\label{aff28}
\and
SISSA, International School for Advanced Studies, Via Bonomea 265, 34136 Trieste TS, Italy\label{aff29}
\and
Dipartimento di Fisica e Astronomia, Universit\`a di Bologna, Via Gobetti 93/2, 40129 Bologna, Italy\label{aff30}
\and
INFN-Sezione di Bologna, Viale Berti Pichat 6/2, 40127 Bologna, Italy\label{aff31}
\and
INAF-Osservatorio Astronomico di Padova, Via dell'Osservatorio 5, 35122 Padova, Italy\label{aff32}
\and
Dipartimento di Fisica, Universit\`a di Genova, Via Dodecaneso 33, 16146, Genova, Italy\label{aff33}
\and
INFN-Sezione di Genova, Via Dodecaneso 33, 16146, Genova, Italy\label{aff34}
\and
Department of Physics "E. Pancini", University Federico II, Via Cinthia 6, 80126, Napoli, Italy\label{aff35}
\and
Dipartimento di Fisica, Universit\`a degli Studi di Torino, Via P. Giuria 1, 10125 Torino, Italy\label{aff36}
\and
INFN-Sezione di Torino, Via P. Giuria 1, 10125 Torino, Italy\label{aff37}
\and
INAF-Osservatorio Astrofisico di Torino, Via Osservatorio 20, 10025 Pino Torinese (TO), Italy\label{aff38}
\and
Institute for Astronomy, University of Edinburgh, Royal Observatory, Blackford Hill, Edinburgh EH9 3HJ, UK\label{aff39}
\and
Leiden Observatory, Leiden University, Einsteinweg 55, 2333 CC Leiden, The Netherlands\label{aff40}
\and
INAF-IASF Milano, Via Alfonso Corti 12, 20133 Milano, Italy\label{aff41}
\and
Centro de Investigaciones Energ\'eticas, Medioambientales y Tecnol\'ogicas (CIEMAT), Avenida Complutense 40, 28040 Madrid, Spain\label{aff42}
\and
Port d'Informaci\'{o} Cient\'{i}fica, Campus UAB, C. Albareda s/n, 08193 Bellaterra (Barcelona), Spain\label{aff43}
\and
Institute for Theoretical Particle Physics and Cosmology (TTK), RWTH Aachen University, 52056 Aachen, Germany\label{aff44}
\and
Deutsches Zentrum f\"ur Luft- und Raumfahrt e. V. (DLR), Linder H\"ohe, 51147 K\"oln, Germany\label{aff45}
\and
INFN section of Naples, Via Cinthia 6, 80126, Napoli, Italy\label{aff46}
\and
Institute for Astronomy, University of Hawaii, 2680 Woodlawn Drive, Honolulu, HI 96822, USA\label{aff47}
\and
Dipartimento di Fisica e Astronomia "Augusto Righi" - Alma Mater Studiorum Universit\`a di Bologna, Viale Berti Pichat 6/2, 40127 Bologna, Italy\label{aff48}
\and
Jodrell Bank Centre for Astrophysics, Department of Physics and Astronomy, University of Manchester, Oxford Road, Manchester M13 9PL, UK\label{aff49}
\and
European Space Agency/ESRIN, Largo Galileo Galilei 1, 00044 Frascati, Roma, Italy\label{aff50}
\and
ESAC/ESA, Camino Bajo del Castillo, s/n., Urb. Villafranca del Castillo, 28692 Villanueva de la Ca\~nada, Madrid, Spain\label{aff51}
\and
Universit\'e Claude Bernard Lyon 1, CNRS/IN2P3, IP2I Lyon, UMR 5822, Villeurbanne, F-69100, France\label{aff52}
\and
Institut de Ci\`{e}ncies del Cosmos (ICCUB), Universitat de Barcelona (IEEC-UB), Mart\'{i} i Franqu\`{e}s 1, 08028 Barcelona, Spain\label{aff53}
\and
Instituci\'o Catalana de Recerca i Estudis Avan\c{c}ats (ICREA), Passeig de Llu\'{\i}s Companys 23, 08010 Barcelona, Spain\label{aff54}
\and
Institut de Ciencies de l'Espai (IEEC-CSIC), Campus UAB, Carrer de Can Magrans, s/n Cerdanyola del Vall\'es, 08193 Barcelona, Spain\label{aff55}
\and
UCB Lyon 1, CNRS/IN2P3, IUF, IP2I Lyon, 4 rue Enrico Fermi, 69622 Villeurbanne, France\label{aff56}
\and
Mullard Space Science Laboratory, University College London, Holmbury St Mary, Dorking, Surrey RH5 6NT, UK\label{aff57}
\and
Departamento de F\'isica, Faculdade de Ci\^encias, Universidade de Lisboa, Edif\'icio C8, Campo Grande, PT1749-016 Lisboa, Portugal\label{aff58}
\and
Instituto de Astrof\'isica e Ci\^encias do Espa\c{c}o, Faculdade de Ci\^encias, Universidade de Lisboa, Campo Grande, 1749-016 Lisboa, Portugal\label{aff59}
\and
Aix-Marseille Universit\'e, CNRS/IN2P3, CPPM, Marseille, France\label{aff60}
\and
INFN-Bologna, Via Irnerio 46, 40126 Bologna, Italy\label{aff61}
\and
University Observatory, LMU Faculty of Physics, Scheinerstr.~1, 81679 Munich, Germany\label{aff62}
\and
Herzberg Astronomy and Astrophysics Research Centre, 5071 W. Saanich Rd. Victoria, BC, V9E 2E7, Canada\label{aff63}
\and
Institute of Theoretical Astrophysics, University of Oslo, P.O. Box 1029 Blindern, 0315 Oslo, Norway\label{aff64}
\and
Jet Propulsion Laboratory, California Institute of Technology, 4800 Oak Grove Drive, Pasadena, CA, 91109, USA\label{aff65}
\and
Felix Hormuth Engineering, Goethestr. 17, 69181 Leimen, Germany\label{aff66}
\and
Technical University of Denmark, Elektrovej 327, 2800 Kgs. Lyngby, Denmark\label{aff67}
\and
Cosmic Dawn Center (DAWN), Denmark\label{aff68}
\and
Max-Planck-Institut f\"ur Astronomie, K\"onigstuhl 17, 69117 Heidelberg, Germany\label{aff69}
\and
NASA Goddard Space Flight Center, Greenbelt, MD 20771, USA\label{aff70}
\and
Department of Physics and Astronomy, University College London, Gower Street, London WC1E 6BT, UK\label{aff71}
\and
Universit\'e de Gen\`eve, D\'epartement de Physique Th\'eorique and Centre for Astroparticle Physics, 24 quai Ernest-Ansermet, CH-1211 Gen\`eve 4, Switzerland\label{aff72}
\and
Department of Physics, P.O. Box 64, University of Helsinki, 00014 Helsinki, Finland\label{aff73}
\and
Helsinki Institute of Physics, Gustaf H{\"a}llstr{\"o}min katu 2, University of Helsinki, 00014 Helsinki, Finland\label{aff74}
\and
Laboratoire d'etude de l'Univers et des phenomenes eXtremes, Observatoire de Paris, Universit\'e PSL, Sorbonne Universit\'e, CNRS, 92190 Meudon, France\label{aff75}
\and
SKAO, Jodrell Bank, Lower Withington, Macclesfield SK11 9FT, UK\label{aff76}
\and
Centre de Calcul de l'IN2P3/CNRS, 21 avenue Pierre de Coubertin 69627 Villeurbanne Cedex, France\label{aff77}
\and
Dipartimento di Fisica "Aldo Pontremoli", Universit\`a degli Studi di Milano, Via Celoria 16, 20133 Milano, Italy\label{aff78}
\and
INFN-Sezione di Milano, Via Celoria 16, 20133 Milano, Italy\label{aff79}
\and
University of Applied Sciences and Arts of Northwestern Switzerland, School of Computer Science, 5210 Windisch, Switzerland\label{aff80}
\and
Universit\"at Bonn, Argelander-Institut f\"ur Astronomie, Auf dem H\"ugel 71, 53121 Bonn, Germany\label{aff81}
\and
INFN-Sezione di Roma, Piazzale Aldo Moro, 2 - c/o Dipartimento di Fisica, Edificio G. Marconi, 00185 Roma, Italy\label{aff82}
\and
Dipartimento di Fisica e Astronomia "Augusto Righi" - Alma Mater Studiorum Universit\`a di Bologna, via Piero Gobetti 93/2, 40129 Bologna, Italy\label{aff83}
\and
Department of Physics, Institute for Computational Cosmology, Durham University, South Road, Durham, DH1 3LE, UK\label{aff84}
\and
Universit\'e Paris Cit\'e, CNRS, Astroparticule et Cosmologie, 75013 Paris, France\label{aff85}
\and
CNRS-UCB International Research Laboratory, Centre Pierre Bin\'etruy, IRL2007, CPB-IN2P3, Berkeley, USA\label{aff86}
\and
Institut d'Astrophysique de Paris, 98bis Boulevard Arago, 75014, Paris, France\label{aff87}
\and
Institut d'Astrophysique de Paris, UMR 7095, CNRS, and Sorbonne Universit\'e, 98 bis boulevard Arago, 75014 Paris, France\label{aff88}
\and
Institute of Physics, Laboratory of Astrophysics, Ecole Polytechnique F\'ed\'erale de Lausanne (EPFL), Observatoire de Sauverny, 1290 Versoix, Switzerland\label{aff89}
\and
Telespazio UK S.L. for European Space Agency (ESA), Camino bajo del Castillo, s/n, Urbanizacion Villafranca del Castillo, Villanueva de la Ca\~nada, 28692 Madrid, Spain\label{aff90}
\and
Institut de F\'{i}sica d'Altes Energies (IFAE), The Barcelona Institute of Science and Technology, Campus UAB, 08193 Bellaterra (Barcelona), Spain\label{aff91}
\and
School of Mathematics and Physics, University of Surrey, Guildford, Surrey, GU2 7XH, UK\label{aff92}
\and
European Space Agency/ESTEC, Keplerlaan 1, 2201 AZ Noordwijk, The Netherlands\label{aff93}
\and
DARK, Niels Bohr Institute, University of Copenhagen, Jagtvej 155, 2200 Copenhagen, Denmark\label{aff94}
\and
Waterloo Centre for Astrophysics, University of Waterloo, Waterloo, Ontario N2L 3G1, Canada\label{aff95}
\and
Department of Physics and Astronomy, University of Waterloo, Waterloo, Ontario N2L 3G1, Canada\label{aff96}
\and
Perimeter Institute for Theoretical Physics, Waterloo, Ontario N2L 2Y5, Canada\label{aff97}
\and
Universit\'e Paris-Saclay, Universit\'e Paris Cit\'e, CEA, CNRS, AIM, 91191, Gif-sur-Yvette, France\label{aff98}
\and
Space Science Data Center, Italian Space Agency, via del Politecnico snc, 00133 Roma, Italy\label{aff99}
\and
Centre National d'Etudes Spatiales -- Centre spatial de Toulouse, 18 avenue Edouard Belin, 31401 Toulouse Cedex 9, France\label{aff100}
\and
Institute of Space Science, Str. Atomistilor, nr. 409 M\u{a}gurele, Ilfov, 077125, Romania\label{aff101}
\and
Dipartimento di Fisica e Astronomia "G. Galilei", Universit\`a di Padova, Via Marzolo 8, 35131 Padova, Italy\label{aff102}
\and
INFN-Padova, Via Marzolo 8, 35131 Padova, Italy\label{aff103}
\and
Institut f\"ur Theoretische Physik, University of Heidelberg, Philosophenweg 16, 69120 Heidelberg, Germany\label{aff104}
\and
Institut de Recherche en Astrophysique et Plan\'etologie (IRAP), Universit\'e de Toulouse, CNRS, UPS, CNES, 14 Av. Edouard Belin, 31400 Toulouse, France\label{aff105}
\and
Universit\'e St Joseph; Faculty of Sciences, Beirut, Lebanon\label{aff106}
\and
Departamento de F\'isica, FCFM, Universidad de Chile, Blanco Encalada 2008, Santiago, Chile\label{aff107}
\and
Universit\"at Innsbruck, Institut f\"ur Astro- und Teilchenphysik, Technikerstr. 25/8, 6020 Innsbruck, Austria\label{aff108}
\and
Satlantis, University Science Park, Sede Bld 48940, Leioa-Bilbao, Spain\label{aff109}
\and
Instituto de Astrof\'isica e Ci\^encias do Espa\c{c}o, Faculdade de Ci\^encias, Universidade de Lisboa, Tapada da Ajuda, 1349-018 Lisboa, Portugal\label{aff110}
\and
Cosmic Dawn Center (DAWN)\label{aff111}
\and
Niels Bohr Institute, University of Copenhagen, Jagtvej 128, 2200 Copenhagen, Denmark\label{aff112}
\and
Universidad Polit\'ecnica de Cartagena, Departamento de Electr\'onica y Tecnolog\'ia de Computadoras,  Plaza del Hospital 1, 30202 Cartagena, Spain\label{aff113}
\and
Caltech/IPAC, 1200 E. California Blvd., Pasadena, CA 91125, USA\label{aff114}
\and
Dipartimento di Fisica e Scienze della Terra, Universit\`a degli Studi di Ferrara, Via Giuseppe Saragat 1, 44122 Ferrara, Italy\label{aff115}
\and
Istituto Nazionale di Fisica Nucleare, Sezione di Ferrara, Via Giuseppe Saragat 1, 44122 Ferrara, Italy\label{aff116}
\and
INAF, Istituto di Radioastronomia, Via Piero Gobetti 101, 40129 Bologna, Italy\label{aff117}
\and
Astronomical Observatory of the Autonomous Region of the Aosta Valley (OAVdA), Loc. Lignan 39, I-11020, Nus (Aosta Valley), Italy\label{aff118}
\and
Universit\'e C\^{o}te d'Azur, Observatoire de la C\^{o}te d'Azur, CNRS, Laboratoire Lagrange, Bd de l'Observatoire, CS 34229, 06304 Nice cedex 4, France\label{aff119}
\and
ICSC - Centro Nazionale di Ricerca in High Performance Computing, Big Data e Quantum Computing, Via Magnanelli 2, Bologna, Italy\label{aff120}
\and
Department of Physics, Oxford University, Keble Road, Oxford OX1 3RH, UK\label{aff121}
\and
Instituto de F\'isica Te\'orica UAM-CSIC, Campus de Cantoblanco, 28049 Madrid, Spain\label{aff122}
\and
Dipartimento di Fisica, Sapienza Universit\`a di Roma, Piazzale Aldo Moro 2, 00185 Roma, Italy\label{aff123}
\and
Aurora Technology for European Space Agency (ESA), Camino bajo del Castillo, s/n, Urbanizacion Villafranca del Castillo, Villanueva de la Ca\~nada, 28692 Madrid, Spain\label{aff124}
\and
Zentrum f\"ur Astronomie, Universit\"at Heidelberg, Philosophenweg 12, 69120 Heidelberg, Germany\label{aff125}
\and
ICL, Junia, Universit\'e Catholique de Lille, LITL, 59000 Lille, France\label{aff126}
\and
CERCA/ISO, Department of Physics, Case Western Reserve University, 10900 Euclid Avenue, Cleveland, OH 44106, USA\label{aff127}
\and
Technical University of Munich, TUM School of Natural Sciences, Physics Department, James-Franck-Str.~1, 85748 Garching, Germany\label{aff128}
\and
Laboratoire Univers et Th\'eorie, Observatoire de Paris, Universit\'e PSL, Universit\'e Paris Cit\'e, CNRS, 92190 Meudon, France\label{aff129}
\and
Departamento de F{\'\i}sica Fundamental. Universidad de Salamanca. Plaza de la Merced s/n. 37008 Salamanca, Spain\label{aff130}
\and
Universit\'e de Strasbourg, CNRS, Observatoire astronomique de Strasbourg, UMR 7550, 67000 Strasbourg, France\label{aff131}
\and
Center for Data-Driven Discovery, Kavli IPMU (WPI), UTIAS, The University of Tokyo, Kashiwa, Chiba 277-8583, Japan\label{aff132}
\and
California Institute of Technology, 1200 E California Blvd, Pasadena, CA 91125, USA\label{aff133}
\and
Department of Physics \& Astronomy, University of California Irvine, Irvine CA 92697, USA\label{aff134}
\and
Department of Mathematics and Physics E. De Giorgi, University of Salento, Via per Arnesano, CP-I93, 73100, Lecce, Italy\label{aff135}
\and
INFN, Sezione di Lecce, Via per Arnesano, CP-193, 73100, Lecce, Italy\label{aff136}
\and
INAF-Sezione di Lecce, c/o Dipartimento Matematica e Fisica, Via per Arnesano, 73100, Lecce, Italy\label{aff137}
\and
Departamento F\'isica Aplicada, Universidad Polit\'ecnica de Cartagena, Campus Muralla del Mar, 30202 Cartagena, Murcia, Spain\label{aff138}
\and
Instituto de F\'isica de Cantabria, Edificio Juan Jord\'a, Avenida de los Castros, 39005 Santander, Spain\label{aff139}
\and
Dipartimento di Fisica - Sezione di Astronomia, Universit\`a di Trieste, Via Tiepolo 11, 34131 Trieste, Italy\label{aff140}
\and
Institute of Cosmology and Gravitation, University of Portsmouth, Portsmouth PO1 3FX, UK\label{aff141}
\and
Department of Computer Science, Aalto University, PO Box 15400, Espoo, FI-00 076, Finland\label{aff142}
\and
 Instituto de Astrof\'{\i}sica de Canarias, E-38205 La Laguna; Universidad de La Laguna, Dpto. Astrof\'\i sica, E-38206 La Laguna, Tenerife, Spain\label{aff143}
\and
Ruhr University Bochum, Faculty of Physics and Astronomy, Astronomical Institute (AIRUB), German Centre for Cosmological Lensing (GCCL), 44780 Bochum, Germany\label{aff144}
\and
INAF-Osservatorio Astrofisico di Arcetri, Largo E. Fermi 5, 50125, Firenze, Italy\label{aff145}
\and
Department of Physics and Astronomy, Vesilinnantie 5, University of Turku, 20014 Turku, Finland\label{aff146}
\and
Finnish Centre for Astronomy with ESO (FINCA), Quantum, Vesilinnantie 5, University of Turku, 20014 Turku, Finland\label{aff147}
\and
Serco for European Space Agency (ESA), Camino bajo del Castillo, s/n, Urbanizacion Villafranca del Castillo, Villanueva de la Ca\~nada, 28692 Madrid, Spain\label{aff148}
\and
ARC Centre of Excellence for Dark Matter Particle Physics, Melbourne, Australia\label{aff149}
\and
Centre for Astrophysics \& Supercomputing, Swinburne University of Technology,  Hawthorn, Victoria 3122, Australia\label{aff150}
\and
Department of Physics and Astronomy, University of the Western Cape, Bellville, Cape Town, 7535, South Africa\label{aff151}
\and
DAMTP, Centre for Mathematical Sciences, Wilberforce Road, Cambridge CB3 0WA, UK\label{aff152}
\and
Kavli Institute for Cosmology Cambridge, Madingley Road, Cambridge, CB3 0HA, UK\label{aff153}
\and
Department of Astrophysics, University of Zurich, Winterthurerstrasse 190, 8057 Zurich, Switzerland\label{aff154}
\and
Department of Physics, Centre for Extragalactic Astronomy, Durham University, South Road, Durham, DH1 3LE, UK\label{aff155}
\and
IRFU, CEA, Universit\'e Paris-Saclay 91191 Gif-sur-Yvette Cedex, France\label{aff156}
\and
Univ. Grenoble Alpes, CNRS, Grenoble INP, LPSC-IN2P3, 53, Avenue des Martyrs, 38000, Grenoble, France\label{aff157}
\and
Centro de Astrof\'{\i}sica da Universidade do Porto, Rua das Estrelas, 4150-762 Porto, Portugal\label{aff158}
\and
Instituto de Astrof\'isica e Ci\^encias do Espa\c{c}o, Universidade do Porto, CAUP, Rua das Estrelas, PT4150-762 Porto, Portugal\label{aff159}
\and
HE Space for European Space Agency (ESA), Camino bajo del Castillo, s/n, Urbanizacion Villafranca del Castillo, Villanueva de la Ca\~nada, 28692 Madrid, Spain\label{aff160}
\and
INAF - Osservatorio Astronomico d'Abruzzo, Via Maggini, 64100, Teramo, Italy\label{aff161}
\and
Theoretical astrophysics, Department of Physics and Astronomy, Uppsala University, Box 516, 751 37 Uppsala, Sweden\label{aff162}
\and
Mathematical Institute, University of Leiden, Einsteinweg 55, 2333 CA Leiden, The Netherlands\label{aff163}
\and
Institute of Astronomy, University of Cambridge, Madingley Road, Cambridge CB3 0HA, UK\label{aff164}
\and
Department of Astrophysical Sciences, Peyton Hall, Princeton University, Princeton, NJ 08544, USA\label{aff165}
\and
Space physics and astronomy research unit, University of Oulu, Pentti Kaiteran katu 1, FI-90014 Oulu, Finland\label{aff166}
\and
International Centre for Theoretical Physics (ICTP), Strada Costiera 11, 34151 Trieste, Italy\label{aff167}
\and
Center for Computational Astrophysics, Flatiron Institute, 162 5th Avenue, 10010, New York, NY, USA\label{aff168}}

\abstract{
The \Euclid Quick Data Release (Q1) is a powerful dataset to study active galactic nuclei (AGN) and their host galaxies. Deriving their physical properties through multi-component spectral energy distribution (SED) fitting is a challenging task for AGN, but it is greatly aided by the \Euclid near-infrared photometry.
Here we present a new method to quantify the reliability of SED-derived parameters, such as AGN bolometric and monochromatic luminosities, host's stellar mass $M_\star$, star-formation rate (SFR) and specific star-formation rate (sSFR), by using mock SEDs of AGN built by combining observed SEDs of QSOs and galaxies. 
We apply this methodology to the ${\sim}1$ million Q1 AGN candidates, constructing a catalogue of AGN and host galaxy properties, alongside their respective reliability values.
With a reliability threshold at 0.5, we find 88\% of sources with robust stellar masses and 76\% with reliable AGN luminosities. 
Moreover, through SED fitting we also measure the AGN fraction $f_{\rm AGN}$ of the total mid-infrared flux and we use its lower-limit to select AGN. A $f_{\rm AGN, \, low} > 0.075$ threshold yields 85\% completeness and purity.
Comparable to colour-colour AGN selections, this method has the advantage of being less affected by redshift evolution and exploring fainter magnitudes.
Additionally, by comparing the AGN and host galaxy parameters across different identification methods, we find that the probed range in stellar mass and AGN luminosity can be quite different. This highlights the importance of combining different approaches and accounting for their selection biases when studying AGN and their role in galaxy evolution. 
Finally, for the X-ray detected sample, we present the X-ray to mid-IR luminosity relation, and the correlation between stellar mass and bolometric luminosity as a function of redshift, in good agreement with previous results.

}


   \titlerunning{AGN and host galaxies properties in \Euclid Q1}
   \authorrunning{Euclid Collaboration:
  B.~Laloux et al.}
   
\maketitle
\nolinenumbers

\section{Introduction}\label{1-Intro}

Lying in the centre of galaxies, active galactic nuclei (AGN) are supermassive black holes (SMBHs) undergoing a phase of intense accretion of matter, converting the gravitational potential into a tremendous amount of radiation \citep{Shakura_1973, Davis_2020}. 
This powerful mechanism makes the most rapidly accreting AGN, the quasars (QSOs), among the brightest objects in the Universe. Although the accretion disc emits from ultraviolet (UV) to optical, it drives various emission processes that extend the AGN spectral energy distribution (SED) throughout the electromagnetic spectrum from gamma rays to radio jets, with characteristic X-ray and infrared features \citep[e.g. ][]{Urry_1995, Ramos_Almeida_2017, Hickox_2018}.

AGN and their host galaxies are believed to be intrinsically connected \citep[e.g. ][]{Kormendy_2013}. The evolution of one impacting the other, as shown by the observed scaling relations between the mass of the bulge \citep[the $M_{\rm BH}{-}M_{\rm bulge}$ relation, e.g. ][]{Magorrian_1998, Haring_2004, McConnell_2013}, the black hole mass $M_{\rm BH}$ and the velocity dispersion of the host galaxy \citep[the $M_{\rm BH}{-}\sigma$ relation, e.g. ][]{Ferrarese_2000, Gebhardt_2000, Haehnelt_2000, Shankar_2016, Shankar_2025}, or the total stellar mass of the host galaxy \citep[the $M_{\rm BH}{-}M_{\star}$ relation, e.g. ][]{Marconi_2003, Bandara_2009, Madau_2014, Suh_2020, Pucha_2025}.
Although widely accepted, this AGN-galaxy co-evolution is still debated and its physical origin is a dynamic research field \citep{Harrison_2024}.
On the one hand, the AGN outflows can compress the gas in the host galaxy and trigger star formation \citep{Silk_2013, Cresci_2015}; on the other hand, powerful AGN-driven outflows could sweep away or excite the star-forming material, thereby quenching the galaxy growth \citep[e.g. ][]{Granato_2004, Shankar_2006, Cano_Diaz_2012, Fabian_2012, King_2015, Wylezalek_2016, Lapi_2018}. Both positive and negative feedback mechanisms have been observed, sometimes even within the same galaxy \citep[e.g. ][]{Bessiere_2022, Mercedes-Feliz_2023}.
Furthermore, galaxy mergers can channel large amounts of material toward the centres of galaxies, simultaneously fuelling star formation and feeding the central AGN \citep[e.g. ][]{Hopkins_2006, Hopkins_2008, Blecha_2018, Pearson_2019, Gao_2020, Andonie_2022, Yutani_2022, La_Marca_2024, Q1-SP013}.

In order to improve our understanding of the co-evolution of AGN and their host-galaxies, accurate measurements of their respective physical properties are essential. One effective approach for this purpose is spectral energy distribution (SED) fitting, which identifies the best-fitting parameters of a combination of models that reproduce the observed photometry \citep[e.g.][]{Conroy_2013}. Each model accounts for different physical processes, and its parameters are linked to the physical properties of the sources. 
However, the presence of an AGN significantly complicates these measurements, since an a priori unknown fraction of the observed flux originates from the AGN rather than the stellar or interstellar components of the host galaxy. Despite this challenge, SED fitting has been successfully applied in numerous studies to investigate the connection between AGN activity and host galaxy properties \citep[e.g. ][]{Burke_2022, Thorne_2022, Yang_2022, Best_2023, Mountrichas_2023, Siudek_2024, Siudek_2025}. 

This work pays particular attention to the impact of AGN on SED fitting, by introducing a novel methodology that provides not only the measurements but also a quantitative reliability characterisation and usage prescriptions, aiming to deliver a transparent, well-characterised reference catalogue of the physical properties of AGN candidates and their host-galaxies detected by \Euclid.
Launched in July 2023, \Euclid is an optical and near-infrared (NIR) space telescope of the European Space Agency (ESA) designed to explore the nature and the evolution of dark matter and dark energy \citep{Laureijs_11, EuclidSkyOverview}. Its unique capabilities enable a wide range of science goals, significantly advancing our understanding of galaxy evolution, including the co-evolution of galaxies with their central SMBH. \Euclid will survey a large fraction of the extragalactic sky with its two instruments: the visible imager \citep[VIS, ][]{Desprez-EP10, EuclidSkyVIS}
and the Near-Infrared Spectrometer and Photometer \citep[NISP, ][]{Maciaszek_22, EuclidSkyNISP}, which provides photometry in three NIR bands (\YE, \JE, and \HE), as well as slitless spectroscopy in the same wavelength range.
These additional NIR photometric bands enable more robust SED fitting-derived constraints (\cref{appendix:compa_bands}) of the AGN-candidates observed in the first \Euclid quick release Q1  \citep{Q1cite,Q1-TP001,Q1-TP002,Q1-TP003}, preceding the mission's first main data release.

This paper is organised as follows. \Cref{2-Sample} presents the  \Euclid Q1 data, including the multi-wavelength photometry, redshift, and AGN selection, while \cref{3-SED fitting} describes in detail the SED fitting methodology.
In \cref{5-Reliability}, we introduce a new approach to quantify the reliability of our physical parameter measurements and to apply the corresponding corrections.
Finally, in \cref{6- Discussion}, we compare our results across different AGN selection methods and discuss potential applications of our catalogue.
Finally, \cref{7-Summary} highlights the main conclusions of this work.

For clarity in the text, we define $\mathcal{M}_\star$ as the logarithmic stellar mass $M_\star$ in solar mass units $M_\odot$ 
such that $\mathcal{M}_\star = \log_{10}(M_\star / M_\odot)$, 
$\mathcal{L}_{\rm AGN}$ as the logarithmic AGN bolometric luminosity $L_{\rm AGN}$ in $\rm erg\,s^{-1}$ 
such that $\mathcal{L}_{\rm AGN} = \log_{10}(L_{\rm AGN} / [\rm erg\, s^{-1}])$, 
$\mathcal{SFR}$ as the logarithmic star formation rate SFR in $\rm M_\odot \, yr^{-1}$ 
such that $\mathcal{SFR} = \log_{10}({\rm SFR} / [\rm M_\odot \, yr^{-1}])$, 
and $\mathit{s}\mathcal{SFR}$ as the logarithmic specific star formation rate sSFR in $\rm yr^{-1}$ 
such that $\mathit{s}\mathcal{SFR} = \log_{10}({\rm sSFR} / [\rm yr^{-1}])$.
We use AB magnitudes throughout the paper and we assume a $\Lambda$CDM cosmology with $H_0 = 70$\,\kmsMpc, $\Omega_{\rm m} = 0.3$, and \mbox{$\Omega_{\Lambda} = 0.7$}.

\section{\Euclid Q1 AGN sample}\label{2-Sample}

This work is based on the photometric catalogue of \Euclid Q1 sources \citep[][hereafter \citetalias{Q1-SP027}]{Q1-SP027}.
The \Euclid Q1 data are described in \cref{subsec_euclid_photometry}, while the additional multi-wavelength data are presented in \cref{subsec_data}. 
Then, \cref{subsec_AGN_selection} lists the different AGN selection approaches applied to Q1 sources, while \cref{subsec_redshift} presents the various redshift qualities and coverage.
Finally, in \cref{subsec_samples}, we describe the specificities of the three source subsamples used in this work.

    \subsection{Euclid photometry}\label{subsec_euclid_photometry}

In this work, we analyse the \Euclid Q1 data \citep{Q1cite}, which contains the first observations of the Euclid Deep Survey (EDS) which includes three deep fields:
(1) Euclid Deep Field North (\mbox{EDF-N}), a roughly circular 20\,deg$^2$ region centred on the well-studied north Ecliptic pole; 
(2) Euclid Deep Field South (EDF-S), a stadium-shaped area of 23\,deg$^2$ with no previous dedicated observations;
and (3) Euclid Deep Field Fornax (EDF-F), roughly circular and of 10\,deg$^2$, centred on the Chandra Deep Field South \citep[CDFS, ][]{Giacconi_2002, Guo_2013, Lacy_2021}, and including the GOODS-S field \citep{Giavalisco_2004} and the Hubble Ultra Deep Field \citep[HUDF][]{Beckwith_2006}.
The effective area of these fields are respectively 22.9, 28.1, and 12.1\,deg$^2$ for a total area of 63.1\,deg$^2$ \citep{Q1-TP001}.
While currently as deep as the Euclid Wide Survey (EWS), with a 5\,$\sigma$ point source depth of 26.7\,mag (AB) in \IE \citep{EuclidSkyOverview}, the EDS fields will reach 2\,mag deeper at the end of the mission.
For more details, refer to \cite{EP-McPartland}.
The Q1 source identification is made in the \IE band and the \IE \mbox{(497--932\,nm)}, \YE \mbox{(0.9--1.2\,\micron)}, \JE \mbox{(1.1--1.6\,\micron)}, and \HE \mbox{(1.5--2.1\,\micron)} photometry is provided for each source \citep{EuclidSkyVIS, EuclidSkyNISP}.

In addition to the EWS and EDS, \Euclid will observe several well-known fields for the calibration of photometric redshifts (\mbox{photo-$z$}), including COSMOS-Wide \citep{Scoville_2007}. We use \Euclid photometry collected in that field for training purpose only (\cref{5-Reliability}).

    \subsection{Multi-wavelength data}\label{subsec_data}

The identification of multi-wavelength counterparts and the compilation of their photometry have been done and described in \citetalias{Q1-SP027}. 
The three EDFs have additional photometric coverage in the following wavebands:
\begin{itemize}
    \item UV: GALEX \citep{Martin_2005} provides shallow far-UV (FUV, 135--175\,nm) and near-UV (NUV 175--280\,nm) photometry;

    \item Optical bands from ground-based telescopes: 
    \mbox{EDF-S} and \mbox{EDF-F} optical photometry has been obtained in the $g$ (480\,nm), $r$ (641\,nm), $i$ (781\,nm), and $z$ (916\,nm) bands by the DECam instrument mounted on the CTIO (Cerro Tololo Inter-American Observatory), as part of the DESI legacy survey DR10 \citep[LS10, ][]{Dey_2019, Saxena_2024}, while for EDF-N, it originates from the UNIONS surveys \citep{Gwyn_2025}: $u$ (368\,nm) and $r$ (638\,nm) by MegaCam on CFHT (Canada-France-Hawai Telescope), $g$ (480\,nm) and $z$ (891\,nm) by Hyper Suprime-Cam (HSC) on Subaru, and $i$ (753\,nm) by Pan-STARRS;
    
    \item Mid-infrared (MIR): WISE \citep{Wright_2010} provides photometry in its four bands W1 (3.4\,$\mu$m), W2 (4.6\,$\mu$m), W3 (12\,$\mu$m), and W4 (22\,$\mu$m) from the AllWISE catalogue but W3 and W4 are only included if their signal-to-noise ratio (SNR) is larger than 3. Additionally, due to systematic photometric discrepancies between {\it Spitzer} \citep{Barmby_2008} and WISE, we only include IRAC\,1 (3.6\,$\mu$m) and IRAC\,2 (4.5\,$\mu$m) forced photometry at \Euclid positions \citep{Q1-SP011} if WISE data are not available.

\end{itemize}

    \subsection{AGN selection}\label{subsec_AGN_selection}

Among the ${\sim} 30$ million sources detected over the three EDFs, only a fraction of them are AGN. \citetalias{Q1-SP027} applies diverse criteria to select AGN-candidates:
\begin{itemize}
    \item B24A: \Euclid-only colour-colour wedge for point-like sources (\texttt{MUMAX\_MINUS\_MAG} < $-2.6$) \citep[][hereafter, \citetalias{Bisigello_2024}]{Bisigello_2024}, with a purity \mbox{$P = 0.166 \pm 0.015$} and completeness \mbox{$C = 0.347 \pm 0.004$} (354\,108 sources); 

    \item B24B: \Euclid, $u$, and $z$ bands colour-colour cut for point-like sources by \citetalias{Bisigello_2024}, only available in \mbox{EDF-N}, with \mbox{$C = 0.813 \pm 0.011$} and \mbox{$P = 0.922 \pm 0.017$} (177\,965 sources but 63\,203 do not have photometric redshift estimation); 
    \item $JH$\_$I_{\text{E}}Y$: QSO \Euclid-only point-like colour-colour selection \citepalias{Q1-SP027}, with \mbox{$P=0.92$} with \mbox{$C=0.63$} for \IE$<21$ at $z<1.6$ (6115 sources); 

    \item $I_{\text{E}}H$\_$gz$: colour-colour cut using \Euclid and optical $g-z$ colour for point-like sources \citepalias{Q1-SP027}, with \mbox{$P=0.93$} with \mbox{$C=0.60$} for \IE$<21$ at $z<1.6$ (1842 sources); %408+619+815

    \item C75 and R90: criteria by \cite{Assef_2018} based on WISE\,1/2 magnitudes with \mbox{$P=0.44$} and \mbox{$C=0.72$}, and \mbox{$P=0.90$} and \mbox{$C=0.09$}, respectively
    (108\,369 and 9019 sources, respectively); 

    \item QSO: this criterion is the combination of various QSO flags from different surveys, namely GAIA\,DR3 \citep{Gaia_2023A, Gaia_2023B}, QUAIA \citep{Storey_Fisher_2024} or DESI\,DR1 \citep{DESI_2025}, selecting 4863 sources;

    \item PRF: Probabilistic random forest approach used in \Euclid's photometric pipeline \citep{Q1-TP005} to classify objects as star, galaxy, or QSO (534\,492 sources); 

    \item X-ray: X-ray-detected sources presented in \cite{Q1-SP003}, hereafter, \citetalias{Q1-SP003}. Due to the limited contamination by other type of sources, this selection method is known to be one of the most reliable to select AGN. \citetalias{Q1-SP003} presents the identification of \Euclid counterparts within the eROSITA survey DR1 \citep{Predehl_2021, Merloni_2024}, XMM-{\it Newton} 4XMM DR13 \citep{Webb_2020}, and {\it Chandra} Source Catalogue 2 \citep[CSC2, ][]{Evans_2024}. 
    However, CSC2 does not include CDFS, one of the deepest X-ray fields, located within \mbox{EDF-F}. We therefore cross-matched our \Euclid sources in \mbox{EDF-F} with the optical counterparts of X-ray emitters presented in \cite{Hsu_2014}, adding 275 new X-ray sources, for a total of 12\,170 sources across the 3 fields. 

    \item BLAGN and NLAGN: respectively, broad- and narrow-line AGN, from optical spectra inspection described in section 4.4.2 of \citetalias{Q1-SP027}. This classification employs several line diagnostics on the spectra of \mbox{EDF-N} sources from the Dark Energy Spectroscopic Instrument survey \citep[DESI, ][]{DESI_2016,DESI_2022} and identifies 2703 and 4623 sources, respectively.
\end{itemize}

    \subsection{Redshift}\label{subsec_redshift}

Accurate redshift estimates are essential for reliable SED fitting measurements. However, the three EDFs differ both in redshift availability and quality. Since some sources have redshift estimates from different methods, we establish the following redshift priority order:

\begin{enumerate}
    \item Spectroscopic redshifts from an updated and expanded version of the validation sample from \cite{Q1-TP005}: this catalogue is the compilation of the different spectroscopic surveys overlapping with the EDFs. A quality criterion has been applied to only select reliable extragalactic spectroscopic redshifts (quality classes 3 or 4 or equivalent in other criteria). In the updated version used here, these publicly available spectroscopic survey catalogues have been matched to \Euclid MER photometry \citep{Q1-TP004} with a \ang{;;0.5} tolerance. \mbox{EDF-S} has only 822 sources with spectroscopic redshift higher than 0.002, almost exclusively (746 sources) from Quaia \citep{Storey_Fisher_2024}. 
    \mbox{EDF-F} benefits from 35\,037 sources with spectroscopic redshifts, a large number of them (21\,331) from the PRIsm MUlti-object Survey \citep[PRIMUS, ][]{Coil_2011,Cool_2013} up to $z<1.2$. The remaining redshift measurements are provided by several surveys, the largest being the Australian Dark Energy Survey \citep[OzDES][]{Yuan_2015} and 3dHST \citep{Momcheva_2016,Brammer_2012}, with 7241 and 3232 sources, respectively. Among the other smaller spectroscopic surveys, some specifically target high-redshift objects such as VANDELS \citep{Pentericci_2018} or the JWST Advanced Deep Extragalactic Survey \citep[JADES, ][]{Eisenstein_2023,DEugenio_2025}. Finally, \mbox{EDF-N} has the best spectroscopic coverage with 47\,465 sources, most of them from DESI \citep{DESI_2025}. The remaining spectroscopic redshifts are provided by various smaller surveys.

    \item \Euclid NISP slitless spectroscopic redshift estimates from \cite{Fu_2025}. The authors provide visually-inspected redshift for ${\sim} 3500$ quasars within the three EDFs, 1936 of which are new: 337 in EDF-N, 1201 in EDF-S, and 389 in EDF-F.
    While previous redshift measurements have higher resolution, the visual inspection makes NISP slitless spectroscopy more reliable, and therefore, for sources with discrepant redshift estimates $|z_{\rm spec} - z_{\rm NISP}|>0.1$, we use instead the NISP redshift value for 24, 37, and 17 sources within \mbox{EDF-N}, \mbox{EDF-S}, and \mbox{EDF-F}, respectively.

    \item Narrow-band template-fitting (TF) photometric redshift from \cite{Hsu_2014}: benefiting from deep and extensive narrow-band coverage, \cite{Hsu_2014} obtained high-quality redshift estimates for 314 X-ray AGN within CDFS. We assume their redshift uncertainties to be null, as for spectroscopic sources.

    \item Machine-learning (ML) photometric redshifts from \cite{Q1-SP003}: by identifying the optical counterparts of X-ray-detected AGN in Q1, the authors measure their photometric redshifts with the \texttt{PICZL} ML algorithm \citep{Roster_2024} from the LS10 de-redenned images in the $g$, $r$, $i$, and $z$ bands, complemented by 4-band WISE aperture photometry. The incorporation of the pixel information from the image greatly improves the redshift estimation, reaching an outlier fraction of $\sim 5$\% for  brighter sources ($r<21.5\,\rm mag$) and $\sim 16$\% for the whole sample. The same process is applied with LS9 images for \mbox{EDF-N}. This approach provides new redshift estimates for 432, 2636, and 3353 sources in \mbox{EDF-N}, \mbox{EDF-S}, and \mbox{EDF-F}, respectively.

    \item TF photometric redshifts, released internally within the \Euclid \enquote{Galaxy Evolution / AGN} Science Working Group (GAEV SWG). 
    The AGN photometric redshift catalogue was constructed using the fully-Bayesian algorithm \texttt{Phosphoros} on the UV-to-MIR photometry described in \cref{subsec_data}.
    \texttt{Phosphoros} can efficiently constrain the redshift of a source by fitting selected SED templates to the observed SED and is deployed within the official \Euclid pipeline \citep{Q1-TP005}. 
    However, due to a set of templates and luminosity priors optimised for inactive galaxies, the pipeline photo-$z$ estimates suffers from a large outlier fraction (56\%) for AGN as shown on \cref{fig:appendix_compa_zphot} in \cref{appendix:zphot}. This fraction goes up to 70\% when excluding the NLAGN whose SEDs are similar to those of inactive galaxies.
    To improve the reliability of the redshift estimates, each source has been run in three configurations, one using galaxy templates \citep[GALAXY; ][]{Ilbert_2009} and two others with AGN templates \citep[QSOV and EXTNV; ][]{Salvato_2022}, each with different luminosity priors. 
    Despite a still relatively high outlier fraction of 40\% (\cref{fig:appendix_compa_zphot}), the results show a significant improvement.
    This approach provides the largest number of redshift estimates with 318\,961, 488\,598, and 210\,304 sources in \mbox{EDF-N}, \mbox{EDF-S}, and \mbox{EDF-F}, respectively.
    The full photometric redshift probability distribution functions ($z$-PDFs) used in this work are publicly available at \url{https://zenodo.org/records/20065887}.

\end{enumerate}

\begin{table}
    \centering
    \caption{Characteristics of the samples A, B, and\,C.}
    \renewcommand{\arraystretch}{1.25}
    \begin{tabular}{|c|c|c|c|c|}
        \cline{2-5}
        \multicolumn{1}{c}{} & \multicolumn{1}{|c|}{Sample}   & A & B & C \\ \cline{2-5}
        \multicolumn{1}{c}{} & \multicolumn{1}{|c|}{Type of}  & AGN \& & X-ray& AGN-candidates \\
        \multicolumn{1}{c}{} & \multicolumn{1}{|c|}{sources}  & non-AGN & AGN & (any selection) \\ \cline{2-5}
        \multicolumn{1}{c}{} & \multicolumn{1}{|c|}{\multirow{2}{*}{Criteria}} & With & \multirow{2}{*}{All}      & \multirow{2}{*}{All} \\
        \multicolumn{1}{c}{} & \multicolumn{1}{|c|}{} & spec-$z$&          &  \\
        \cline{2-5} \noalign{\vskip 2pt}
        \hline
              & {\small spec-$z$}        & 49\,418 & 212  & 9123   \\
        {\small EDF-N} & {\small ML phz} &  --     & 432  & 432     \\
              & {\small TF phz}          &  --     & 342  & 318\,961 \\
        \hline
              & {\small spec-$z$}        &  2031   & 868  & 1947   \\
        {\small EDF-S} & {\small ML phz} &  --     & 2636 & 2636    \\
              & {\small TF phz}          &  --     & 699  & 488\,598 \\
        \hline
              & {\small spec-$z$}        & 36\,062 & 2130  & 3233   \\
        {\small EDF-F} & {\small ML phz} &  --     & 3353  & 3353    \\
              & {\small TF phz}          &  --     & 1603  & 210\,522 \\
        \hline\hline
              & {\small spec-$z$}     & 87\,511 & 3210 & 14\,303   \\
        {Total} & {\small ML phz}     &   --    & 6421 & 6421       \\
              & {\small TF phz}       &   --    & 2644 & 1\,018\,081 \\
        \hline
    \end{tabular}
    \tablefoot{The table indicates the number of sources per field per type of redshift: spectroscopic, machine-learning (ML) photometric, or template-fitting (TF) photometric.}
    \label{tab:redshift_origins}
\end{table}

    \subsection{Our samples}\label{subsec_samples}

We divide the \Euclid sources into three different samples that partially overlap, each having different science objectives. Their respective characteristics are summarised in \cref{tab:redshift_origins}, in which the number of sources per field and per redshift quality class is also indicated.

        \subsubsection{Sample A: the spectroscopic sample \label{subsubsec:specz}}

This sample contains all sources detected by \Euclid with spectroscopic redshift measurements, irrespective of their AGN classification.
Having both active and inactive galaxies allows us to compare and assess the validity of our methodology for both populations.
Moreover, as this sample benefits from the best possible quality redshift, our conclusion are even more robust. We do not include the sources with high-quality narrow band TF photo-$z$, even if their redshift uncertainty is set to 0.

This sample contains 87\,511 sources including 14\,293 AGN-candidates. Within \mbox{EDF-N}, \mbox{EDF-S}, and \mbox{EDF-F}, there are 49\,418, 2031, and 36\,062 sources, including 9123, 1947, and 3233 AGN-candidates, respectively.

        \subsubsection{Sample B: the X-ray-detected sample}\label{subsubsec:Xray_photoz}

This sample is constructed from all the \Euclid X-ray-detected source sample presented by \cite{Q1-SP003}. 
The special status of this particular AGN-selection method is justified by its high reliability, combined with well defined X-ray selection functions, providing a powerful tool to study AGN intrinsic populations.
Moreover, as mentioned in \cref{subsec_redshift}, in addition to the spectroscopic redshift available for approximately a quarter of the sources, 61 other sources have  high-quality TF photo-$z$ from \cite{Hsu_2014} in \mbox{EDF-F} while most of the remaining ones (6421) benefit from robust ML photo-$z$.
Finally, since the remaining sources are too faint to be in LS10, the ML photo-$z$ approach cannot be conducted, therefore, we use the TF photo-$z$ estimates derived within the Euclid GAEV SWG (as detailed in point 5 in \cref{subsec_redshift}).
This sample includes a total of 12\,275 X-ray-detected AGN, of which 986, 4203 and 7086 are located in \mbox{EDF-N}, \mbox{EDF-S}, and \mbox{EDF-F}.

        \subsubsection{Sample C: the optical/NIR-selected AGN}\label{subsubsec:photoz}

The last sample in our catalogue is the largest one because it contains all the sources in the three EDFs selected as AGN-candidates by any approach presented in \cref{subsec_AGN_selection}, including X-ray detection.
As shown in \cref{tab:redshift_origins}, most sources have TF photometric redshifts, for which accuracy is known to be limited for AGN-candidates (see \cref{appendix:zphot}).
Sample\,C contains 328\,516, 493\,181, and 217\,108 AGN-candidates within \mbox{EDF-N}, \mbox{EDF-S}, and \mbox{EDF-F}, for a total of 1\,038\,805 sources.

\section{SED fitting procedure}\label{3-SED fitting}

This section describes the SED fitting approach adopted in this work to derive the physical properties of our AGN-candidates.
Two examples of best-fit SEDs are displayed in \cref{fig:best-fit_SED}, one for a galaxy with high SFR and no AGN emission, and one with clear AGN activity.

    \subsection{The SED fitting algorithm}\label{subsec: SED fitting algorithm}

SED fitting consists of combining different theoretical and/or empirical model templates to reproduce the observed multi-wavelength photometry of our sources, providing information on the underlying physical properties \citep{Conroy_2013}.
In the case of AGN, the nuclear emission must be carefully separated from the stellar population emission from the host galaxy in order to estimate crucial galaxy properties such as stellar mass \citep[e.g. ][]{Ciesla_2015, Toba_2022, Buchner_2024}. 
In this work, we decided to use the SED fitting algorithm \texttt{CIGALE} V2022.1 \citep[Code Investigating GALaxy Emission, ][]{Boquien_2019, Yang_2022} as it is extensively tested and computationally efficient to analyse the large \Euclid sample.

\texttt{CIGALE} benefits from a large choice of SED-generating modules representing different physical mechanisms that can be combined together to reproduce the complex SEDs of both inactive galaxies and AGN. As \texttt{CIGALE} is based on an energy balance principle, the radiative energy absorbed in the UV and optical by dust is re-emitted in the mid-to-far IR regime, linking realistically some parameters of the different modules. 
\texttt{CIGALE} then compares modelled and observed fluxes and uses $\chi^2$ statistics to determine the best-fit parameters and associated uncertainties.

~

    \subsection{The different SED modules}\label{subsec:SED_modules}

The choice of modules and their parameter range values is a delicate matter, as it requires enough flexibility to reproduce well the population diversity, while being computationally efficient. In this work, with the exception of the AGN model, we follow the prescription of \cite{Siudek_2024} that analysed the physical properties of 1.3 million galaxies from the DESI Survey \citep{DESI_2016, DESI_2022}.
Below, we briefly present the modules used in this work. More details on the parameter values and ranges are available in \cref{tab:cigale_params} in \cref{appendix:SED_fitting}.
We also describe additional alternative model setups, displayed in \cref{tab:alternative_setup}, which are used later in \cref{appendix:compa_model} to quantify the impact of our model assumptions.

    \subsubsection{Stellar population}
For the stellar emission of the host galaxy, we adopt the stellar population template models of \cite{Bruzual_2003} with the initial mass function (IMF) of \cite{Chabrier_2003} and fixed solar metallicity ($Z=0.02$). The star-formation history (SFH) is defined as a delayed SFH with exponential decrease with a potential recent star-forming burst. This SFH prescription provides flexibility to reproduce the SED of both star-forming and quiescent galaxies \citep{Ciesla_2015, Siudek_2024}, i.e. sources with high and low SFRs, respectively.

We propose two different alternative model setups to assess the impact of our stellar population modelling. 
One of these setups uses instead the stellar population template models presented by \cite{Maraston_2005}. Similarly, the choice of IMF is known to shift the estimated stellar mass, and therefore, we defined a model setup using the IMF of \cite{Salpeter_1955} instead of \cite{Chabrier_2003}.

\subsubsection{Nebular emission}
Massive stars can ionise large quantities of interstellar gas, which in return emit some strong narrow emission lines. Therefore, the modelling of these lines can offer valuable information regarding the young stellar population of galaxies \citep{Anders_2003, De_Barros_2014, Mobasher_2015}.
The nebular line templates used in \texttt{CIGALE} were generated by \cite{Inoue_2011}, and in our case, we set the parameters following the prescription of \cite{Lopez_2023}.

\subsubsection{Dust absorption and emission}

Interstellar dust significantly impacts the observed SED of galaxies by absorbing short-wavelength radiation and re-emitting it at longer wavelengths. As mentioned previously, \texttt{CIGALE} relies on the energy balance principle, i.e. the energy absorbed by dust must equal the energy it re-emits. Consequently, \texttt{CIGALE} models both components simultaneously to fit the observed SED.
For our baseline model, we adopt the modified attenuation curves of \cite{Calzetti_2000}, while those from \cite{Charlot_2000} are used in an alternative model.

Among the various dust emission modules provided by \texttt{CIGALE}, we chose the \cite{Draine_2014} for the baseline setup and \cite{Dale_2014} for an alternative setup.

\subsubsection{AGN emission}

In this work, we use the SKIRTOR AGN module \citep{Stalevski_2012, Stalevski_2016}, a physically motivated model that proposes a two-phase clumpy dusty toroidal structure in which high-density dusty clouds are embedded in a less dense smooth component. This type of clumpy torus agrees better with observational constraints than a smooth distribution, but may cause parameter degeneracies.
The model is highly parametrisable, but only a few parameters are left free to vary, notably the AGN fraction scaling parameter which is defined as the ratio of the AGN integrated luminosity to the total luminosity within a given wavelength interval, 5 to 20\,\micron\, \citep{Thorne_2022}.
This flexible model allows the reproduction of a variety of AGN, including type\,1 and type\,2 AGN.

To compare the impact of our choice of AGN model, we propose an alternative setup using the \cite{Fritz_2006} AGN model as in \cite{Siudek_2024}. Unlike the clumpy SKIRTOR model, this  model assumes a uniform distribution of dust surrounding the torus, as pictured in the standard unified AGN model \citep{Urry_1995}. We discuss in later sections our choice of AGN model within our baseline setup.

We also use another alternative model in which we simply do not include any AGN component. This setup allows us to investigate how biased SED fitting results are when tailored for inactive galaxies but applied to AGN.

    \subsection{Accounting for photometric redshift uncertainty}\label{subsec:SED_zphot}

Template fitting photometric redshift uncertainties can significantly impact SED fitting results. Therefore, in our analysis we account for the entire $z$-PDF provided for both ML and TF photo-$z$ estimates (\cref{subsec_redshift}).
However, \texttt{CIGALE} requires the redshift of each source to be fixed but this limitation can be addressed by following the approach of \cite{Laloux_2024}.
For a given source, we sample its $z$-PDF into 0.05 redshift intervals up to 1 and 0.1 intervals at higher redshift and for each interval, we fit the SED of the source with the corresponding fixed redshift. It results in multiple estimations for each physical parameter, such as the stellar mass, and their corresponding uncertainties. To obtain the final physical constraints, we combine these multiple estimations in a Gaussian mixture: for each, we create a Gaussian centred on their best-fit value, with the deviation and amplitude fixed to their uncertainty and corresponding redshift bin weight, respectively. The sum of all the Gaussians is the final parameter posterior distribution of the source from which the median, the $16^{\rm th}$ percentile and $84^{\rm th}$ percentile values are considered, respectively, as the best-fit, lower-limit and upper-limit values. For more details, we refer to Section 3.2 and Figure 3 in \cite{Laloux_2024}.

\section{Reliability of the physical parameter measurements}\label{5-Reliability}

Using our SED fitting approach and baseline model (\cref{3-SED fitting}), we constrained for all the sources in our samples several key physical parameters, including host galaxy properties, such as stellar mass, age, SFR, sSFR, and dust attenuation $E(B-V)$, as well as AGN properties, such as bolometric luminosity, monochromatic luminosities at 2500\,\AA\, and 6\,\micron, torus viewing angle, and the AGN fraction in the 5--20\,\micron\, range.
Here we focus on the reliability associated to these parameters.

Indeed, given the challenges associated with estimating these physical properties via SED fitting, arising from well-known parameter degeneracies, such as age–metallicity \citep{Worthey_1994} and colour–redshift \citep{Masters_2015}, and, in AGN hosts, nuclear emission that contaminates and can dominate the galaxy SED, we have developed a novel method to quantify the impact of these effects on the derived parameters in order to derive prescriptions to account for them in scientific analyses.
Briefly, we construct controlled mock SEDs by linearly combining quasar and galaxy observed SEDs, which set the ground truth parameter values, across a wide range of redshift, AGN/host contrasts, and photometric coverages. 
By fitting these mock SEDs we derive physical parameter estimates to compare with ground truth values. We then map the parameter bias and scatter due to AGN activity in the $L_{\rm AGN}{-}M_\star$ plane, and derive the per-parameter reliability functions that accompany the catalogue.
Below we describe the method step by step.

    \subsection{The adapted Chimera benchmark}
    \label{subsec:chimera}
    
To quantify the reliability of SED-derived parameters, one possible approach is to simulate the SED of an AGN with its host and quantify the goodness of the fit. However, the results are inherently model-dependent, since both the simulated and fitting SEDs rely on the same underlying models. 
To overcome this limitation, \cite{Buchner_2024} developed the Chimera benchmark.
Rather than modelling the SEDs of AGN and host galaxies, it uses real observations of both pure galaxies (GALs) and pure quasars (QSOs). Pure GALs are defined as galaxies with no detectable AGN activity, while pure QSOs are sources in which the AGN dominates the emission across all wavelengths, rendering the host galaxy contribution negligible. 
The pure QSO sample has been constructed from the SDSS Data Release\,7 \citep{Schneider_2010} consisting of sources brighter than\,$-22$ in absolute magnitude in the $i$-band and exhibiting at least one broad emission line in their spectra. 
The pure GAL sample has been selected within the COSMOS field \citep{Scoville_2007} which offers deep and extensive multi-wavelength coverage \citep[COSMOS2020, ][]{Weaver_2022}. This coverage includes photometric bands not available for the pure QSO sample, notably in the MIR (IRAC\,3/4) and far-IR (FIR) range \citep[MIPS\,24\,\micron, PACS\,100/160\,\micron, and SPIRE\,250/350/500\,\micron,][]{Shirley_2021}, which allow to reliably rule out AGN activity via SED fitting. 
Further details on the construction of the samples are provided in \cite{Buchner_2024}.

\begin{figure*}
    \centering
    \includegraphics[width=0.99\linewidth]{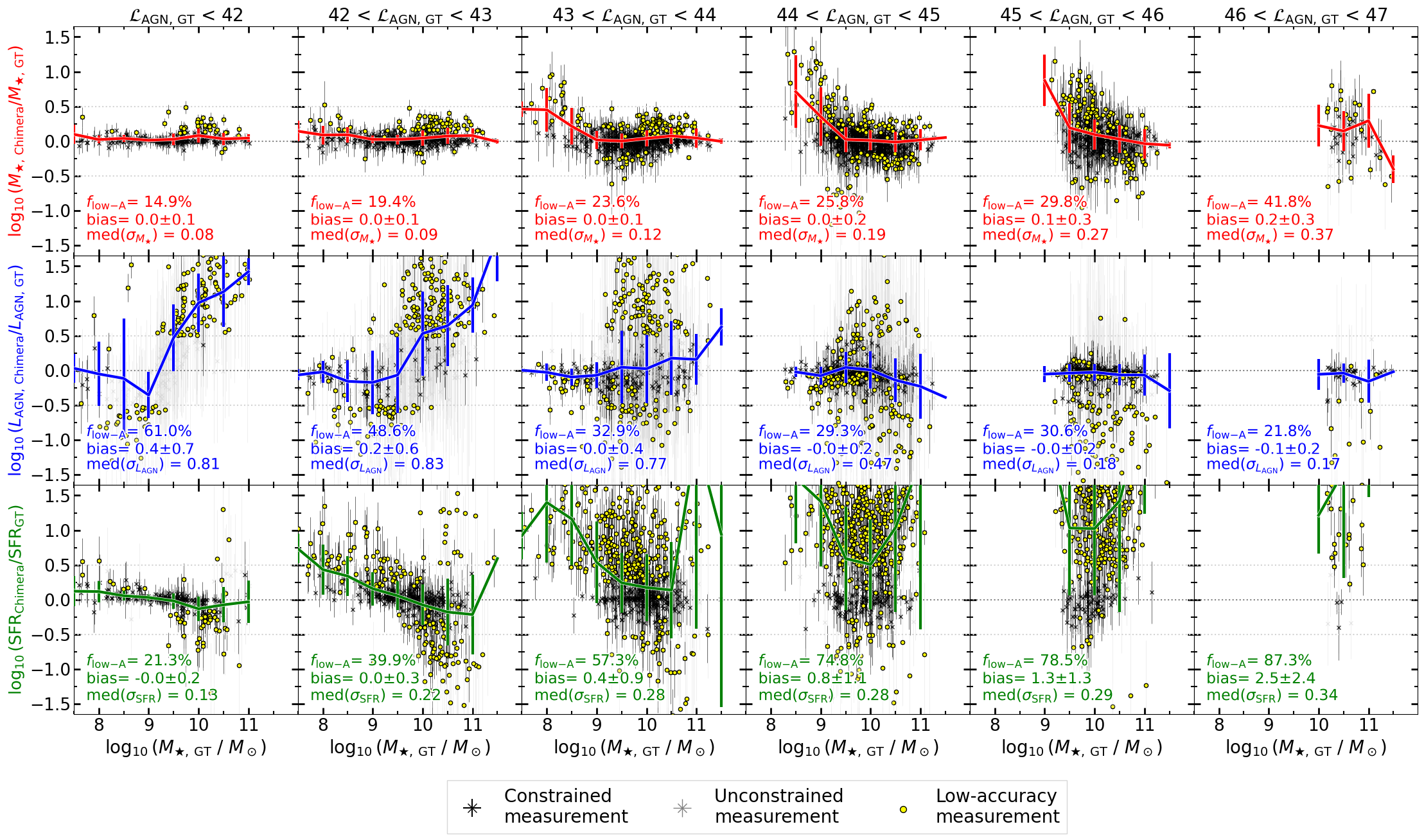}
    \caption{
    {\it Top row:} $\mathcal{M}_{\star,\,\rm Chimera} {-} \mathcal{M}_{\star,\,\rm GT}$ as a function of $\mathcal{M}_{\star,\,\rm GT}$ for different $\mathcal{L}_{\rm AGN, \, GT}$ intervals.
    For clarity, sources with unconstrained measurements ($\sigma_{M_{\star, \, \rm Chimera}}>0.5$) are displayed in light grey.
    Additionally, sources with low-accuracy measurements (\cref{eq:correct_Mstar}) are highlighted as yellow-filled black circles.
    The solid line in each panel indicates the stellar mass average bias and scatter as a function of the true stellar mass.
    The mean and standard deviation of the stellar mass difference, weighted by the inverse uncertainty, is indicated at the bottom of each panel, along with the median uncertainty ${\rm med}(\sigma_{M_\star})$.
    Additionally, the fraction of low-accuracy sources $f_{\rm low-A}$ is indicated for each ground truth AGN luminosity interval. 
    The grey dotted horizontal lines illustrates the 0.5\,dex margin from the ground truth value.
    {\it Middle and bottom rows:} Same as the top row but for the AGN luminosity difference $\mathcal{L}_{\rm AGN,\, Chimera} {-} \mathcal{L}_{\rm AGN,\, GT}$ and SFR difference $\mathcal{SFR}_{\rm Chimera} {-} \mathcal{SFR}_{\rm GT} $, respectively. Unconstrained measurements are defined as $\sigma_{L_{\rm AGN, \,  Chimera}}>0.5$ and $\sigma_{{\rm SFR}_{\rm Chimera}}>0.5$, respectively.
    }
    \label{fig:compa_AGN_models}
\end{figure*}

From this benchmark, we construct \enquote{Chimera AGN} objects by attributing a pure quasar to each pure galaxy with the only condition that their spectroscopic redshift lies within a 0.01 interval, to ensure a similar rest-frame wavelength coverage of the observed bands for each pair of objects. 
No additional properties of the QSO or the host
are considered for the pairing to not bias our sample.
Once paired, the flux of the galaxy and the quasar in each photometric band can be combined as follows
\begin{equation}
    f_{\rm Chimera}({\rm band}) = f_{\rm GAL}({\rm band}) + w_{\rm QSO} \; f_{\rm QSO}({\rm band}) \, ,
    \label{eq:flux_sum}
\end{equation}
\noindent with the QSO weight factor $w_{\rm QSO}$ randomly chosen from \{0.0001, 0.0003, 0.001, 0.003, 0.01, 0.03, 0.1, 0.3, 1\}, allowing a large variety of relative power between the central AGN and its host galaxy. The same method is applied to the flux uncertainties.
The photometric bands available for the Chimera AGN are only those in common between galaxies and QSOs: SDSS {\it uriz} in the optical, 2MASS {\it JHK} in the NIR and IRAC\,1/2 in the MIR.

However, as presented in \cref{subsec_euclid_photometry,subsec_data}, the EDFs also benefit from the $g$ band and from \Euclid photometry, requiring us to adapt the Chimera benchmark to this standard.
Since the pure GAL sample is located in the COSMOS field, which is also used as a calibration field for \Euclid, the fluxes in the \IE, \YE, \JE, and \HE bands are simply retrieved by positional cross-matching.
In opposition, the pure QSO sample is spread across the entire sky, therefore, we restrict our sample to the sources having at least two positive fluxes within the $JHK$ bands and obtain their fluxes in the \Euclid bands via SED fitting using the baseline model with the AGN fraction fixed to 0.99.

The flux of the Chimera AGN objects in these bands is then obtained by summation (see \cref{eq:flux_sum}). On the other hand, their respective flux uncertainties are randomly drawn as a function of their flux, following the \Euclid flux-dependent uncertainty distribution observed in the COSMOS calibration field.
The same method is applied to obtain the $g$-band photometry of our adapted Chimera sample.
Once the Chimera sample is fully constructed, for each \Euclid band, the fluxes below their detection thresholds in the EDFs are fixed to the threshold value and considered as upper limits, while for optical bands below the detection threshold, the fluxes are suppressed and not included in the SED fitting. Sources with fluxes below the detection threshold in all the \Euclid bands are discarded. 
Hereafter, we refer to this sample of objects as the \enquote{adapted Chimera AGN} sample.

While the flux measurement via SED fitting partially reintroduces model dependence, in contrast with the original objective of the Chimera benchmark, this step is nevertheless necessary to adapt it to the \Euclid mission. We tested the impact of this assumption by varying the fixed AGN fraction and by adopting an alternative AGN model \citep{Fritz_2006}. These variations result in a typical flux scatter of $\sim$5\% across the relevant bands of the Chimera AGN. 
When propagated to the our subsequent analysis on the goodness of SED fitting measurements (see \cref{subsec:reliability_maps}), these differences have a negligible impact and do not affect our conclusions.
In the future, as the area covered by \Euclid will increase, the fluxes of the pure QSO sample in the \Euclid bands will be directly available, rendering this step unnecessary.

    \subsection{Ground truth versus measured stellar mass, AGN luminosity, and SFR}
    \label{subsec:Mstar_compa}

Once we adapted the Chimera sample for our purpose, we use \texttt{CIGALE} to perform the SED fitting on the pure GAL, the pure QSO, and the adapted Chimera AGN samples with the baseline model setup in \cref{tab:cigale_params}. 
The only difference in the fit of the three samples is the AGN fraction parameter of the model, which is left free to vary for the Chimera AGN, but is set to 0.99 for the pure QSO sample, and to 0 for the pure GAL sample.

By fitting the SEDs of the pure GAL sample, we measure the \enquote{ground truth} values of the host galaxy properties, notably the stellar mass $M_{\star,\,\rm GT}$, star formation rate $\rm SFR_{GT}$, and specific star formation rate $\rm sSFR_{GT}$, of the corresponding Chimera AGN. To increase the robustness of these measurements, we include in the SED fit the additional NIR and FIR photometric bands that are only available for the pure GAL sample. 
Similarly, the SED fit of the pure QSO sample provides the \enquote{ground truth} measurements of the AGN properties of the corresponding Chimera AGN, including the AGN luminosity after rescaling by the attributed QSO weight $w_{\rm QSO}$, i.e.   $\mathcal{L}_{\rm AGN,\, GT} = \mathcal{L}_{\rm QSO} + \log_{10}(w_{\rm QSO})$.
This process then allows us to properly compare the outputs of the SED fit of the Chimera AGN with the ground truth values within the assumed framework.
For instance, it is known that the somewhat arbitrary choice of the IMF between \cite{Salpeter_1955} and \cite{Chabrier_2003} leads to a ${\sim} 0.2$\,dex shift in the estimated stellar mass. However, as the stellar masses of the pure GAL sample and Chimera AGN are derived using the same SED fitting setup, they are treated consistently, shifting similarly the ground truth value and the one from the Chimera AGN, therefore, offering robust insights on the impact of AGN activity on physical parameter estimation.

With these ground truth property values, we can derive for each Chimera AGN its Eddington rate following
\begin{equation}
    \lambda_{\rm Edd} = \frac{L_{\rm bol, \,AGN}}{L_{\rm Edd}} \approx \frac{L_{\rm AGN, \,GT}}{1.3 \times 10^{38} \; \left(0.002\, \; M_{\star,\,\rm GT}\right)}\, ,
\end{equation}
\noindent where $L_{\rm AGN, \,GT}$ is considered equal to the AGN bolometric luminosity $L_{\rm bol, \,AGN}$, and the black hole mass $M_{\rm BH}$ is scaled to the host-galaxy stellar mass using the scaling relationship \mbox{$M_{\rm BH} = 0.002\, M_\star$} \citep{Marconi_2003, Haring_2004}.
Due to the random QSO weight factor $w_{\rm QSO}$ attribution, numerous Chimera AGN appear in the super-Eddington regime, $\lambda_{\rm Edd} >1$. 
To obtain a more realistic AGN population and account for the ${\sim}0.5$\,dex scatter of observed $M_{\rm BH}{-}M_{\star}$ scaling relations \citep{Suh_2020, Lopez_2023}, we restricted our sample to AGN with $\log_{10}(\lambda_{\rm Edd}) < 0.5$.

The top row of \cref{fig:compa_AGN_models} compares the estimated stellar mass using the Chimera AGN sample $\mathcal{M}_{\star,\,\rm Chimera}$ with the ground truth stellar mass $\mathcal{M}_{\star,\,\rm GT}$ across different intervals of ground truth AGN luminosity.
The middle and bottom rows of \cref{fig:compa_AGN_models} show a similar comparison but for the AGN luminosity, $\mathcal{L}_{\rm AGN,\, Chimera} {-} \mathcal{L}_{\rm AGN,\, GT}$, and SFR, $\mathcal{SFR}_{\rm Chimera} {-} \mathcal{SFR}_{\rm GT}$, respectively. 
For the clarity of the plot, we represent with black and light grey error bars the sources with \enquote{constrained} ($\sigma_\xi \leq 0.5$) and \enquote{unconstrained} ($\sigma_\xi>0.5$) measurements $\xi$ ($\mathcal{M}_\star$, $\mathcal{L}_{\rm AGN}$, or $\mathcal{SFR}$). 
The decreasing number of low-mass AGN at high AGN luminosities in \cref{fig:compa_AGN_models} is due to our high-Eddington-luminosity cut, which prevents the brightest QSOs from being hosted in the least massive galaxies.

For the rest of the analysis, a $\xi$ measurement is defined as \enquote{high-accuracy} if

\begin{equation}
    |\xi_{\rm Chimera} - \xi_{\rm GT}| < {\rm min}\left(0.5, \, \sqrt{\sigma_{\xi_{\rm Chimera}}^2 + \sigma_{\xi_{\rm GT}}^2}\right) \,,
    \label{eq:correct_Mstar}
\end{equation}

with $\sigma_{\xi_{\rm Chimera}}$ and $\sigma_{\xi_{\rm GT}}$ being the uncertainty of the Chimera and ground-truth measurements, respectively.
This definition selects measurements that satisfy two complementary conditions: the estimate must not deviate from the ground-truth by more than 0.5\,dex and the ground-truth value must lie within the 1\,$\sigma$ quadratic uncertainty interval. 
The first condition guarantees the accuracy of the measurements independently of their uncertainties, ensuring scientifically meaningful results. 
Although the 0.5\,dex threshold is somewhat arbitrary, it provides a practical criterion requiring an order-of-magnitude agreement with the ground-truth \citep{Buchner_2024, Cochrane_2025}.
The second condition ensures that the reported uncertainties are realistic. As illustrated by the yellow-filled black circles in \cref{fig:compa_AGN_models}, several measurements can satisfy the 0.5\,dex accuracy criterion while still failing to include the ground truth within their 1\,sigma uncertainty interval, indicating that their uncertainties are underestimated.

\begin{figure*}
    \centering
    \includegraphics[width=1\linewidth]{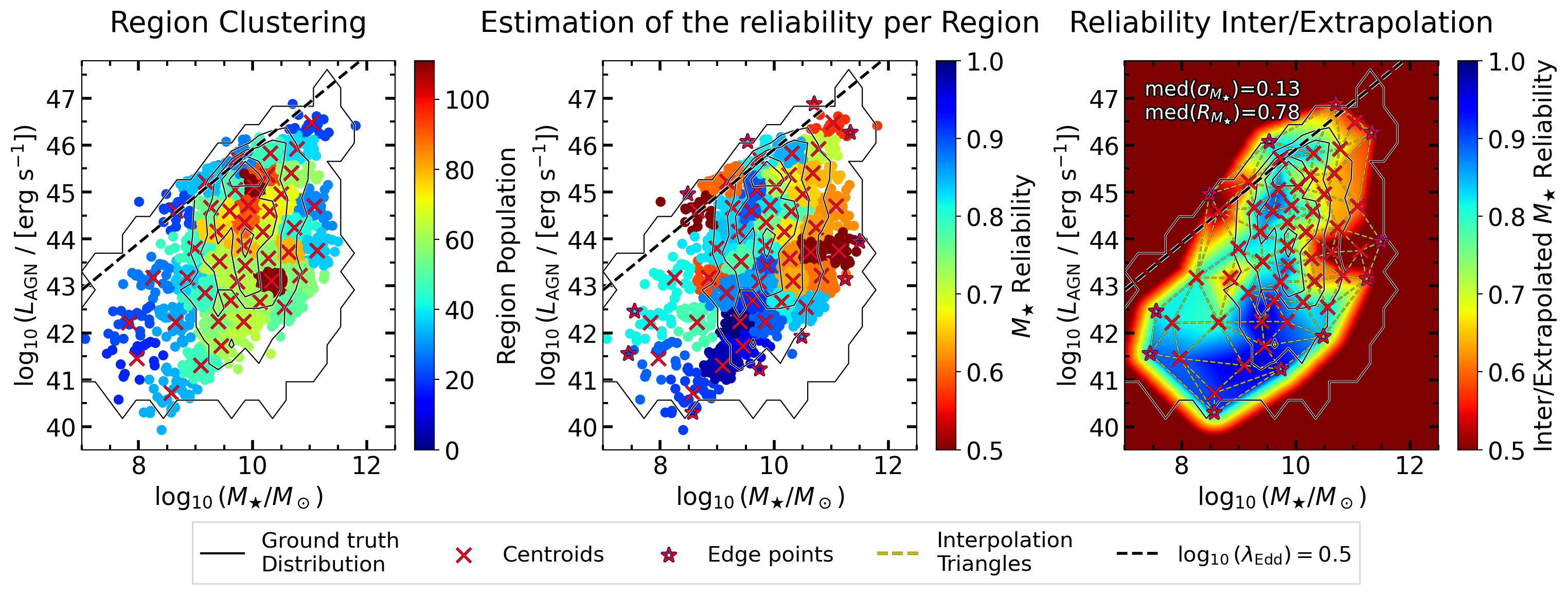}
    \caption{{\it Left panel}: distribution of the sources in the $\mathcal{L}_{\rm AGN} {-} \mathcal{M}_\star$ plane when using all available bands. The red crosses mark the centre of the clusters while the colour of each dot indicates the number of sources in their respective cluster. The black contours represent the ground truth distribution of these sources. The black dashed line corresponds to the $\log_{10}(\lambda_{\rm Edd})=0.5$ limit.
    {\it Middle panel}: Similar to the left panel but the colour of the dots represent the reliability $R_{M_\star}$ within each cluster defined as the fraction that are not outliers in $M_\star$.
    The red stars correspond to edge points of the source distribution to expend the interpolation region. 
    {\it Right panel}: Interpolated and extrapolated $M_\star$ reliability map. 
    The total interpolation region is subdivided into triangles (yellow dashed lines) in which the reliability is interpolated linearly. 
    Outside the interpolation region, the reliability is set to the nearest non-null value divided by a factor of its distance. The median uncertainty $\sigma_{M_\star}$ and median $R_{M_\star}$ are indicated.
    }
    \label{fig:reliability_map_construction}
\end{figure*}

From the upper panels of \cref{fig:compa_AGN_models}, there are several indication that the accuracy of the stellar mass estimates decreases with increasing AGN luminosity, the fraction of low-accuracy measurement $f_{\rm low-A}$ and the median uncertainty ${\rm med}(\sigma_{M_\star})$ progressively increase with increasing $\mathcal{L}_{\rm AGN, \, GT}$, while the bias, i.e. the weighted deviation from the ground truth value as a function of $M_{\star \rm ,\, GT}$ (solid line), is becoming more important \citep{Buchner_2024}.
This bias indicates that at higher $L_{\rm AGN,\, GT}$, stellar masses tend to be overestimated for low-mass galaxies. Nevertheless, the bias when averaged across the stellar mass range, remains small, with only a modest increase in its average scatter as a function of the AGN luminosity.

The middle row of \cref{fig:compa_AGN_models} shows, as expected, an opposite trend for the AGN luminosity: the accuracy of $L_{\rm AGN}$ measurements appears to improve with increasing $L_{\rm AGN, \, GT}$, as shown by the decreasing fraction of low accuracy sources, the decreasing mean bias, and the decreasing median uncertainty ${\rm med}(\sigma_{L_{\rm AGN}})$. 
The AGN luminosity is significantly over-estimated in massive galaxies at low $L_{\rm AGN, \, GT}$, while at higher $L_{\rm AGN, \, GT}$, this bias is much smaller.

The lower row of \cref{fig:compa_AGN_models} shows a more critical situation for the SFR, which can only be accurately estimated for weak AGN, i.e. $\mathcal{L}_{\rm AGN, \, GT}<42$ \citep{Buchner_2024} when the FIR photometry is not available. Indeed, we observe an increasing SFR offset with increasing true AGN luminosity, suggesting that \texttt{CIGALE} overestimates the SFR to reproduce the IR photometry produced in reality by AGN activity while returning relatively small uncertainties. Consequently, $f_{\rm low-A}$ is also increasing with AGN luminosity.

In \cref{appendix:compa_model} and \cref{appendix:compa_bands}, we conduct a similar analysis for alternative model setups (\cref{tab:alternative_setup}) and using different photometric bands, respectively. We conclude that our choice of baseline setup is slightly better than the alternative ones. However, we emphasise that all three wavebands (optical, NIR, and MIR) need to be combined to obtain the most accurate measurements.

    \subsection{Reliability of the measurements for AGN}\label{subsec:reliability_maps}

Based on the above SED fitting results on the Chimera AGN, we develop a new probabilistic approach to estimate the reliability of the SED fitting measurements in AGN host galaxies. We define the reliability $R_{\mathbf{\xi}}$ of a measurement $\mathbf{\xi}$ as the probability for it to be of high-accuracy (\cref{eq:correct_Mstar}) as a function of its $M\star$ and $L_{\rm AGN}$ values. 
To estimate this probability empirically, we partition the Chimera AGN in the $\mathcal{L}_{\rm AGN}$–$\mathcal{M}_\star$ plane into clusters, each containing at least ten sources to ensure sufficient statistics.
Then, for each cluster, this probability, i.e. the measurement reliability, is computed as the fraction of sources with a high-accuracy $\mathbf{\xi}$ measurement.

To obtain a continuous function, we assign each cluster’s reliability to its centroid and interpolate these values onto a finer grid in the $\mathcal{L}_{\rm AGN}$–$\mathcal{M}_\star$ plane. 
To enlarge the region of validity, sources located on the periphery of the source distribution, the \enquote{edge points}, are used as additional interpolation nodes, each inheriting the reliability of its parent cluster. 
Finally, we extrapolate by propagating the value from the nearest point within the interpolation region weighting by the inverse of the distance, and we set the reliability to zero beyond the range $7<\mathcal{M}_\star <13$ and $39<\mathcal{L}_{\rm AGN}<48$. 
The resulting surface defines a reliability function $\mathbf{R_\xi}$ that
can be applied to any real source using its estimated parameters $\mathcal{M}_\star$ and $\mathcal{L}_{\rm AGN}$ from SED fitting.

This reliability analysis can theoretically be applied to any physical property $\mathbf{\xi}$ derived by \texttt{CIGALE} or by any other SED-fitting algorithm. In practice, we apply it to $M_\star$, AGN luminosities, SFR, and sSFR, i.e. to parameters for which trustworthy ground-truth estimates can be established from the pure samples. Indeed, parameters that are strongly model-dependent and affected by severe degeneracies (e.g. the e-folding time of the main stellar population or the torus opening angle) are poorly constrained in the pure GAL or pure QSO samples, rendering comparisons of Chimera AGN measurements uninformative for these quantities. For $L_{\rm AGN}$, fitting the pure QSO sample provides a reliable ground truth, allowing us to define $R_{L_{\rm AGN}}$ analogously to $R_{M_\star}$. For SFR (and sSFR), FIR photometry in the pure GAL sample yields meaningful ground truths, from which we derive $R_{\rm SFR}$ (and $R_{\rm sSFR}$).

\subsubsection{Stellar Mass \texorpdfstring{($R_{M_{\star}}$)}{(R\_M*)}}

\Cref{fig:reliability_map_construction} illustrates the full steps described above to derive the stellar mass reliability function: the source clustering in the $\mathcal{L}_{\rm AGN}$–$\mathcal{M}_\star$ plane (left panel), the reliability calculation for each cluster (central panel), and the interpolation and extrapolation of the $R_{M_{\star}}$ function (right panel).
As reliability values below 0.5 are considered unreliable, we set the colour bar range from 0.5 to 1, to highlight the variation of reliability within the parameter space.

We can notice on the right panel of \cref{fig:reliability_map_construction} that the stellar mass reliability is relatively high in most of the interpolated region of the $\mathcal{L}_{\rm AGN}$–$\mathcal{M}_\star$ plane.  The $R_{M_\star}$ distribution of the Chimera AGN, when using all available photometric band is centred on $0.78 \pm 0.12$. In detail, the $M_\star$ measurements are reliable up $\mathcal{L}_{\rm AGN} \leq 42$, at higher luminosities, only AGN hosted in intermediate-mass galaxies $\mathcal{M}_\star \sim 10$ have $R_{M_\star}>0.9$. 
While at lower stellar masses the decrease of $R_{M_\star}$ is mostly due to a larger difference with the ground truth value, at higher stellar masses, underestimated uncertainties impact the reliability. Moreover, there are two regions with  $R_{M_\star}<0.5$. The first one is located above the $\log_{10}(\lambda_{\rm Edd})=0.5$ limit for which SED fitting underestimate the stellar mass of their host $\mathcal{M}_\star<9$. The second one concerns relatively bright AGN $\mathcal{L}_{\rm AGN} {\sim} 44$ in over-massive galaxies $\mathcal{M}_\star{\sim}11$. However, this offset with the ground truth appears to originate from the lack of FIR photometry rather than from the AGN contamination. 
Indeed, fitting the corresponding pure galaxies without AGN model and without including the FIR photometry yield consistent $M_\star$ values to the Chimera AGN ones, but present an offset to the ground truth values.

To account for the various combinations of photometric band available in real observations, we reproduced this methodology but using the SED fitting results obtained with different photometric bands. 
The $R_{M_\star}$ functions using only the \Euclid bands, or in addition with optical bands, including or not the $u$ band are shown on the first row of \cref{fig:relia_multiband} in appendix. 
We see that with only \Euclid photometry, the stellar mass measurements are relatively reliable ($\mathbf{\sim 0.8}$) only in galaxies with $\mathcal{M}_\star>9.5$. 
Adding optical bands expands the $R_{M_\star}$ interpolation region but counter-intuitively decreases slightly $R_{M_\star}$ due to smaller uncertainties. Adding MIR photometry improves moderately both the uncertainties and $R_{M_\star}$ and on a larger area of the $\mathcal{M}_\star {-} \mathcal{L}_{\rm AGN} $ plane.

\begin{figure}
    \centering
    \includegraphics[width=0.95\linewidth]{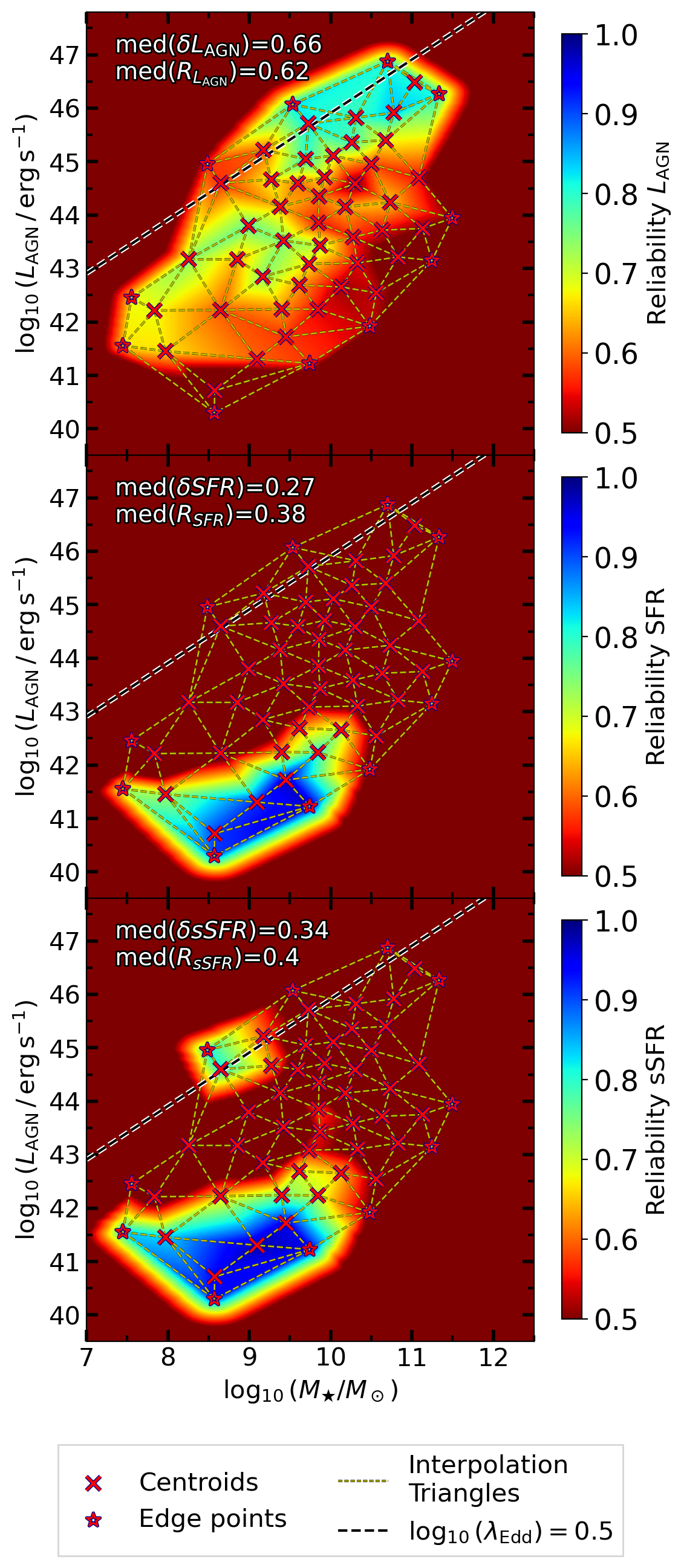}
    \caption{{\it Top panel}: AGN luminosity reliability $R_{L_{\rm AGN}}$ of the adapted Chimera sample in the $\mathcal{L}_{\rm AGN}{-}\mathcal{M}_\star$ plane. The median $L_{\rm AGN}$ uncertainty is indicated.
    {\it Central panel}: Same as top panel but for $R_{\rm SFR}$.
    {\it Bottom panel}: Same as top panel but for $R_{\rm sSFR}$.}
    \label{fig:Reliability_LAGN_SFR}
\end{figure}

The reliability values should be interpreted as a data-driven estimates of the SED-fitting performance, rather than formally calibrated probabilities.
Nevertheless, to assess the predictive power of our methodology, we randomly selected 80\% of the Chimera sample to construct another $M_\star$ reliability function. We then apply this function onto the remaining 20\% of the Chimera sample, and for each 0.1\,dex $R_{M_\star}$ interval, we measure the true fraction of high-accuracy measurement. The strong agreement between the predicted reliability and observed fraction,
shows that the reliability definition is well  calibrated within the parameter space explored and can be interpreted as the probability of obtaining a high accuracy measurement under similar observational conditions.

\subsubsection{AGN Luminosity \texorpdfstring{($R_{L_{\rm AGN}}$, $R_{L_{6 \micron}}$, and $R_{L_{2500\,\text{\AA}}}$)}{(R\_L\_AGN, R\_L\_6micron, and R\_L\_2500A)}}

The reliability function for $L_{AGN}$ when using all the available bands is shown in the top panel of \cref{fig:Reliability_LAGN_SFR}. 
$R_{L_{\rm AGN}}$ is significantly lower than $R_{M_{\star}}$, with a distribution centred on $0.62 \pm 0.09$ for the Chimera AGN sample. The $L_{\rm AGN }$ reliability is maximal at the highest AGN luminosities $\mathcal{L}_{\rm AGN} \geq 45.5$. As expected, the overall trend shows a $R_{L_{\rm AGN}}$ decrease with decreasing $L_{\rm AGN}/M_\star$ ratio, but with a band of reduced reliability around $\mathcal{L}_{\rm AGN}\sim 44.5$. 
Since the AGN luminosities are large, with ${\rm med}(\sigma_{L_{\rm AGN}})=0.66$\,dex, the low reliability regions are mainly due to a large scatter around the ground truth rather than underestimated uncertainties as it is partially the case for $R_{M_\star}$.

We then repeat this process to define the reliability functions of the AGN monochromatic luminosities, $R_{L_{6 \micron}}$ and $R_{L_{2500\,\text{\AA}}}$. 
We observe a similar shape of the reliability map with high reliability values at high $L_{\rm AGN}/M_\star$ ratios with a decrease around \mbox{$\mathcal{L}_{\rm AGN}\sim 44.5$}. While $R_{L_{6 \micron}}$ is similar to $R_{L_{\rm AGN}}$ with a median reliability distribution centred on $0.63\pm0.12$, $R_{L_{2500\,\text{\AA}}}$ is lower with \mbox{$0.55\pm0.16$}.

Finally, we also compute $R_{L_{\rm AGN}}$, $R_{L_{6 \micron}}$, and $R_{L_{2500\,\text{\AA}}}$ for various combinations of available photometric bands. In appendix, the middle row of \cref{fig:relia_multiband} shows that $L_{\rm AGN}$ measurements are not reliable without MIR photometry, except, moderately, in low mass galaxies ($\mathcal{M}_\star <9 $) thanks to the combination of optical and \Euclid NIR coverage.

\subsubsection{Star Formation Rate \texorpdfstring{($R_{\rm SFR}$ and $R_{\rm sSFR}$)}{(R\_SFR and R\_sSFR)}}
The central and bottom panels of \Cref{fig:Reliability_LAGN_SFR} show the $R_{\rm SFR}$ and $R_{\rm sSFR}$ functions, respectively.
Both reliability functions are significantly lower, with a median distribution for the Chimera AGN sample centred on $0.38\pm 0.20$ and $0.40\pm 0.20$, respectively.
They have a similar shape, with reliable measurements only in galaxies with $\mathcal{M}_\star<10.5$ and relatively low AGN luminosity $\mathcal{L}_{\rm AGN} \leq 42$. 
Interestingly, the $R_{\rm sSFR}$ map shows an additional small region with moderate reliability ($R_{\rm sSFR}\sim 0.75$) at high accretion rates ($\log_{10}(\lambda_{\rm Edd})\sim 0.5$) for $\mathcal{M}_\star \sim8.5$ galaxies. In this regime, both $M_\star$ and SFR are underestimated, compensating each others and resulting in a moderately reliable sSFR measurement.

The $R_{\rm SFR}$ and $R_{\rm sSFR}$ functions are also constructed for each combination of photometry. As shown in the lower row of \cref{fig:relia_multiband}, only the inclusion of MIR photometry significantly improve the reliability of the measurements. 
Moreover, even with MIR coverage, the low $R_{\rm SFR}$ values are due to the absence of FIR photometry bands in our Chimera AGN. Indeed, while the AGN contribution dominates in the mid-IR which can be probed by IRAC\,1/2 (or WISE\,1/2), the star-formation contribution dominates at longer wavelength and only the FIR can efficiently constrain SFR.
However, in our real observations in the EDFs, some sources have W3/W4 photometry which should improve the reliability and accuracy of both SFR and $L_{\rm AGN}$. Due to the absence of bands at longer wavelengths than IRAC\,2 for the Chimera AGN, it is impossible to quantify this improvement and, consequently, we provide instead a flag indicating good WISE\,3/4 photometry as an indicator of $L_{\rm AGN}$ and SFR reliability.

        \subsection{Reliability of the measurements for inactive galaxies \label{subsubsec:R_FT}}

As no AGN selection method is completely pure \citepalias{Q1-SP027}, we expect that a fraction of our AGN sample are misidentified inactive galaxies.
However, as the nature of the sources is not know {\it a priori}, we always fit their SEDs with galaxy+AGN models. To quantify how it impacts the measurements of the physical properties in inactive galaxies, we fit the pure GAL sample using our baseline galaxy+AGN model. 
Then, with the same method used for AGN in \cref{subsec:reliability_maps}, we construct the $R_{M_{\star}, \, \rm GAL}$, $R_{\rm SFR, \,  GAL}$, and $R_{\rm sSFR, \,  GAL}$ reliability functions for different combinations of available bands. 
Obviously, due to the absence of AGN activity in the pure GAL sample, there is no ground truth $L_{\rm AGN}$, and therefore, no $R_{L_{\rm AGN, \, GAL}}$ functions.

When using all available photometric bands, the $R_{M_{\star}, \, \rm GAL}$, $R_{\rm SFR, \,  GAL}$, and $R_{\rm sSFR, \,  GAL}$ distributions for the pure GAL sample are centred on $0.88 \pm 0.18$, $0.82 \pm 0.20$, and $0.82 \pm 0.21$, respectively.
These results show that the use of an AGN model in the SED fit of an inactive galaxy has little impact on the host galaxy physical measurements, even when \texttt{CIGALE} mistakenly returns high AGN luminosities, up to $\mathcal{L}_{\rm AGN}\sim45$.
Similarly to the $R_{M_\star}$ function shown in the right panel of \cref{fig:reliability_map_construction}, there are two regions of low $R_{M_{\star}, \, \rm GAL}$. The first one concerns the sources for which SED fitting claims wrongly high AGN luminosities $\mathcal{L}_{\rm AGN}\sim 45$ in host galaxies with underestimated stellar masses $\mathcal{M}_\star \leq 9.5$.
The second one is due to sources for which the inclusion of FIR photometry shifts $M_{\star, \rm \, GT}$ to lower values. Fitting with or without AGN models the pure GAL SEDs, excluding the FIR, provides consistent results, in agreement with the Chimera AGN estimates.
Interestingly, despite having no true AGN contribution in their SED, these sources are located in a relatively well-constrained range of SED-fitting-derived AGN luminosity.

\subsection{Reliability values in the catalogue}

For each source in the Q1 sample (i.e. samples A, B, C, \cref{subsec_samples}), we provide two reliability values for each physical parameters, one using the functions derived for AGN ($R_{M_{\star}}$, $R_{L_{\rm AGN}}$, $R_{L_{\rm 6\micron}}$, $R_{L_{2500\,\text{\AA}}}$, $R_{\rm SFR}$, and $R_{\rm sSFR}$) and one using those for inactive galaxies ($R_{M_{\star}, \, \rm GAL}$, $R_{\rm SFR, \, GAL}$, and $R_{\rm sSFR, \, GAL}$). 
These values are computed by applying the adequate reliability function set on the SED-fitting-derived $\mathcal{M}_\star$ and $\mathcal{L}_{\rm AGN}$ values.
The choice of function set depends on the available photometric bands for the source, particularly the $u$ band (only available in \mbox{EDF-N}) and/or the presence of MIR band (WISE/IRAC). Although the reliability functions were constructed using SDSS (optical) and IRAC (MIR) bands, they are not tied to a specific filter set. Instead, they depend only on the presence or absence of coverage in a given wavelength regime. For instance, MIR coverage provided by WISE\,1/2 or IRAC\,1/2 is treated equivalently. Similarly, \mbox{EDF-N} optical photometry comes from a mix of surveys, whereas \mbox{EDF-S} and \mbox{EDF-F} uses only DESI photometry, yet all three fields use the same reliability functions because the relevant information is the band coverage, not the survey of origin.
Finally, if the reduced $\chi_2$ is greater than 10, all reliability values are set to\,0. This empirical threshold separates meaningful SED fits from catastrophic failures, often resulting from incoherent photometry.
The reliability values provided in the catalogue can then be used to define source samples according to the user's requirements for measurement robustness or sample completeness.

We tested the sensitivity of the reliability functions to the adopted methodology. For instance, varying the number of clusters used to sample the $\mathcal{L}_{\rm AGN}$–$\mathcal{M}\star$ plane, as well as adopting an alternative nearest-neighbour approach, yields consistent reliability estimates, with only minor differences. Similarly, replacing the discrete $w_{\rm QSO}$ values with a continuous distribution does not introduce any significant systematic variation. These tests indicate that the derived reliability functions do not strongly depend on the methodological choices.

The set of reliability functions for the different properties and photometric band combinations, for active and inactive galaxies, is publicly available (\url{https://zenodo.org/records/20065887}). These functions can be applied to any previous and future SED fitting results obtained with \texttt{CIGALE}, and are aimed to be employed to analyse \Euclid DR1 data. The use of these functions requires nonetheless a reasonably similar SED fitting setup as the one described in \cref{subsec:SED_modules} and a similar band coverage, from optical to mid-IR. The impact of additional photometric bands, notably UV, FIR, or optical narrow-bands, cannot be investigated with the current Chimera benchmark, but one could expect the reliability functions used in this paper to be lower limits.
Unfortunately, for significantly different \texttt{CIGALE} setups or for other SED fitting algorithms, the reliability functions should be re-computed altogether from the Chimera benchmark \citep{Buchner_2024}.

One limitation of the reliability functions is incapacity to be extrapolated outside of the ground truth parameter ranges, \mbox{$7.5<\mathcal{M}_{\star , \, \rm GT} < 11.3$}, \mbox{$41.0<\mathcal{L}_{\rm AGN, \, GT} < 46.4$}, and \mbox{$-5 <\mathcal{SFR}_{\rm GT} < 2.1$}. This means that any measurement located outside of these ranges would be attributed a low reliability value, while not fundamentally inaccurate. To avoid automatically discarding these sources, we also provide a flag indicating if whether they lie within the interpolation region of the reliability functions.
In addition, the Chimera AGN sample, designed to span a wide parameter space rather than to reproduce a fully physical AGN population, may include non-physical, or not yet-observed, AGN-host configurations. 
The adopted Eddington ratio cut suppresses the most extreme cases for which SED fitting poorly constrains the host galaxy physical properties, but constrains relatively well the AGN ones.
Applying such a cut therefore increases the $M_\star$ reliability primarily in the highest $L_{\rm AGN}$ and $M_\star$ regime, while decreases the $L_{\rm AGN}$ reliability in the highest accretion rate regime.
We consider that these reliability differences are justified since the Eddington ratio cut produces a more realistic AGN population, better suited to apply on real observations.

\subsection{Measurement corrections} \label{subsec:corrections}

In the previous sections, we present a method to compute the reliability of various parameter measurements. Here, we propose two possible approaches to improve the reliability of the measurements: either by increasing the uncertainties or by shifting the best-fit values.

    \subsubsection{Correction of the uncertainties}\label{subsubsec:needed_sigma}

From the top row of \cref{fig:compa_AGN_models}, we notice that a large fraction (55\%) of the low-accuracy  $M_\star$ measurements lie within 0.3\,dex from the ground truth. This suggests that \texttt{CIGALE} underestimates the $\sigma_{M_\star}$ uncertainties \citep{Buchner_2024}.
Therefore, to improve {\it a posteriori} the reliability, we propose to increase the output uncertainties, making them more realistic.

Similarly to the construction of reliability functions, we split the sources into clusters as a function of their output $\mathcal{M}_\star$ and $\mathcal{L}_{\rm AGN}$.
Then, for each source, we estimate an uncertainty multiplicating factor $n_{\sigma, \, M_\star}$ such that $M_{\star, \, \rm GT}$ lies within the 1\,$\sigma$ quadratic uncertainty (used in \cref{eq:correct_Mstar}). 
For sources with high-accuracy $M_\star$ measurements $n_{\sigma, \, M_\star} = 1$, while for those with low-accuracy,
\begin{equation}
    n_{\sigma, \, M_\star} = \sqrt{\frac{(\mathcal{M}_{\star,\,\rm Chimera} - \mathcal{M}_{\star,\,\rm GT})^2 - \sigma_{\mathcal{M}_{\star,\,\rm GT}^2}}{\sigma_{\mathcal{M}_{\star,\,\rm Chimera}^2}}} \, ,
    \label{equa:nb_sigma}
\end{equation}
\noindent with a fixed maximum value of $n_{\sigma, \, M_\star}=4$.
For a Gaussian distribution, we expect to have 68\% of the measurements within the 1\,$\sigma$ interval, analogue to a 0.68 reliability value. Therefore, we attribute to each cluster of sources the value of the 68$^{\rm th}$ percentile of their $n_{\sigma, \, M_\star}$ distribution.
Clusters with less than 68\% of measurements with $|\xi_{\rm Chimera} - \xi_{\rm GT}|<0.5$ have systematically a reliability value lower than 0.68, regardless on the potential multiplicative factor. These clusters are therefore discarded when interpolating $n_{\sigma, \, M_\star}$. Similarly, as its behaviour is unknown outside of the interpolation region, $n_{\sigma, \, M_\star}$ is not extrapolated.

The same process is repeated for all the band combinations, but also for the inactive galaxies.
For $L_{\rm AGN}$ and more importantly SFR, low-accuracy sources mostly originate from a large offset with the ground truth value, rather than from underestimated uncertainties. 
Consequently, in the catalogue, we only provide $n_{\sigma, \, M_\star}$ which can be applied to guaranty a minimum of 68\% reliability for each source.

\subsubsection{Mean offset correction}\label{subsubsec:mean_offset}

Another approach that can be adopted to improve the reliability of our measurements is to measure the mean offset with the ground truth value as a function of the SED-derived $\mathcal{M}_\star$ and $\mathcal{L}_{\rm AGN}$ values. 
Similarly to the reliability functions, we compute for each source cluster the median and standard deviation of the distribution of offsets. These values are then interpolated between cluster, but not extrapolated because their behaviour outside of the interpolation region is unknown.
The offset-corrected values are obtained by adding the interpolated median offset to the SED-derived value, while the interpolated standard deviation of the offset is added in quadrature to the measurement uncertainties.

This correction approach is worth applying only to the SFR, because, as shown in \cref{fig:compa_AGN_models}, the bias for both $M_\star$ and $L_{\rm AGN}$ is not significant with respect to the scatter, while SFR is strongly shifted to higher values than the ground truth. 
Consequently, in the catalogue, we provide offset-corrected values and uncertainties only for the SFR.
However, we caution that due to the larger uncertainties, we recommend using the bias-corrected values only for statistical purposes, when analysing the SFR properties of large samples. For studies of individual objects, the best-fit SFR values should be used carefully, and preferably if the WISE\,3/4 photometry is available.

\section{Results and discussion}\label{6- Discussion}

With our baseline SED fitting model setup presented in \cref{3-SED fitting}, we use the photometry of our three samples in the three EDFs to constrain the physical parameters of AGN (bolometric and monochromatic luminosities, AGN fraction, and torus viewing angle) and host galaxies ($M_\star$, age, SFR, sSFR, and dust attenuation) and to compute the reliability values of several of these parameters.

\begin{figure}
    \centering
    \includegraphics[width=0.95\linewidth]{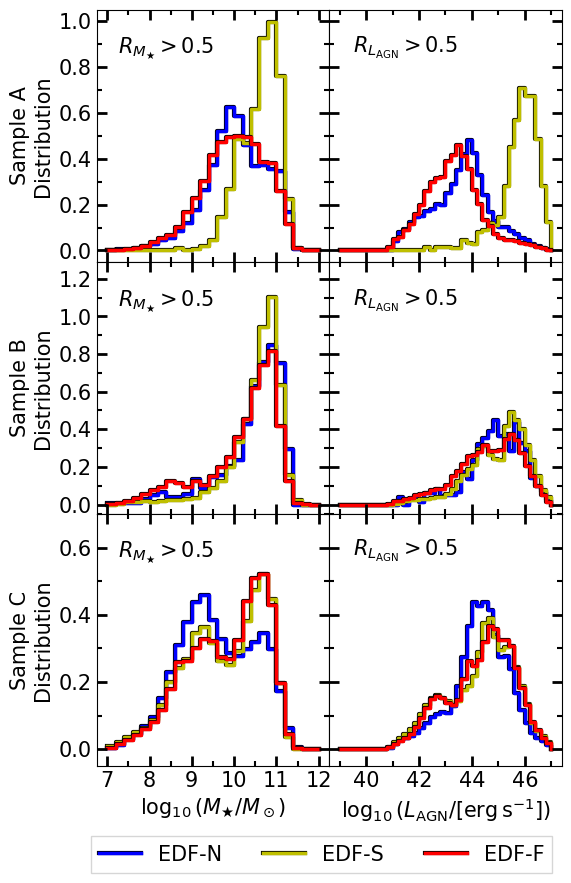}
    \caption{Reliable stellar mass ({\it left column}) and AGN luminosity ({\it right column}) normalised distributions for sample\,A (sources with spec-$z$, {\it top row}), B (X-ray detected AGN, {\it middle row}), and C (all AGN-candidates, {\it bottom row}) for the three EDFs.}
    \label{fig:sample_distribs}
\end{figure}

 \subsection{\Euclid stellar masses and AGN luminosities}\label{subsec:results}

\Cref{fig:sample_distribs} shows the stellar mass (left panel) and AGN luminosity (right panel) distribution in each field for the three analysed samples, sample\,A containing all sources with spectroscopic redshift, sample\,B including all the X-ray detected AGN, and sample\,C containing all AGN-candidates (\cref{subsec_samples}).
Here, we considered only sources with a reliability greater than 0.5 in the respective parameters.
For $M_\star$, it represents 85\%, 71\%, and 87\% of the sources in \mbox{EDF-N}, \mbox{EDF-S}, and  \mbox{EDF-F}, respectively, in sample\,A, 73\%, 68\%, and 77\%, respectively, in sample\,B, and 88\%, 88\%, and 88\%, respectively, in sample\,C.
For $L_{\rm AGN}$, it represents 70\%, 80\%, and 82\%  of the sources in \mbox{EDF-N}, \mbox{EDF-S}, and  \mbox{EDF-F}, respectively, in sample\,A, 68\%, 68\%, and 79\%, respectively, in sample\,B, and 70\%, 78\%, and 80\%, respectively, in sample\,C.
Looking at the distribution of the stellar masses (left panels of \cref{fig:sample_distribs}), we note that  sample\,B, shows a significantly narrower distribution located at higher stellar masses for all EDFs compared to the two other samples, indicating that X-ray AGN are on average found in more massive galaxies. Interestingly, in sample\,C, all fields show a bi-modality in their distribution, while both \mbox{EDF-S} and \mbox{EDF-F} are peaking at $\mathcal{M}_{\star}=9.3$ and $\mathcal{M}_{\star}=10.7$, the strength of the two peaks is inverted for \mbox{EDF-N}.
This observation is a consequence of the various AGN selection methods, notably the PRF, which is discussed further in \cref{subsec:selections}.

The right panels of \cref{fig:sample_distribs} reveal that the sources of the sample\,A within \mbox{EDF-N} and \mbox{EDF-F} have an AGN luminosity significantly lower than those of sample\,B and\,C, which is not the case within \mbox{EDF-S}. This is due to the fact that samples\,B and\,C mostly consist of AGN while sample\,A is dominated by inactive galaxies with the exception of \mbox{EDF-S}, where most sources are QSOs from QUAIA \citep{Storey_Fisher_2024}.

The median values and the 1\,$\sigma$ uncertainties for each distribution are presented in \cref{tab:results_fields}, along with the redshift, AGN fraction, SFR, stellar mass reliability, and AGN luminosity reliability. 
\Cref{fig:compa_phosphoros} compares the stellar masses from our analysis with those derived by the \Euclid photo-$z$ pipeline for sample\,C sources, revealing a significant scatter across the entire mass range, and a systematic offset of approximately $-1$\,dex at low stellar masses $\mathcal{M}_\star<9.5$.
These discrepancies highlight the importance of proper AGN modelling when deriving physical properties and underscore the advantage of using our dedicated physical parameter catalogue over the photo-$z$ pipeline results optimised for inactive galaxies.

\subsection{AGN fraction and identification}\label{4-AGN_fraction}

By decomposing the photometric contributions of the different SED components, we define the AGN fraction $f_{\rm AGN}$ of a source as the ratio of the AGN flux to the total flux within a wavelength range. Following previous works \citep[][]{Dale_2014, Thorne_2022, Siudek_2025}, we identify AGN by measuring their AGN fraction $f_{\rm AGN}$ in the rest-frame MIR band (5--20\,\micron), in which the AGN contribution is the strongest with respect to the host galaxy emission, with stellar population in the optical and star-formation in the FIR.

A similar approach is presented in \citetalias{Q1-SP027}, using $f_{\rm AGN} >0.25$, they find the best compromise between purity and completeness on the spectroscopic sample of EDF-N.
However, we notice that many galaxies wrongly identified as AGN with this method have large AGN fraction uncertainties $\sigma_{f_{\rm AGN}}$, indicating that \texttt{CIGALE} is unable to robustly constrain the AGN component in those sources. 
Instead, in our analysis, we take a more conservative approach by inspecting the 1\,$\sigma$ lower limit on the AGN fraction estimate, defined as \mbox{$f_{\rm AGN,\, low} = {\rm max}(f_{\rm AGN} - \sigma_{f_{\rm AGN}}, \, 0)$}, and selecting as AGN any source above a certain $f_{\rm AGN,\, low}$ threshold.

\begin{figure}
    \centering
    \includegraphics[width=0.95\linewidth]{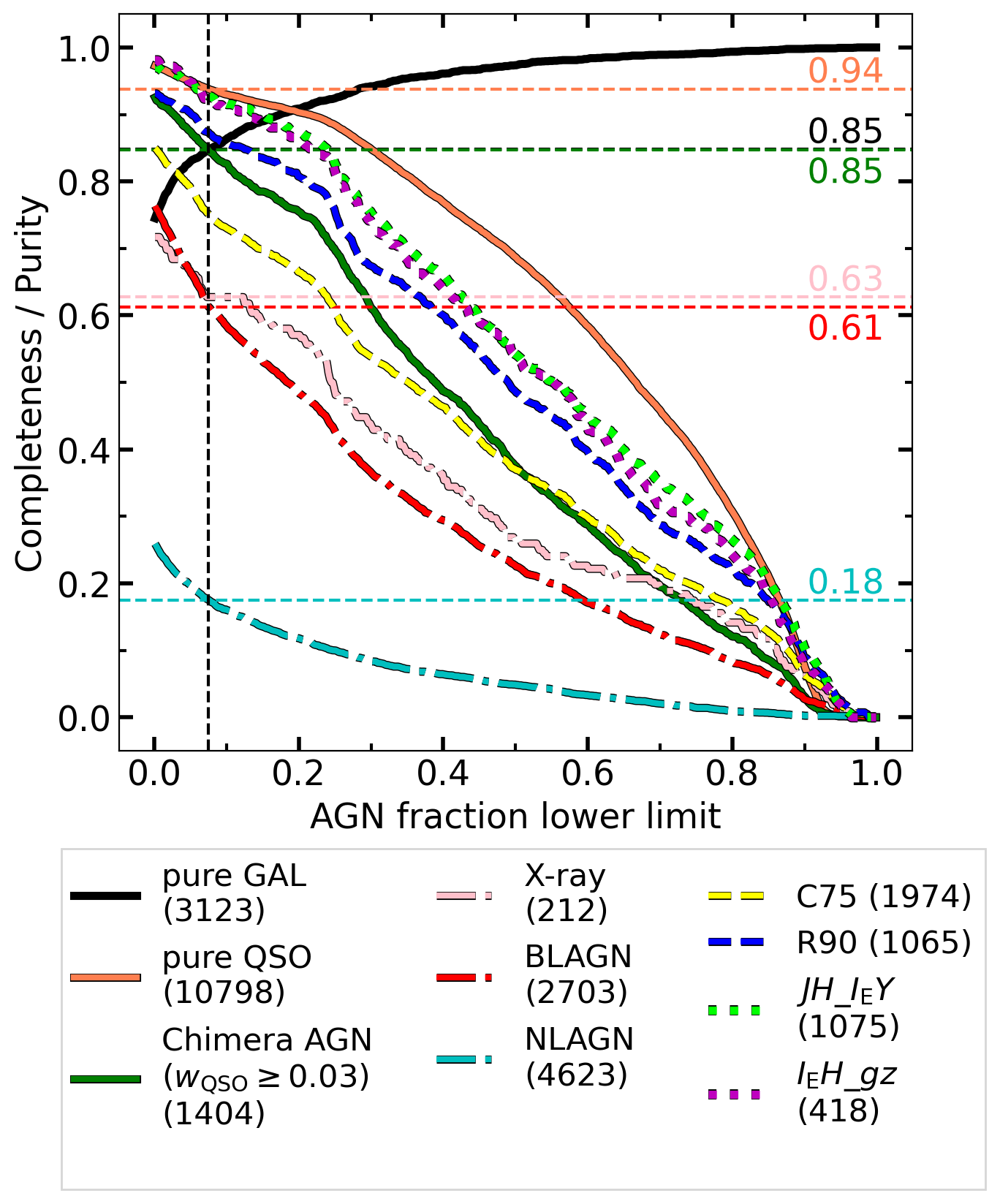}
    \caption{Completeness and purity of the SED-fitting AGN selection as a function of the AGN fraction lower limit.
    The black solid line shows the cumulative distribution of the pure GAL sample (\cref{subsec:chimera}), representing the purity of the SED-fitting approach, while the orange and green solid lines represent the completeness of the SED fitting method to identify pure QSOs and Chimera AGN, respectively. We only consider Chimera AGN with $w_{\rm QSO} \geq 0.03$.
    Dot-dashed lines represent the completeness of our approach to select robustly identified AGN (BLAGN, NLAGN, X-ray), while dotted and dashed lines represent respectively the completeness of NIR and MIR colour-colour-selected AGN.
    The number of sources in each sample is indicated in parenthesis.
    The horizontal lines highlight the corresponding completeness/purity when using the  $f_{\rm AGN,\, low} >0.075$ threshold.
    }
    \label{fig:AGN_fraclowlim}
\end{figure}

We analyse the AGN fraction lower limits of the three subsamples of the Chimera benchmark: the pure GAL, the pure QSO, and the Chimera AGN (\cref{subsec:chimera}). The pure GAL sample is used to measure the purity of our SED-based approach, while the pure QSO and Chimera AGN sample indicate the completeness of the approach to select sources from the respective samples. To guarantee a minimum of AGN activity with respect to the host galaxy emission, we only select Chimera AGN with a QSO weight \mbox{$w_{\rm QSO} \geq 0.03$}.
The purity, QSO completeness, and Chimera AGN completeness of the SED-based approach as a function of $f_{\rm AGN,\, low}$ threshold are displayed with solid black, orange, and green lines in \cref{fig:AGN_fraclowlim}. 
We suggest using \mbox{$f_{\rm AGN,\, low}>0.075$} (black vertical dashed line), the intersection of the purity and Chimera AGN completeness curves, as a threshold to select AGN, i.e. sources with a non-negligible AGN fraction, even accounting for uncertainties. This results in a 85\% purity and AGN completeness, and a 94\% completeness for QSOs as indicated by the corresponding horizontal dashed lines.
Nevertheless, users can adjust the $f_{\rm AGN, \, low}$ threshold based on their specific requirements for purity or completeness. For instance, with \mbox{$f_{\rm AGN,\, low}>0$}, the purity decreases to 75\% while the pure QSO and Chimera AGN completeness increase to 97\% and 93\%, respectively. Conversely, fixing the threshold to \mbox{$f_{\rm AGN,\, low}>0.2$}, increases the purity to 91\% but reduces the pure QSO and Chimera AGN completeness to 90\% and 76\%, respectively.
These purity and completeness values are estimated using the Chimera sample, therefore having IRAC\,1/2 coverage. With W3/W4 photometry, this AGN selection method could improve, and conversely, without any MIR coverage, the purity and completeness would decrease.

However, these purity and completeness values concern mostly BLAGN since the Chimera sample is constructed exclusively with broad-line QSO. This is further illustrated on \cref{fig:AGN_fraclowlim} when applying this analysis to the \Euclid observed sources benefiting from DESI's robust spectroscopic classification \citep{DESI_2025}, i.e. sample\,A within \mbox{EDF-N}.
Indeed, while BLAGN (red dot-dashed line) have a good completeness of 61\%, almost no NLAGN (cyan dash-dotted line) are identified as AGN with our SED-fitting-based method, resulting in a completeness of only 18\%.
This observation indicates, as mentioned by \citetalias{Q1-SP027}, that the SED fitting selection is not well-suited to select NLAGN because their AGN contribution is mostly absorbed, revealing an inactive galaxy-like SED.
\Cref{fig:AGN_fraclowlim} also shows that the completeness of X-ray AGN (pink do-dashed line) is relatively good, indicating 63\% at $f_{\rm AGN, \, low}>0.075$.
The comparison with these reliable AGN selection methods reveals that SED fitting is less efficient, but it has the advantage of not requiring specific observations, and it can be enhanced by additional photometric coverage, especially FIR which can improve the detection of obscured AGN \citep{Andonie_2022}.

\begin{figure}
    \centering
    \includegraphics[width=0.8\linewidth]{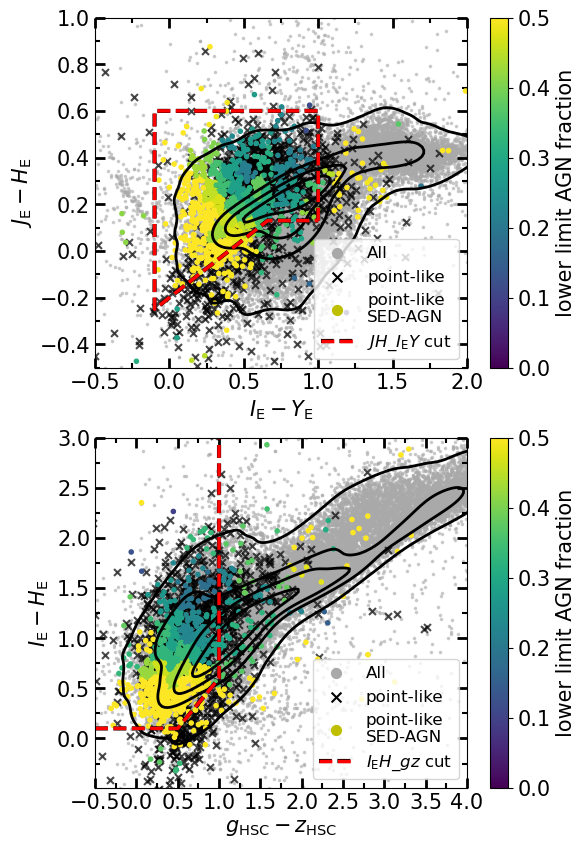}
    \caption{SED-based selection vs \Euclid colour-colour selection of AGN. {\it Top panel}: The grey points and black contours correspond to the sources of our sample\,A in \mbox{EDF-N} in the ($\JE-\HE$) versus ($\IE-\YE$) colour diagram and their respective density distribution. 
    The red dashed region is the AGN colour-colour cut defined in equation\,6 of \citetalias{Q1-SP027}.
    The black crosses are the point-like sources, while the coloured dots are the point-like AGN identified via SED-fitting ($f_{\rm AGN, \, low}>0.075$) and coloured as a function of the local average $f_{\rm AGN, \, low}$ value.
    {\it Bottom panel}: Same colour codes as in the upper panel but in the ($\IE-\HE$) versus ($g_{\rm HSC}-z_{\rm HSC}$) colour plane. The red dashed colour cut is defined by equation\,7 in \citetalias{Q1-SP027}.
    }
    \label{fig:colour_selection}
\end{figure}

\Cref{fig:AGN_fraclowlim} also shows the completeness of our SED-based approach to select AGN identified using MIR, (C75 and R90, yellow and blue dashed lines), yielding 75\% and 87\%, respectively.
Moreover, the lime and magenta dotted lines show respectively a completeness of 93\% and 92\% for the $JH$\_$I_{\text{E}}Y$ and $I_{\text{E}}H$ \Euclid colour-colour selection approaches presented in \citetalias{Q1-SP027} (see equations\,6 and\,7).
Moreover, in \cref{fig:colour_selection}, also for sample\,A within the \mbox{EDF-N}, we compare this NIR selections with our SED-based AGN approach.
The point-like (\texttt{MUMAX\_MINUS\_MAG} < $-2.6$) SED-selected AGN are split in clusters and their colour represent their mean $f_{\rm AGN,\,low}$ value. 
We can notice that most of them fell within the two colour-colour wedges, but also that the mean $f_{\rm AGN,\,low}$ varies gradually within that region. 
Nevertheless, we observe a few sources outside of the wedge with high $f_{\rm AGN, \, low}$ values and the visual inspection of their SED support their AGN nature in most cases.
Additionally, we inspect the redshift evolution of the $f_{\rm AGN,\,low}$ values within the two colour-colour planes. We find that, in both planes, the lobe with high AGN fraction, shifts progressively to the bottom right, getting closer to the edge of the respective colour-colour selections. This reflects the redshift evolution of the AGN colours and the advantage of using an SED fitting selection over a redshift-independent colour cut.

Our results in Sample\,C support the AGN nature of 179\,577, 182\,930, and 78\,558 AGN candidates within \mbox{EDF-N}, \mbox{EDF-S}, and \mbox{EDF-F}. This indicates that, if considering SED-based AGN selection entirely pure and complete, the combination of all AGN-selections has a purity of 42\%. This number is dominated by the PRF AGN selection method, constituting approximately half of sample\,C but for which only 13\% have a significant AGN contribution in their SED.

    \subsection{Comparison of the different AGN selections}\label{subsec:selections}

\begin{figure}
    \centering
    \includegraphics[width=0.95\linewidth]{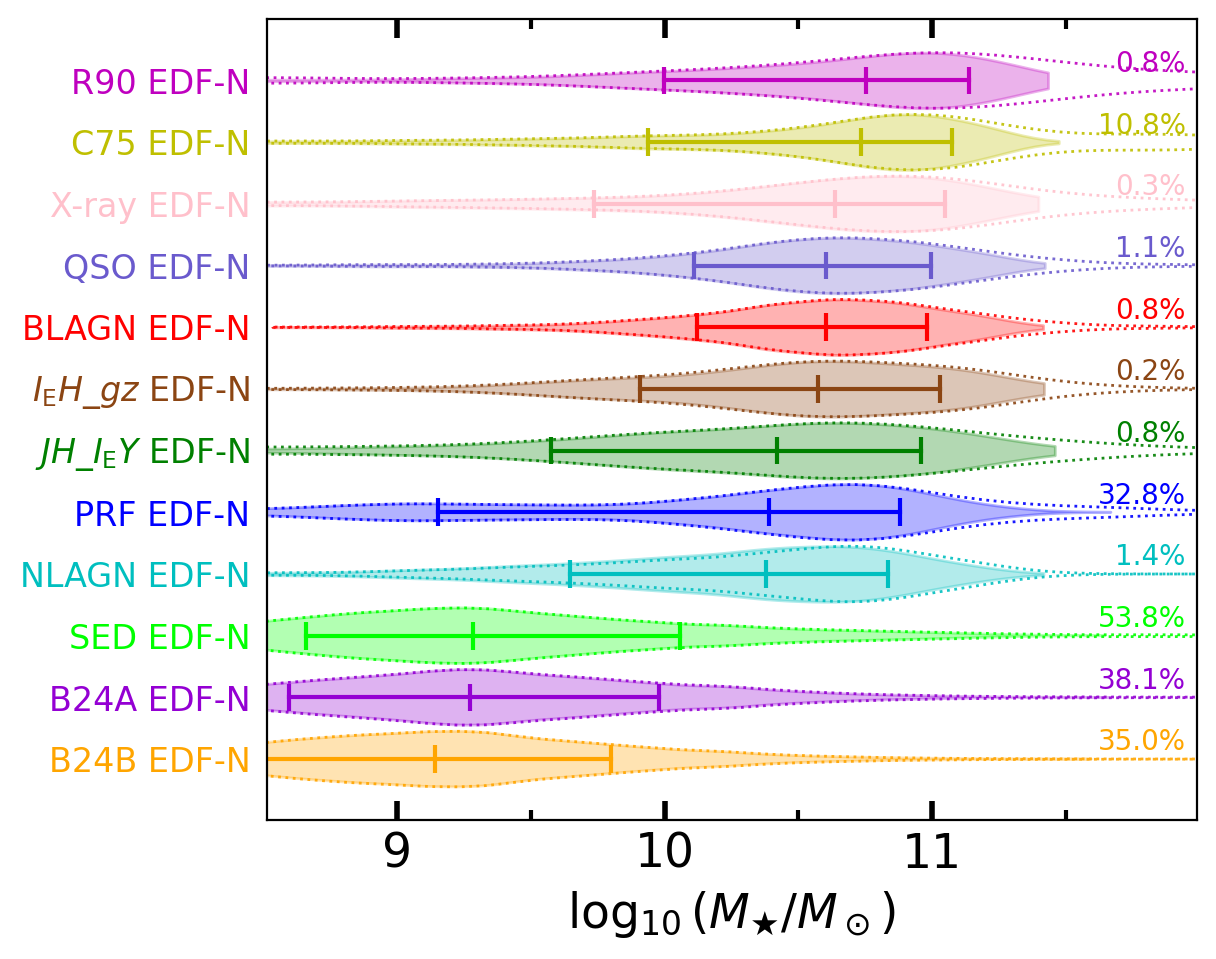}
    \caption{Stellar mass distribution for different AGN selections in \mbox{EDF-N}. The empty dotted violin plots and the filled ones correspond respectively to the entire samples and to the reliable subsamples ($R_{M_{\star}}>0.5$).
    The distributions are sorted from top to bottom by decreasing median reliable stellar mass. 
    The error bars represent the median and 1\,$\sigma$ scatter of each reliable distribution. The percentage value on the right of each violin plot indicates the fraction of the sample selected by the corresponding AGN selection.
    }
    \label{fig:Mstar_distrib_selections}
\end{figure}

\Cref{fig:Mstar_distrib_selections} compares the stellar mass distributions
of sample\,C within \mbox{EDF-N} from different AGN selections.
For each, both the full sample (empty dotted violins) and the reliable subsample ($R_{M_\star} > 0.5$, filled violins) are shown. The order of the violin plots is given by the median $M_\star$ of the reliable subsamples.

We notice differences in the stellar mass distribution between the different selections. For instance, MIR AGN  (C75 and R90) from \cite{Assef_2018} are hosted in massive galaxies $\mathcal{M}_\star>10.5$ while colour-colour selected AGN (B24A, B24B) from \citetalias{Bisigello_2024} are located on average in less massive hosts $\mathcal{M}_\star \leq 9.5$.  It is worth mentioning that some of these AGN selections target specifically QSO, the brightest AGN, by applying magnitude cuts, such as $\IE<21$ for $JH$\_$I_{\text{E}}Y$ and $I_{\text{E}}H$\_$gz$ \citepalias{Q1-SP027}, while B24A and B24B do not apply such a cut.
Additionally, our SED-based approach has the broadest $M_\star$ distribution, selecting AGN regardless of the mass of their host.
We also notice that the $R_{M_\star}>0.5$ cut excludes AGN hosted in the most massive galaxies $\mathcal{M}_\star \geq 11.5$. As mentioned in \cref{5-Reliability}, due to the absence of such massive galaxies in the Chimera benchmark, their reliability cannot be estimated properly and a low value is consequently attributed.

\begin{figure}
    \centering
    \includegraphics[width=0.95\linewidth]{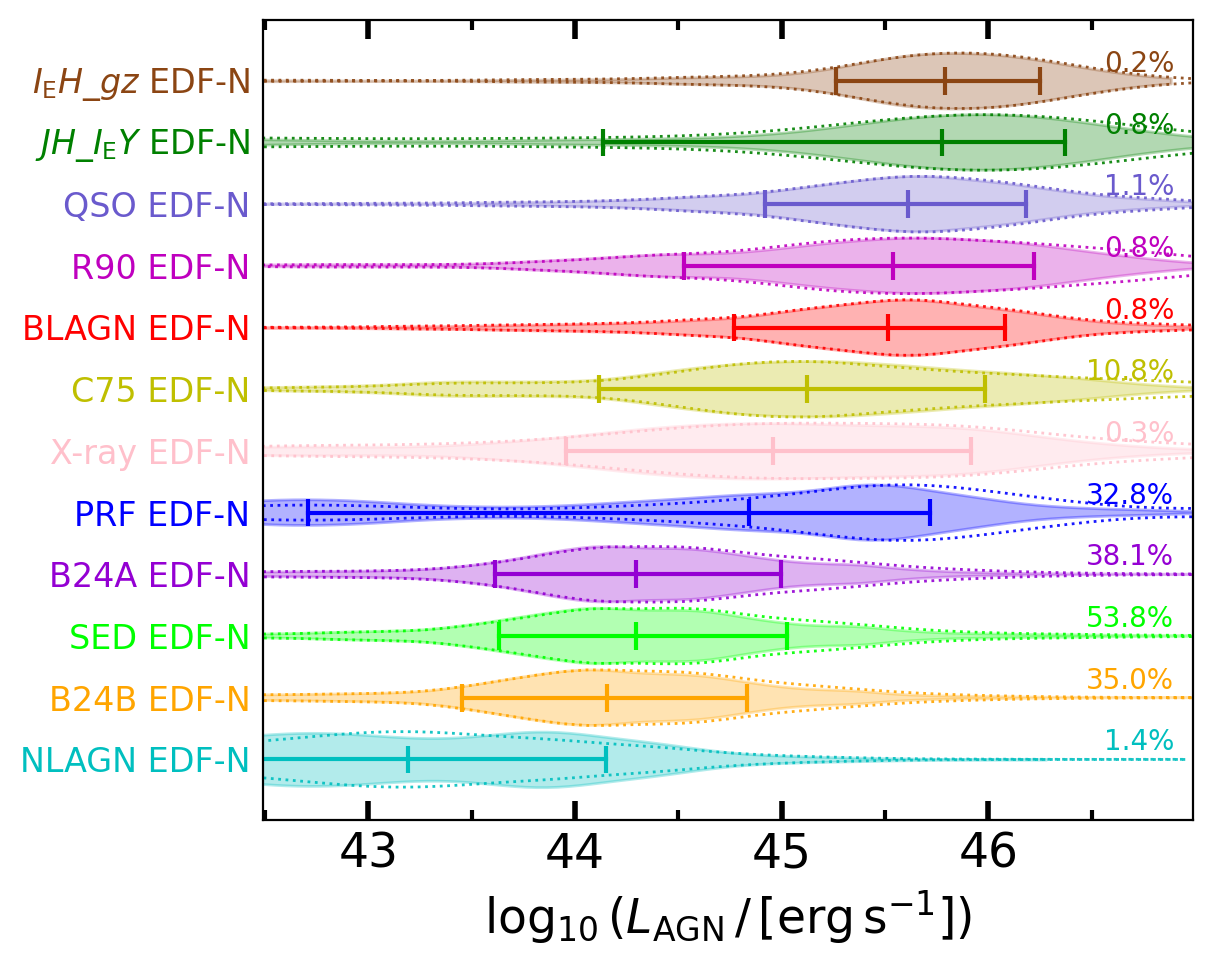}
    \caption{AGN luminosity distribution for different AGN selections of sample\,C within \mbox{EDF-N}.
    The empty dotted violin plots and the filled ones correspond respectively to the entire samples and to the reliable subsamples ($R_{L_{\rm AGN}}>0.5$).
    The distributions are sorted from top to bottom by decreasing median reliable AGN luminosity. 
    The error bars represent the median and 1\,$\sigma$ scatter of each reliable distribution. The percentage value on the right of each violin plot indicates the fraction of the sample selected by the corresponding AGN selection.}
    \label{fig:LAGN_distrib_selections}
\end{figure}

Similarly, \cref{fig:LAGN_distrib_selections} shows the $L_{\rm AGN}$ distributions as a function of the AGN-selection approach for both full and reliable ($R_{L_{\rm AGN}}>0.5$) samples.
Differences are notable, with median $\mathcal{L}_{\rm AGN}$ values ranging from 43 for the narrow line AGN to 46 for $I_{\text{E}}H$\_$gz$ AGN. Moreover, the shape of distribution varies too. For instance, the X-ray selection has the broadest distribution because, given the weak contamination by the host galaxy in the X-ray band, it can select AGN of various luminosities, even weak in the MIR with respect to their host. 
The AGN luminosity distribution for sources selected by SED fitting (\cref{4-AGN_fraction}) is also broad, not showing a specific luminosity bias. 
Additionally, for the PRF approach, the $\mathcal{L}_{\rm AGN}$ distribution displays a tail towards lower luminosity, potentially due to non-AGN contamination.

Interestingly, when investigating the full sample\,C across the three deep fields simultaneously, both the median values and overall shapes of the $M_\star$ and $L_{\rm AGN}$ distributions are remarkably consistent from field to field for each selection. This preserves approximately the order of the violin plots seen in \cref{fig:Mstar_distrib_selections,fig:LAGN_distrib_selections}. This consistency suggests that, rather than field-specific effects, these $M_\star$ and $L_{\rm AGN}$ difference between the AGN selections reflects primarily their intrinsic properties, notably their AGN purity.
Therefore, this implies that the various AGN-identification approaches do not sample uniformly the AGN population, introducing biases in the observed $M_\star$ and $L_{\rm AGN}$ distributions. 
As a consequence, AGN population studies, e.g. AGN incidence in galaxies, must properly account for the selection bias induced by each AGN identification methods in order to investigate the intrinsic property distribution.

\subsection{Relations to X-ray properties}\label{Xray_analysis}

The properties of X-ray AGN provide valuable information on their accretion properties and the torus line-of-sight column density.
The correlation between the monochromatic luminosity at 6 or 12\,\micron\, and the 2--10\,keV X-ray luminosity $L_{\rm X, \, 2-10\,keV}$ \citep{Gandhi_2009, Fiore_2009, Stern_2015} can be employed to study the multi-wavelength AGN emission mechanisms \citep{Mateos_2015}, but can also be used to convolve SED fitting with X-ray spectroscopy to break down parameter degeneracies \citep{Laloux_2023}. 
Our sample\,B of X-ray detected AGN is perfectly designed to investigate this X-ray/mid-IR relationship.

\begin{figure}
    \centering
    \includegraphics[width=0.95\linewidth]{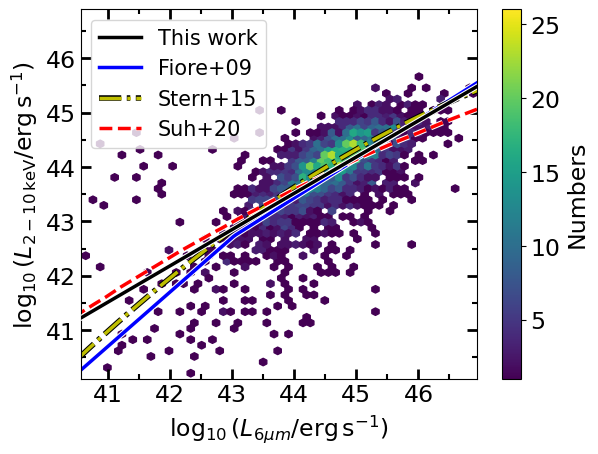}
    \caption{Distribution of the sources of sample B with reliable measurements in the $L_{\rm X, \, 2-10\,keV}$ and $L_{6\micron}$ plane.
    The mid-IR/X-ray relations from \protect\cite{Fiore_2009}, \protect\cite{Stern_2015}, and \protect\cite{Suh_2020} are over-plotted along the linear fit of this work (black solid line).}
    \label{fig:Lx_L6um}
\end{figure}

\Cref{fig:Lx_L6um} presents the distribution of sample\,B sources in the $L_{\rm X, \, 2-10\,keV}${-}$L_{6\micron}$ plane. For this analysis, we only consider the sources with reliable and well-constrained mid-IR measurements ($R_{L_{6\micron}}>0.5$ and $\sigma_{L_{6\micron}}<\log_{10}(3)\,{\rm dex}$), and with available photometry in each wavebands (optical, near-IR, and mid-IR). The $L_{\rm X, \, 2-10\,keV}$ value is derived from the 0.5--2\,keV X-ray luminosity from \citetalias{Q1-SP003}, assuming a simple power law spectrum with an effective photon index of $\Gamma_{\rm eff}=2.18$, $\Gamma_{\rm eff}=1.65$, and $\Gamma_{\rm eff}=1.51$ for the sources from eROSITA DR1 \citep{Merloni_2024}, CSC2 \citep{Evans_2024}, and 4XMM \citep{Webb_2020}, respectively.

The linear fit of the source distribution yields 
\begin{equation}
\begin{split}
\log_{10}\left(\frac{L_{\rm X,\, 2-10\,keV}}{\rm erg\,s^{-1}}\right) 
    &= (42.10\pm0.04) \\
    &\quad+ (0.69\pm 0.01) 
      \log_{10}\left(\frac{L_{6\micron}}{10^{42} \,\rm erg\,s^{-1}}\right)
\end{split} \, .
\end{equation}
\noindent This result is in agreement with previously determined relations \citep{Fiore_2009, Stern_2015, Suh_2020}, all of them based only on COSMOS field data. While our results cannot discriminate between these relations, they allow us to generalise the MIR-X-ray relation to a different and larger area.
Moreover, by extending this work to \Euclid DR1 data which will observe very luminous and rare objects, we will be able to provide better constraints on the high-luminosity tail of such relation.

\begin{figure}
    \centering
    \includegraphics[width=0.95\linewidth]{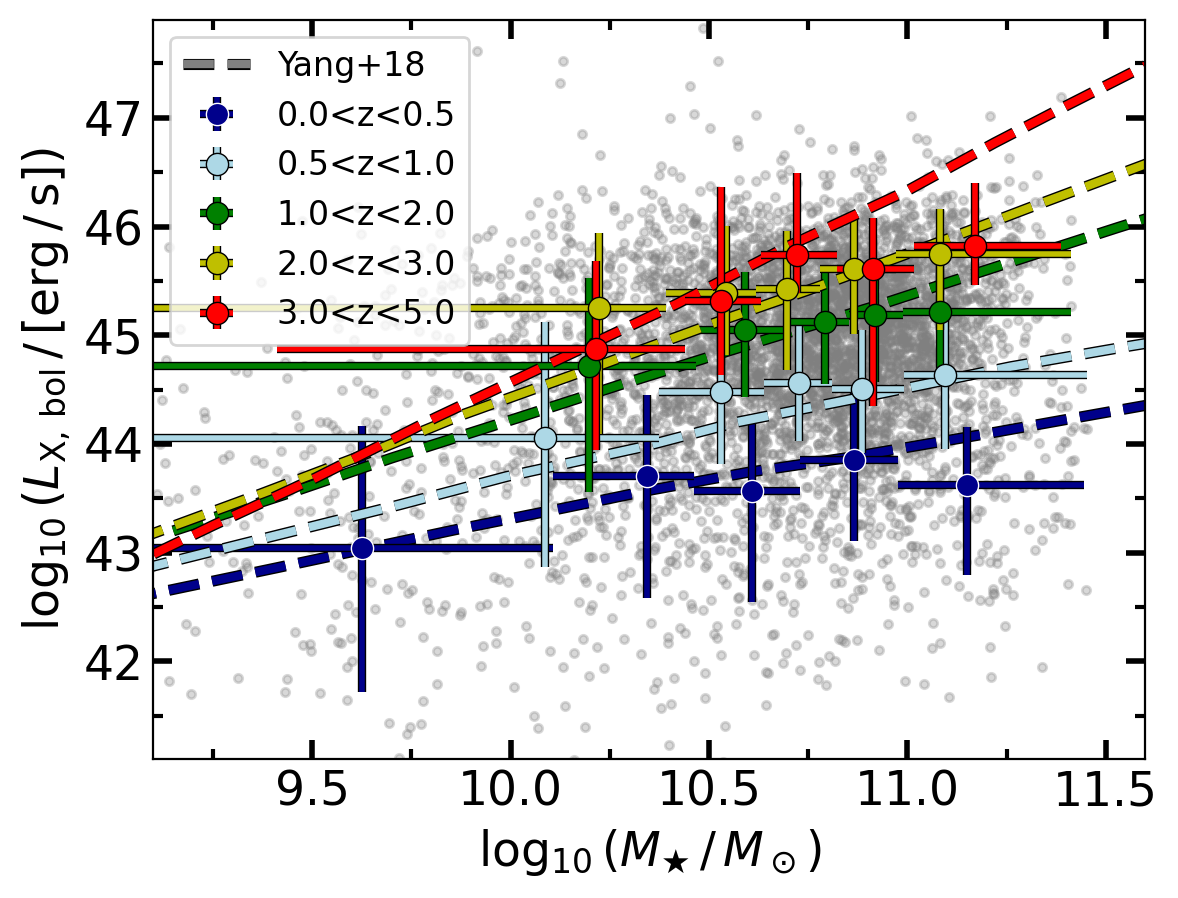}
    \caption{Distribution of the sources of sample\,B in the $M_\star{-}L_{\rm X, \, bol}$ plane. The dashed lines represent the redshift-dependent relations from \protect\cite{Yang_2018}, while the coloured error-bars correspond to the median $L_{\rm X, \, bol}$ as a function of stellar mass for different redshift intervals.}
    \label{fig:Mstar_Lbol}
\end{figure}

Using the same sample\,B, we can also study the relation between the bolometric luminosity derived from the X-ray luminosity, which is a good indicator of the accretion rate of matter into the SMBH, and the host galaxy stellar mass and its evolution with redshift. This provides information on the fuelling mechanisms of AGN \citep{Yang_2018, Suh_2020}.
\Cref{fig:Mstar_Lbol} shows the distribution of  sample\,B sources in the  $M_\star$ and $L_{\rm X, \, bol}$ plane. Again, we limit our sample to the sources with constrained and reliable $M_\star$ measurements. The $L_{\rm X, \, bol}$ is derived from the previously derived 2{--}10\,keV X-ray luminosity using the bolometric conversion presented in \cite{Duras_2020}.
The dashed lines represent the $M_\star{-}L_{\rm X, \, bol}$ relations presented in \cite{Suh_2020}, derived from \cite{Yang_2018}, for different redshift intervals up to $z=5$. 
For the same redshift intervals, our sample is subdivided in same-size stellar mass bins for which the median and 1\,$\sigma$ dispersion are represented by error bars. 
\cite{Yang_2018} claims that black hole accretion, and consequently AGN bolometric luminosity, correlates with the stellar mass, with a relation steepening with redshift.
While our results agree well up to $z<3$, they appear to differ in the last bin with a flatter relation at high stellar masses.
The large scatter of the data weakens the correlation, as reflected by the low Pearson coefficients (0.2--0.4) across redshift bins. 
Not applying the reliability cut allows us, as previously mentioned, to probe higher stellar masses ($\mathcal{M}_\star>11.5$) but the correlation remains weak.

More detailed analyses of these X-ray/optical-IR relations can be carried on to unveil nuclear-to-large scale AGN mechanisms. For instance, by extending this work to the future \Euclid DR1, the number of sources with reliable physical properties will increase ten-folds, providing larger statistical significance, especially at highest luminosities. 
This unique sample will particularly synergise with eROSITA DR1 and its future data releases, allowing an unprecedented multi-wavelength analysis of AGN physical properties, inspecting simultaneously their evolution and parameter co-dependence.

\section{Summary and conclusions}\label{7-Summary}

In this paper, we have exploited the \Euclid Q1 AGN catalogue \citepalias{Q1-SP027} to derive AGN and host galaxies physical properties through multi-component SED fitting. We introduced a new reliability parameter $R_\xi$ that quantify the robustness of these measurements and used the derived AGN fraction as an additional method to select AGN.
Our main results can be summarised as follows:

\begin{itemize}

    \item Using mock AGN SEDs, built from the combination of observed SEDs of galaxies and QSO, we investigated the impact of AGN activity on SED fitting results (\cref{fig:compa_AGN_models,subsec:chimera,subsec:Mstar_compa});

    \item We define reliability functions $R_\xi$ to estimate the probability to have a high-accuracy measurement (\cref{eq:correct_Mstar}) for various parameters (\cref{subsec:reliability_maps,fig:reliability_map_construction,fig:Reliability_LAGN_SFR}). Adopting a $R>0.5$, we find that 88\%, 76\%, 8\%, and 14\% of the Q1 AGN-candidates have reliable $M_\star$, $L_{\rm AGN}$, SFR, and sSFR estimates, respectively. The reliability depends on both AGN luminosity and host stellar masses: host-galaxy properties are better constrained at low-to-moderate AGN dominance while AGN luminosities are more reliable at high $L_{\rm AGN}/M_\star$ ratios;

    \item The reliability functions have been defined for various combinations of photometric coverage and can be applied to any sample under the condition of using reasonably similar SED fitting setup, notably upcoming \Euclid DR1;
    
    \item We propose a strategy to improve the reliability of $M_\star$ measurements by increasing their underestimated uncertainties (\cref{subsubsec:needed_sigma}). 
    To improve the low SFR reliability of large populations, we suggest to shift the derived values (\cref{subsubsec:mean_offset}).

    \item We present a criterion selecting AGN within a galaxy sample based on the SED-derived AGN 5--20\,\micron\,  flux fraction $f_{\rm AGN}$ that accounts for its uncertainties (\cref{4-AGN_fraction}). We propose to use the AGN selection threshold $f_{\rm AGN, \, low}>0.075$ to obtain a completeness and purity of 85\% (\cref{fig:AGN_fraclowlim,fig:colour_selection});

    \item By comparing with other methods, we find that using the AGN fraction misses AGN reliably identified in the X-ray or by optical spectroscopy, especially for NLAGN. This method is comparable to other colour-colour selections, with the advantage to be less affected by $z$ evolution and to explore fainter magnitude.

    \item We find that the AGN populations selected by various methods have different average properties in terms of $M_\star$, for instance, mid-IR selected AGN are brighter and hosted in more massive galaxies compared to those selected by optical/near-IR colour-colour cuts. SED fitting appears to select a broader, more unbiased AGN population. This result highlights the importance of combining different approaches and accounting for their selection biases when studying AGN in order to really capture the full diversity of AGN and their role in galaxy evolution (\cref{fig:Mstar_distrib_selections,fig:LAGN_distrib_selections,subsec:selections});

    \item For the X-ray detected sample, we analysed the X-ray to MIR relation and the  $M_\star{-}L_{\rm X, \, bol}$ relation as a function of redshift, finding good agreement with previous results. In particular, we find a clear evidence of a redshift evolution of the $M_\star{-}L_{\rm X, \, bol}$ relation, although we found a weak correlation in the individual $z$ bins (\cref{fig:Lx_L6um,fig:Mstar_Lbol,Xray_analysis}).
    
\end{itemize}

Our work demonstrates that accounting for AGN contribution is a challenging but essential step to derive robust physical AGN and galaxies properties. 
We release a publicly available catalogue (\url{https://zenodo.org/records/20065887}) containing the physical properties of 1\,111\,908 sources included in the three samples defined in \cref{subsec_samples}.
The catalogue include the physical properties of the AGN: $L_{\rm AGN}$, $L_{2500\,\text{\AA}}$, $L_{6\,\mu{\rm m}}$, $f_{\rm AGN}$, and torus viewing angle; as well as of their host galaxies: $M_\star$, age, SFR, sSFR, and $E(B-V)_{\rm dust}$ (\cref{fig:sample_distribs}).
Additionally, for the stellar mass, various AGN luminosities, SFR, and sSFR, we provide the reliability of their measurements using the new methodology presented in this paper. We also provide the $n_{\sigma,\, M_\star}$ coefficients and the offset-corrected SFR values.

The main objective of this work is to provide a more critical point of view on SED-derived physical properties of AGN and host galaxies. The reliability functions used in our analysis are publicly available and can be applied to any work using similar SED fitting setups.
Notably, applying this methodology to the much wider \Euclid DR1 catalogue would have multiple potential applications, from studying black hole-stellar mass scaling relation to measurements of the specific accretion rate distribution, leading to a better understanding of the AGN and host-galaxy co-evolution.

\section{Data availability}\label{data_availability}

The catalogue of the physical properties of the ${\sim}1$ million sources analysed in this work and the catalogue of the full photometric $z$-PDF are both publicly available at \textcolor{blue}{link\_to\_CDS}.
Additionally, along these two catalogues, the data and the script to build the reliability functions presented in \cref{5-Reliability} are available for use at \url{https://zenodo.org/records/20065887}.

\begin{acknowledgements}
 
The authors thank the referee for their insightful comments that ultimately improved the quality of this work.
B.L., A.B. and V.A. acknowledge the support from the INAF Large Grant \enquote{AGN \& \Euclid: a close entanglement} Ob. Fu. 01.05.23.01.14.
A.B. acknowledge the hospitality of the University of Geneva.
This research was supported by the International Space Science Institute (ISSI) in Bern, through ISSI International Team project \#23-573 ``Active Galactic Nuclei in Next Generation Surveys''.
M.M. acknowledges support from the Spanish Ministry of Science and Innovation through the project PID2021-124243NBC22. This work was partially supported by the program Unidad de Excelencia Mar\'ia de Maeztu CEX2020-001058-M.
ELSA: Euclid Legacy Science Advanced analysis tools" (Grant Agreement no. 101135203) is funded by the European Union. Views and opinions expressed are however those of the author(s) only and do not necessarily reflect those of the European Union or Innovate UK. Neither the European Union nor the granting authority can be held responsible for them. UK participation is funded through the UK Horizon guarantee scheme under Innovate UK grant 10093177. The authors acknowledge the use of computational resources from the parallel computing cluster of the Open Physics Hub (https://site.unibo.it/openphysicshub/en) at the Physics and Astronomy Department in Bologna.
\AckQone
\AckEC  
\end{acknowledgements}

\bibliography{bib}

\begin{appendix}
  \onecolumn 

\section{Photometric redshift for AGN}\label{appendix:zphot}
\begin{figure}[ht]
    \centering
    \includegraphics[width=1\linewidth]{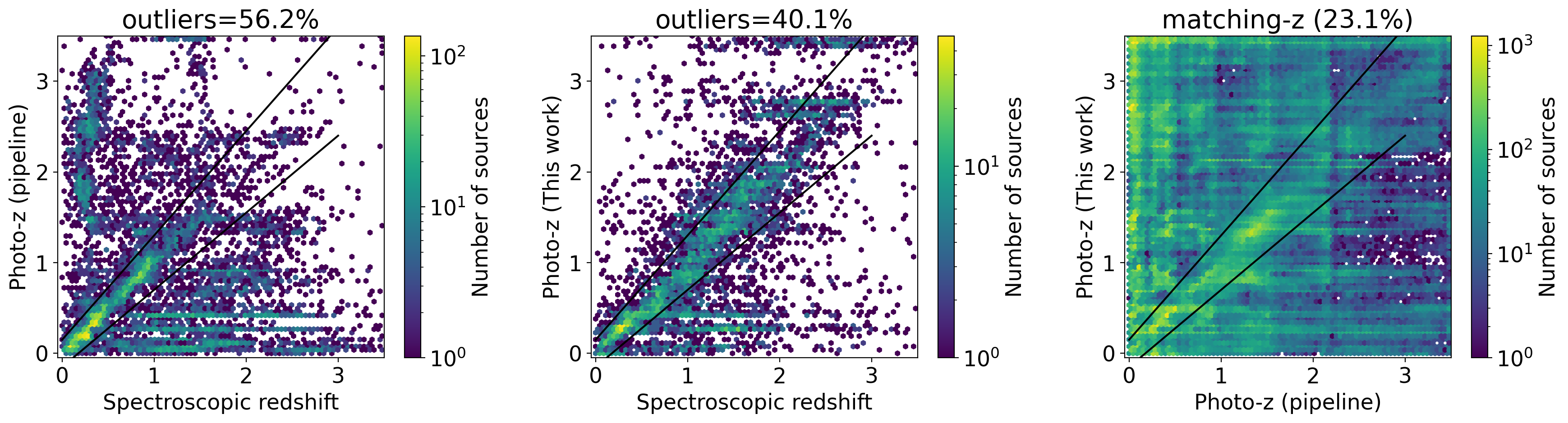}
    \caption{{\it Left panel}: Comparison of the photometric redshift from the pipeline against the spectroscopic redshift values for AGN-candidates. The black solid line separate the outliers defined as $|\frac{z - z_{\rm spec}}{z_{\rm spec} + 1}|>0.15$. {\it Central panel}: Similar to the left panel but with the photometric redshift estimates used in this work.
    {\it Right panel}: Comparison of the two photometric redshift estimates.}
    \label{fig:appendix_compa_zphot}
\end{figure}

\Cref{fig:appendix_compa_zphot} compares the photometric redshift estimates from the pipeline \citep[left panel, ][]{Q1-TP005} and the ones used in this work (central panel) with the spectroscopic redshift for the AGN-candidates of the three EDFs.
The outlier fraction, defined as the fraction of sources with $|\frac{z - z_{\rm spec}}{z_{\rm spec} + 1}|>0.15$, is respectively 56.2\% and 40.1\%. 
However, when inspecting the redshift accuracy for different AGN-selection samples, we notice that the outlier fraction for the pipeline photo-z is systematically above 75\%, reaching up to 95\% (for $JH$\_$I_{\text{E}}Y$ and $I_{\text{E}}H$\_$gz$ selected AGN). The only exception are of X-ray and NLAGN, which show an outlier fraction of 65.8\% and 30.1\%, respectively. Indeed, the SED of NLAGN look similar to inactive galaxies, while the AGN contribution to the SED can be weak for X-ray AGN.
In contrast, the outlier fraction of the photo-$z$ used in this work never exceeds 45\%.
Excluding NLAGN increases the pipeline outlier fraction to 69.8\% while our estimates remain at 40.1\%.
The right panel of \cref{fig:appendix_compa_zphot} compares the photometric redshift estimates from both methods, showing that only 23.1\% of the sources have matching redshifts. 
Overall, these results emphasise the need to account for AGN when measuring the photometric redshift. Our photo-$z$ estimates show a strong improvement compared to the pipeline estimates, and since the large majority of our AGN-candidate sources only benefits from photometric redshift, this has a significant impact on the SED fitting analysis. Our full photometric $z$-PDF are available at \url{https://zenodo.org/records/20065887}.

\section{SED fitting parameters}\label{appendix:SED_fitting}

\Cref{tab:cigale_params} presents the parameters of the different \texttt{CIGALE} SED modules in our baseline model setup presented in \cref{subsec:SED_modules}. It also indicates their fixed values or their range when set free to vary. 
On the other hand, \cref{tab:alternative_setup} summarises the different alternative setups. For each of them, only one SED module is changed from the baseline setup in order to quantify the impact of the change.

\begin{table}[ht]
    \centering
    \caption{Choice of modules and parameter range values for our SED fitting baseline model.}
    \begin{tabular}{ccc}
        \hline\hline
        & {\bf Module} & \\
         Parameter & Symbol & Value  \\
         \hline\hline
         
          & {\bf Star-formation history} & \\
            & Delayed SFH with recent burst & \\
          Age of the main population & $t_{\rm main}$ & 0.5, 1, 3, 4.5, 6, 8, 10, 13 Gyr\\
           e-folding time main population & $\tau_{\rm main}$ & 0.1, 0.5, 1, 3, 5, 8 Gyr\\
            Age of the burst & $t_{\rm burst}$ & 0.05 Gyr\\
             e-folding time of the burst & $\tau_{\rm burst}$ & 10 Gyr\\
              Burst stellar mass fraction & $f_{\rm burst}$ & 0.0, 0.001, 0.1, 0.2\\
        \hline \hline
        
          & {\bf Simple Stellar population} & \\
           & \protect\cite{Bruzual_2003} & \\ 
           Initial mass function & IMF & \protect\cite{Chabrier_2003} \\
            Metallicity & $Z$ & 0.02 \\
        \hline \hline
        
         & {\bf Galactic dust extinction} & \\
          & \protect\cite{Calzetti_2000} & \\
          Colour excess of nebular emission & $E(B-V)$$_{\rm line}$ & 0, 0.05, 0.15, 0.3, 0.5, 0.75, 0.9, 1.1, 1.3, 1.6\\
          Centre of UV bump & UV$_{\rm bump}^{\lambda}$ & 217.5 nm \\
        \hline \hline
        
        & {\bf Galactic dust emission} & \\
        & \protect\cite{Draine_2014} & \\
        Mass fraction of PAH & $q_{\rm PAH}$ & 0.47, 1.12, 2.5, 3.19 \\
        Minimum radiation field & $U_{\rm min}$ & 15 \\
        Powerlaw slope & $\alpha$ & 2.0 \\
        Illuminated fraction & $\gamma$ & 0.02 \\
        \hline \hline
        
        & {\bf Nebular emission} & \\
        & \protect\cite{Inoue_2011} & \\
        Ionisation parameter  & $\log_{10} U$ & $-2.0$ \\
        Width lines &  & 300 km\,s$^{-1}$ \\
        \hline \hline
        
        & {\bf AGN emission}  &   \\
        & \protect\cite{Stalevski_2016} & \\
        Optical depth at 9.7\,$\mu$m & $\tau_{\rm AGN}$ & 3, 7\\
        Torus viewing angle & $\psi$ & 0, 10, 20, 40, 70, 90\\
        AGN fraction & $f_{\rm AGN}$ & 0, 0.01, 0.1, 0.3, 0.5, 0.7, 0.9, 0.99\\

        \hline \hline
        
        Number of models per redshift & & 737 280\\
        \hline \hline
        
    \end{tabular}
    \label{tab:cigale_params}
\end{table}

\begin{table}[ht]
    \centering
    \caption{Summary of the alternative model setups.}
    \begin{tabular}{|c|c|c|}
        \hline \rule{0pt}{10pt}
        Changed & \multicolumn{1}{|c|}{\multirow{2}{*}{Baseline Model}} & \multicolumn{1}{|c|}{\multirow{2}{*}{Alternative Setup}} \\
        Module &  & \\\hline\hline \rule{0pt}{10pt}
        IMF & \cite{Chabrier_2003} & \cite{Salpeter_1955}\\\hline\hline
        \multirow{2}{*}{SP} &  \multirow{2}{*}{\parbox{2.3cm}{\cite{Bruzual_2003}}} & \multirow{2}{*}{\cite{Maraston_2005}} \\
        &&\\\hline\hline\rule{0pt}{10pt}
        Dust  & \multirow{2}{*}{\cite{Calzetti_2000}} & \multirow{2}{*}{\cite{Charlot_2000}} \\
        attenuation & & \\\hline\hline \rule{0pt}{10pt}
        Dust  & \multirow{2}{*}{\cite{Draine_2014}} & \multirow{2}{*}{\cite{Dale_2014}} \\
        emission & & \\\hline\hline \rule{0pt}{10pt}
        \multirow{2}{*}{AGN} & \multirow{2}{*}{\cite{Stalevski_2016}} & \cite{Fritz_2006} \\\cline{3-3} \rule{0pt}{10pt}
         &  & no AGN \\\hline
    \end{tabular}
    \label{tab:alternative_setup}
    \tablefoot{For each alternative setup, only one SED module is changed from the baseline setup.}
\end{table}

\begin{figure}[ht]
    \includegraphics[width=0.99\linewidth]{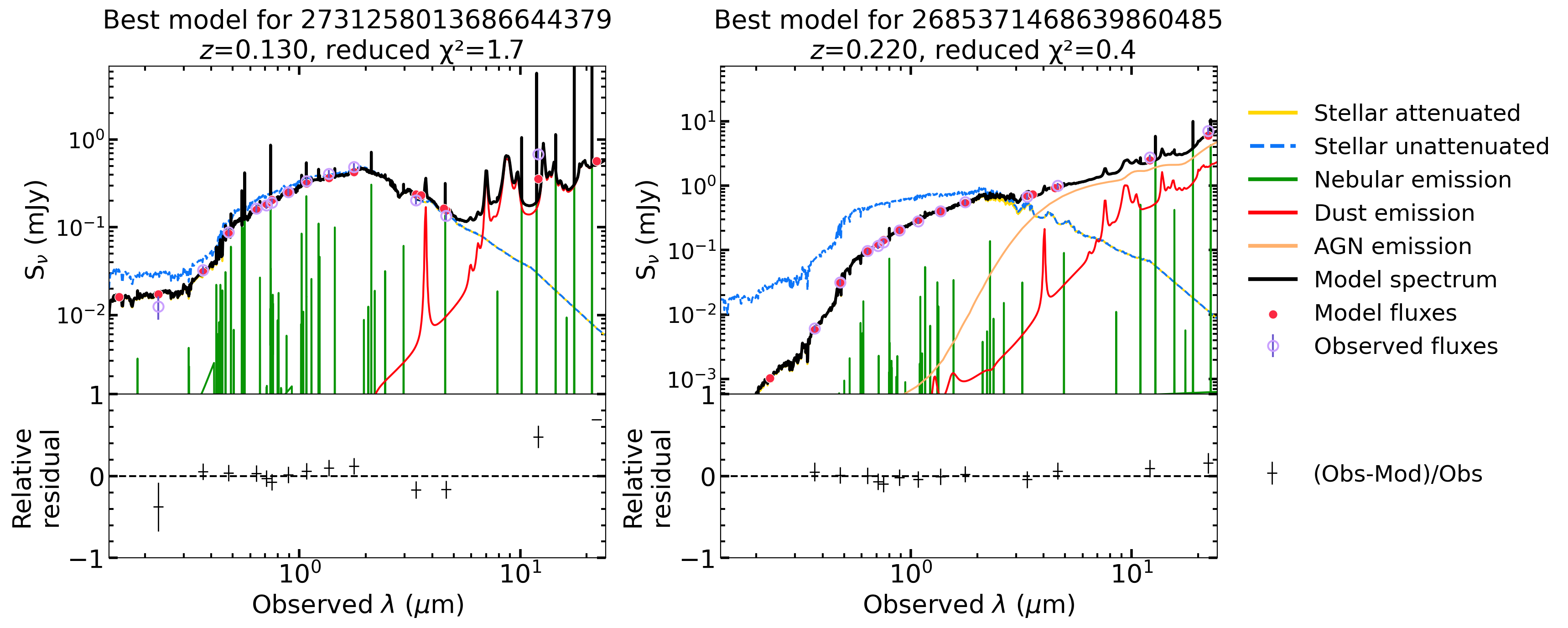}
    \caption{Example of best-fit SED models (black line) of a star-forming galaxy ({\it left panel}) and an AGN ({\it right panel}).
    The observed fluxes in the different bands are shown by purple circles, while the best-fit modelled fluxes in these bands are represented by red dots. The different coloured lines correspond to the various SED components presented in \cref{tab:cigale_params}.}
    \label{fig:best-fit_SED}
\end{figure}

\Cref{fig:best-fit_SED} shows two examples of best-fit SED using the baseline setup. At short wavelength ($\lambda \lesssim 2\,\mu{\rm m}$), both SEDs are dominated by the stellar emission (blue dashed line), which, once attenuated by the dust (yellow solid line), results in the best-fit model spectrum (black solid line).
At longer wavelengths, the SED differences reveal the nature of the two sources. For 2731258013686644379 (left panel), there is only dust emission (red solid line) from the star formation, indicating a star-forming galaxy, while for 2685371468639860485 (right panel), the AGN emission (orange solid line) dominates, revealing its AGN nature.
The best-fit SED on the left panel indicates a galaxy with a stellar mass $\mathcal{M}_\star=10.01 \pm 0.05$. Its AGN fraction $f_{\rm AGN}=0.03 \pm 0.07$, suggesting the absence of AGN (see \cref{4-AGN_fraction}). Following the methodology described in \cref{5-Reliability}, the reliability of this stellar mass measurement is $R_{M_\star}=0.91$.
On the other hand, the best-fit SED on the right panel yields a stellar mass and AGN luminosity $\mathcal{M}_\star=10.69 \pm 0.13$ and $\mathcal{L}_{\rm AGN} = 44.50 \pm 0.07$ with an AGN fraction $f_{\rm AGN}=0.77 \pm 0.10$. The stellar mass and AGN luminosity reliability are $R_{M_\star}=0.64$ and $R_{L_{\rm AGN}}=0.82$, respectively.

\section{Comparison of the model setups}\label{appendix:compa_model}

\begin{figure*}[ht]
    \centering
    \includegraphics[width=1\linewidth]{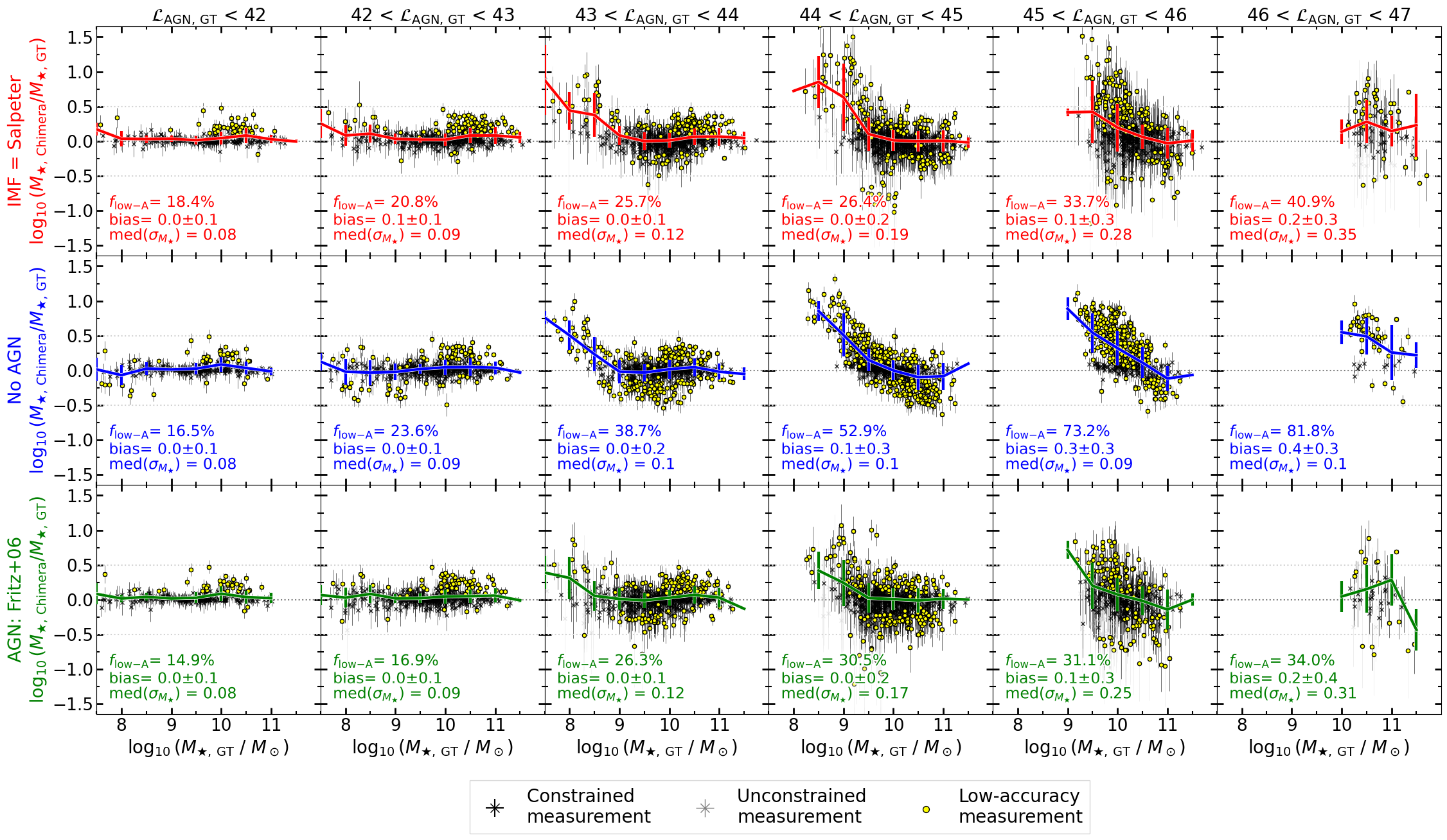}
    \caption{$\mathcal{M}_{\star,\,\rm Chimera} {-} \mathcal{M}_{\star,\,\rm GT}$ as a function of the ground truth stellar mass for different $\mathcal{L}_{\rm AGN, \, GT}$ intervals. This plot is similar to \cref{fig:compa_AGN_models}, but uses alternative SED-fitting setups. The \emph{top row} uses the IMF from \cite{Salpeter_1955}, the \emph{middle row} does not have an AGN model, and the \emph{bottom row} uses the AGN model by \cite{Fritz_2006}.}
    \label{fig_app:compa_Mstar_model}
\end{figure*}

Here, we test the alternative SED model setups to quantify the impact of our choice of baseline model.
The different setups are mentioned in \cref{subsec:SED_modules} and summarised in \cref{tab:alternative_setup}. 
We notably perform a similar analysis as in \cref{subsec:Mstar_compa}, studying the evolution of $\mathcal{M}_{\star,\,\rm Chimera} {-} \mathcal{M}_{\star,\,\rm GT}$ and $\mathcal{L}_{\rm AGN,\, Chimera} {-} \mathcal{L}_{\rm AGN,\, GT}$ as a function of the ground truth stellar mass for different $\mathcal{L}_{\rm AGN, \, GT}$ intervals. 
\Cref{fig_app:compa_Mstar_model} compares the $M_\star$ performance between three different setups other than the baseline. On the top row, we use the IMF of \cite{Salpeter_1955}, the middle row does not have an AGN model, and the bottom row uses another AGN model \citep{Fritz_2006}.
By comparing with \cref{fig:compa_AGN_models}, one can notice small difference in the fitting performance when using a different IMF or a different AGN model. On the other hand, the middle row demonstrates the need of an AGN module in the fitting process. Indeed, its absence reduces the stellar mass uncertainties unrealistically, increases significantly the bias and multiplies the fraction of low-accuracy measurements.

We noticed that the $M_\star$ bias is increasing with AGN luminosity for any alternative setups.
Various comparison metrics for both $M_\star$ and $L_{\rm AGN}$ are shown in \cref{tab:compa_setups02}, including the bias and scatter (weighted by the inverse uncertainty), the median uncertainty and the fraction of low-accuracy measurements $f_{\rm low-A}$ (\cref{eq:correct_Mstar}).
One can notice that the setup without AGN component is significantly worse than the other setups to estimate $M_\star$ with a low accuracy fraction of 44.6\%. This highlight the necessity to include an AGN component for our AGN-candidate sources to obtain reliable $M_\star$ measurements.
As the bias values varies as a function of the ground truth AGN luminosity, the averaged value presented in \cref{tab:compa_setups02} might not be representative.
\cref{tab:compa_setups02} shows that the alternative setups provide similar results to the baseline setup, suggesting the independence of our results with respect to the  choice of SED fitting model.

\begin{table*}[ht]
    \centering
    \caption{Metric comparison for $M_\star$ and $L_{\rm AGN}$ estimates between the different \texttt{CIGALE} model setups. }
    \setlength{\extrarowheight}{3pt}
    \begin{tabular}{|l|c|c|c|c|c|c|}
        \cline{2-7} 
        \multicolumn{1}{c}{} & \multicolumn{3}{|c}{$M_\star$} & \multicolumn{3}{|c|}{$L_{\rm AGN}$} \\
        \hline \rule{0pt}{11pt}
        Setup & bias & ${\rm med}(\sigma)$ & $f_{\rm low-A}$ [\%] & bias & ${\rm med}(\sigma)$ & $f_{\rm low-A}$ [\%]\\[2pt]
        \hline \rule{0pt}{11pt}
        Baseline                             & 0.05 $\pm$ 0.17 & 0.16 & 24.0 & $-$0.00 $\pm$ 0.32 & 0.70 & 37.1 \\
        IMF: \cite{Salpeter_1955}            & 0.06 $\pm$ 0.19 & 0.17 & 26.5 & $-$0.00 $\pm$ 0.30 & 0.69 & 36.0 \\
        SP: \cite{Maraston_2005}             & $-$0.01 $\pm$ 0.22 & 0.18 & 31.2 & 0.06 $\pm$ 0.33 & 0.56 & 38.3 \\
        Dust absorption: \cite{Charlot_2000} &  0.06 $\pm$ 0.22 & 0.16 & 22.7 & $-$0.02 $\pm$ 0.29 & 0.69 & 36.7 \\
        Dust emission: \cite{Dale_2014}      & 0.04 $\pm$ 0.17 & 0.16 & 24.1 & 0.00 $\pm$ 0.31 & 0.71 & 35.0 \\
        No AGN                               &  0.09 $\pm$ 0.25 & 0.12 & 44.6 & - & - & - \\
        AGN model: \cite{Fritz_2006}         & 0.03 $\pm$ 0.17 & 0.16 & 25.5 & $-$0.01 $\pm$ 0.36 & 1.11 & 36.8 \\

        \hline
    \end{tabular}
    
    \label{tab:compa_setups02}
\end{table*}

\section{Comparison of the available bands}\label{appendix:compa_bands}

\begin{figure*}[ht]
    \centering
    \includegraphics[width=0.95\linewidth]{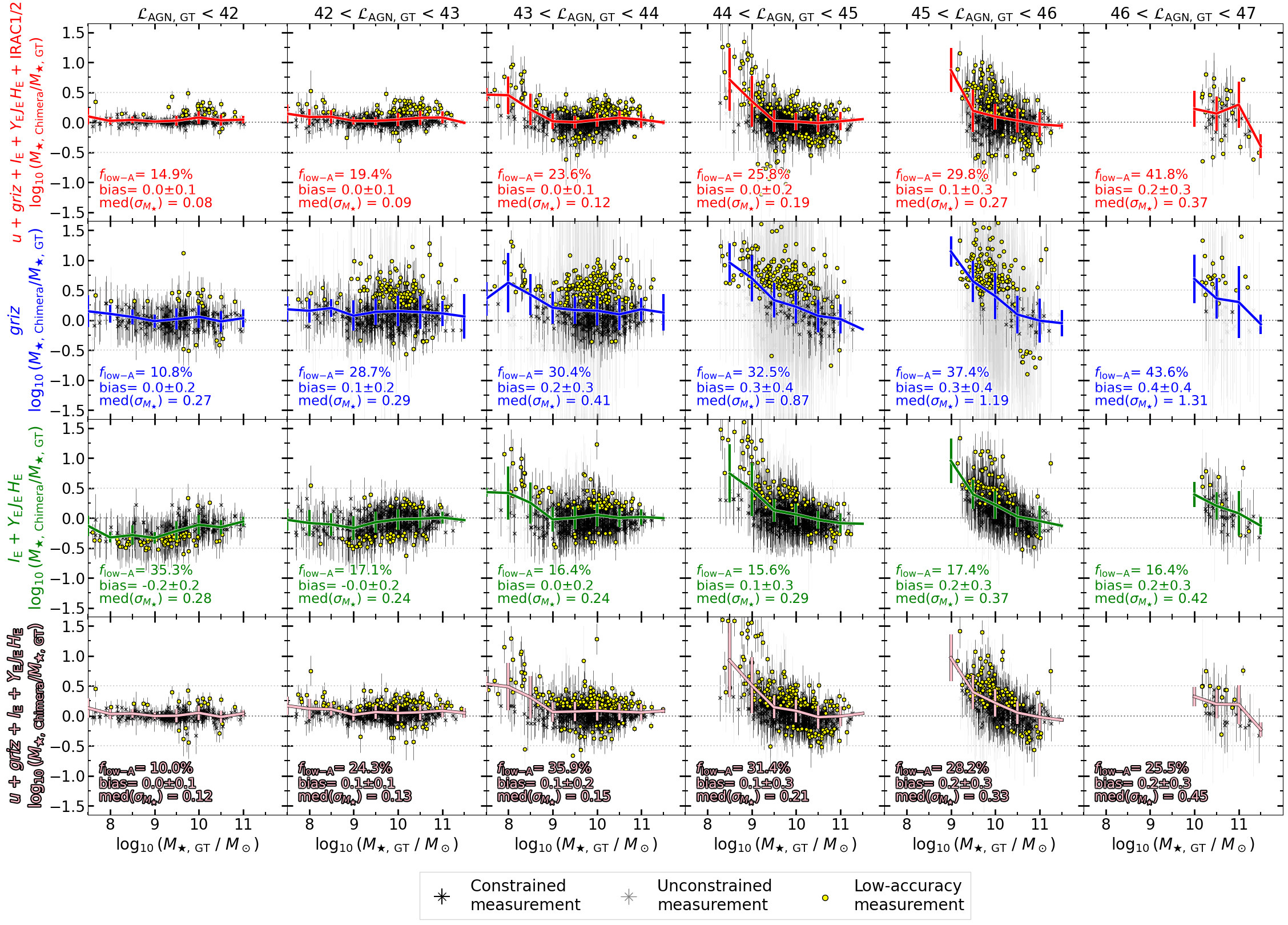}
    \caption{$\mathcal{M}_{\star,\,\rm Chimera} - \mathcal{M}_{\star,\,\rm GT}$ as a function of $\mathcal{M}_{\star,\,GT}$ for different $\mathcal{L}_{\rm AGN, \, GT}$ intervals. The plot is similar to \cref{fig:compa_AGN_models} but each row represent a different combination of available photometry bands for the SED fit: $u$ + {\it griz} + \IE + \YE\JE\HE + IRAC\,1/2, {\it griz}, \IE + \YE\JE\HE, $u$ + {\it griz} + \IE + \YE\JE\HE from top to bottom, respectively.
    }
    \label{fig:compa_dif_photometry}
\end{figure*}

In this section, we investigate the impact of available photometric bands on the estimated physical properties. For that, we perform SED fitting of the Chimera AGN using the baseline setup and different band combinations.
\Cref{fig:compa_dif_photometry} shows the $\mathcal{M}_\star$ difference as a function of the ground truth stellar mass for different ground truth AGN luminosity bins, similarly to \cref{fig_app:compa_Mstar_model}. The different rows represent different combinations of available bands.
We notice the same trend as when using all the photometric bands, i.e. lower stellar mass galaxies get progressively over-estimated with increasing AGN luminosity. However, as indicated in \cref{tab:compa_bands}, some differences are noticeable, e.g. with only optical bands, the median stellar mass uncertainties $\sigma_{M_\star}$ significantly increase.
Having only NIR (\Euclid) photometry significantly bias the stellar mass measurements at low AGN luminosity. The low bias value refereed in the paper is due to the averaging over all $L_{\rm AGN, \, GT}$.
Combining both optical and NIR photometry reduces significantly the scatter around the ground truth. Adding the MIR photometry does not significantly impact the $M_\star$ accuracy but improves considerably the $L_{\rm AGN}$ reliability as shown in  \cref{tab:compa_bands}.
This exercise demonstrates the necessity of \Euclid NIR photometry to obtain both reliable stellar mass and AGN luminosity measurements.
This analysis was conducted on more band combinations that shown in \cref{fig:compa_dif_photometry} and \cref{tab:compa_bands}.

\begin{table*}[ht]
    \centering
    \caption{Metric comparison for $M_\star$ and $L_{\rm AGN}$ estimates using different photometric bands and using the baseline model. }
    \setlength{\extrarowheight}{3pt}
    \begin{tabular}{|l|c|c|c|c|c|c|}
        \cline{2-7} 
        \multicolumn{1}{c}{} & \multicolumn{3}{|c}{$M_\star$} & \multicolumn{3}{|c|}{$L_{\rm AGN}$} \\
        \hline \rule{0pt}{11pt}
        Setup & bias & ${\rm med}(\sigma)$ & $f_{\rm low-A}$ [\%] & bias & ${\rm med}(\sigma)$ & $f_{\rm low-A}$ [\%]\\[2pt]
        \hline \rule{0pt}{11pt}
        {\it griz} & 0.17 $\pm$ 0.31 & 0.58 & 30.2 & 1.09 $\pm$ 1.16 & 1.92 & 68.5 \\
        \IE + \YE\JE\HE & 0.02 $\pm$ 0.25 & 0.29 & 18.5 & 0.88 $\pm$ 0.92 & 1.96 & 60.7 \\
        {\it griz} + \IE + \YE\JE\HE & 0.09 $\pm$ 0.20 & 0.20 & 25.2 & 0.65 $\pm$ 0.87 & 1.98 & 58.8 \\
        \IE + \YE\JE\HE + IRAC1/2 & 0.01 $\pm$ 0.19 & 0.21 & 21.7 & 0.04 $\pm$ 0.32 & 0.78 & 34.3 \\
        {\it griz} + IRAC1/2 & 0.01 $\pm$ 0.23 & 0.22 & 25.8 & 0.04 $\pm$ 0.35 & 0.72 & 34.6 \\
        $u$ + {\it griz} + \IE + \YE\JE\HE + IRAC1/2 & 0.05 $\pm$ 0.17 & 0.16 & 24.0 & $-$0.00 $\pm$ 0.32 & 0.70 & 37.1 \\[2pt]
        \hline
    \end{tabular}

    \label{tab:compa_bands}
\end{table*}

\section{Reliability functions for different photometric bands}\label{appendix:reliability_bands}

\begin{figure*}[ht]
    \centering
    \includegraphics[width=0.95\linewidth]{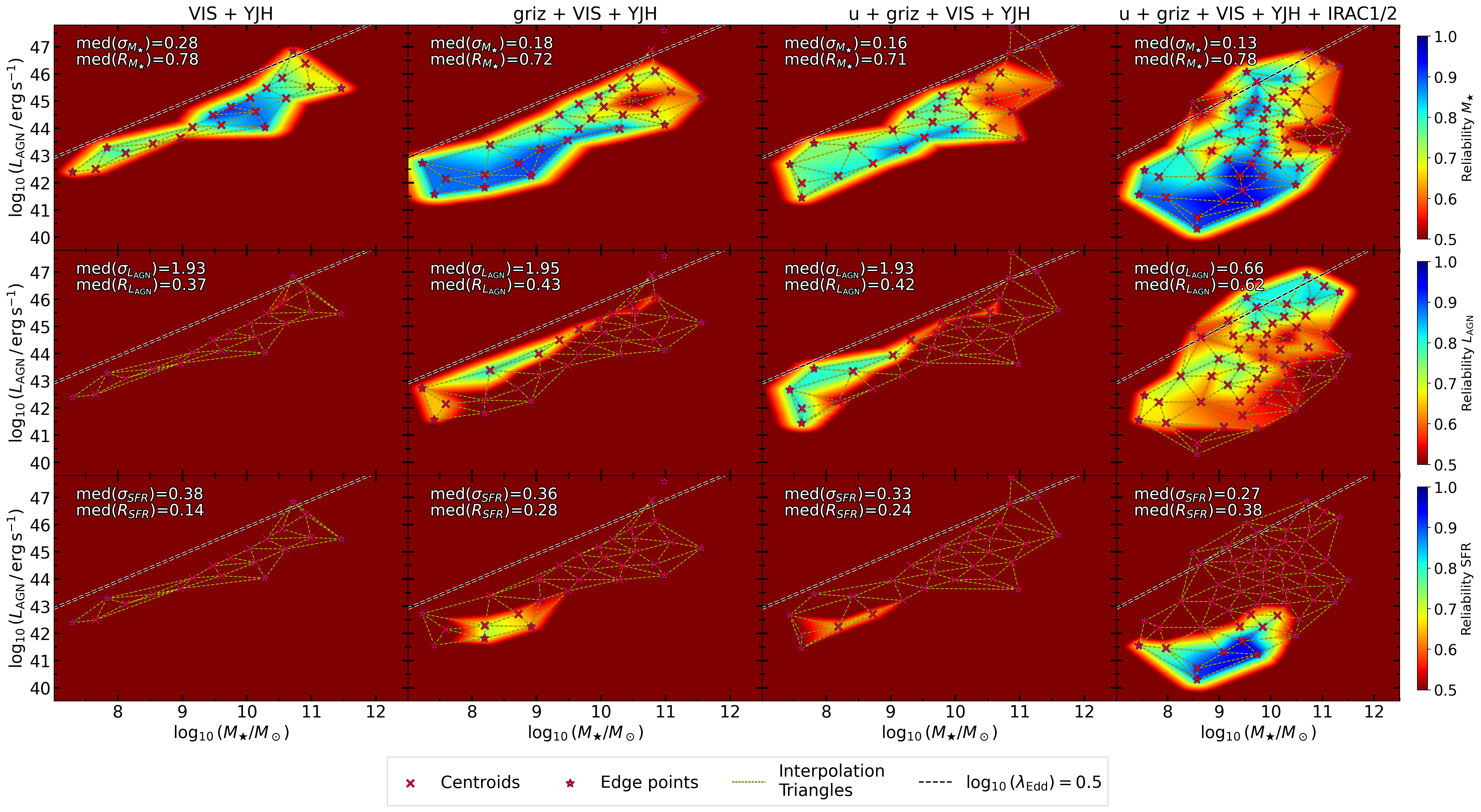}
    \caption{Reliability functions for $M_\star$, $L_{\rm AGN}$, and SFR for different photometric band combinations, from left to right: \Euclid (\IE, \YE, \JE, and \HE) only, optical ($griz$) + \Euclid, optical ($u+griz$) + \Euclid, optical ($u+griz$) + \Euclid + IRAC\,1/2.
    }
    \label{fig:relia_multiband}
\end{figure*}

\Cref{fig:relia_multiband} shows how the $R_{M_\star}$ (upper panels), $R_{L_{\rm AGN}}$ (middle panels), and $R_{\rm SFR}$ (lower panels) reliability functions vary as a function of the available photometry coverage.
It highlights the necessity of MIR photometry to separate $L_{\rm AGN}$ and SFR emission and increase their reliability in their respective regions of the $\mathcal{L}_{\rm AGN} {-} \mathcal{M}_\star$ plane. 
The lack of photometric band is compensated by the larger $\sigma_{M_\star}$ uncertainties making the median $R_{M_\star}$ approximately constant.
The reliability functions have been constructed for more band combinations than illustrated in \cref{fig:relia_multiband}.

\section{Results tables}

In this section, we present a table summarising our SED fitting results in the three fields for the three samples.
\cref{tab:results_fields} yields for each subsample and each field, the number of sources, the median and 1\,$\sigma$ scatter of the redshift, AGN fraction, stellar mass, AGN luminosity, SFR,  stellar mass reliability, AGN luminosity reliability, and SFR reliability.

\begin{table*}[ht]
    \centering
    \caption{Summary of the SED fitting results.}
    \renewcommand{\arraystretch}{1.5}
    \begin{tabular}{|c|c|c|c|c|c|c|c|c|c|c|}
        \hline
        Sample & Field & Number & Redshift & $f_{\rm AGN}$ & $\mathcal{M}_\star$ & $R_{M_{\star}}$ & $\mathcal{L}_{\rm AGN}$ & $R_{L_{\rm AGN}}$ & SFR & $R_{\rm SFR}$ \\
        \hline
            & EDF-N & 49\,418 & $0.73^{+0.51}_{-0.53}$ & $0.20^{+0.15}_{-0.13}$ & $10.15^{+0.92}_{-0.76}$ & $0.73^{+0.13}_{-0.22}$ & $43.50^{+1.07}_{-1.38}$ & $0.56^{+0.11}_{-0.12}$ & $0.53^{+0.76}_{-1.56}$ & $0.37^{+0.19}_{-0.18}$ \\ 
        A & EDF-S & 2031 & $1.35^{+0.71}_{-0.74}$ & $0.63^{+0.29}_{-0.36}$ & $10.84^{+0.47}_{-0.63}$ & $0.61^{+0.13}_{-0.20}$ & $45.87^{+0.59}_{-0.88}$ & $0.70^{+0.15}_{-0.24}$ & $2.12^{+0.63}_{-0.78}$ & $0.13^{+0.15}_{-0.09}$ \\ 
            & EDF-F & 36\,062 & $0.62^{+0.38}_{-0.34}$ & $0.19^{+0.12}_{-0.10}$ & $10.17^{+0.75}_{-0.84}$ & $0.72^{+0.13}_{-0.20}$ & $43.17^{+1.00}_{-1.02}$ & $0.60^{+0.07}_{-0.11}$ & $0.19^{+0.83}_{-1.14}$ & $0.42^{+0.27}_{-0.14}$ \\ \hline
            & EDF-N & 986 & $1.58^{+1.72}_{-1.02}$ & $0.34^{+0.24}_{-0.16}$ & $10.79^{+0.51}_{-0.85}$ & $0.62^{+0.13}_{-0.26}$ & $45.11^{+1.02}_{-1.35}$ & $0.58^{+0.15}_{-0.32}$ & $1.81^{+0.90}_{-1.44}$ & $0.17^{+0.15}_{-0.11}$ \\ 
        B & EDF-S & 4203 & $1.22^{+0.93}_{-0.77}$ & $0.34^{+0.34}_{-0.19}$ & $10.82^{+0.48}_{-0.65}$ & $0.59^{+0.14}_{-0.25}$ & $45.15^{+0.86}_{-1.48}$ & $0.59^{+0.22}_{-0.29}$ & $1.83^{+0.76}_{-1.43}$ & $0.19^{+0.15}_{-0.14}$ \\ 
            & EDF-F & 7086 & $1.48^{+1.03}_{-0.89}$ & $0.31^{+0.21}_{-0.13}$ & $10.64^{+0.53}_{-1.12}$ & $0.66^{+0.15}_{-0.20}$ & $44.82^{+1.03}_{-1.54}$ & $0.62^{+0.17}_{-0.16}$ & $1.44^{+0.96}_{-1.39}$ & $0.23^{+0.14}_{-0.16}$ \\ \hline
            & EDF-N & 328\,516 & $2.05^{+1.71}_{-1.35}$ & $0.40^{+0.10}_{-0.13}$ & $9.60^{+1.21}_{-0.96}$ & $0.70^{+0.10}_{-0.15}$ & $44.58^{+1.13}_{-1.36}$ & $0.60^{+0.12}_{-0.23}$ & $1.03^{+1.06}_{-1.08}$ & $0.20^{+0.14}_{-0.08}$ \\ 
        C & EDF-S & 493\,181 & $1.76^{+1.98}_{-1.36}$ & $0.33^{+0.16}_{-0.09}$ & $10.05^{+0.86}_{-1.27}$ & $0.72^{+0.10}_{-0.17}$ & $44.69^{+1.08}_{-1.91}$ & $0.63^{+0.11}_{-0.19}$ & $1.21^{+0.95}_{-1.44}$ & $0.24^{+0.17}_{-0.13}$ \\ 
            & EDF-F & 217\,108 & $1.76^{+2.04}_{-1.29}$ & $0.32^{+0.16}_{-0.08}$ & $10.11^{+0.80}_{-1.29}$ & $0.72^{+0.10}_{-0.17}$ & $44.70^{+1.04}_{-1.78}$ & $0.64^{+0.10}_{-0.18}$ & $1.25^{+0.89}_{-1.37}$ & $0.24^{+0.16}_{-0.13}$ \\ \hline
    \end{tabular}
    \label{tab:results_fields}
    \tablefoot{ The results are shown for the samples\,A, B, and\,C within the three different deep fields.
    For each property, the median and the 1\,$\sigma$ interval of the distribution is indicated.}
\end{table*}

\begin{figure*}[ht]
    \centering
    \includegraphics[width=0.95\linewidth]{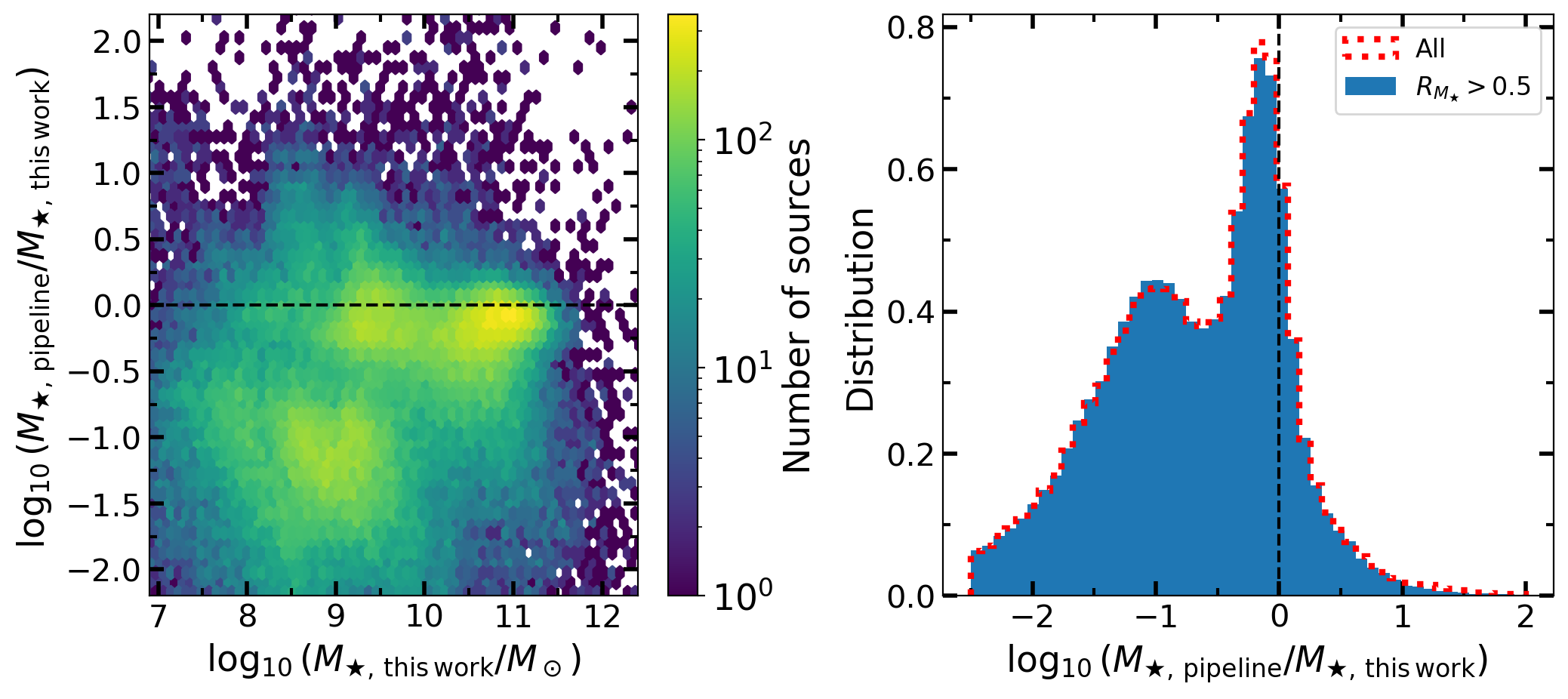}
    \caption{
    $\mathcal{M}_\star$ difference with \Euclid pipeline value for sample\,C sources with consistent redshift. The left panel shows the difference distribution as a function of $\mathcal{M}_\star$ measured in this work. The right panel shows the histogram of the difference for all sources and for the sources with $R_{\mathcal{M}_\star}>0.5$.
    }
    \label{fig:compa_phosphoros}
\end{figure*}

\Cref{fig:compa_phosphoros} shows the $\mathcal{M}_\star$ difference between the values of the \Euclid photo-$z$ pipeline with the values obtained in this work for the sample\,C sources. We restrict the sample to the sources with consistent redshift $|z_{\rm pipeline}-z| / (z+1) <0.15$, representing approximately 20\% of the sample (\cref{appendix:zphot}).
We notice a large scatter but more importantly, a lobe at low stellar masses $\mathcal{M}_\star <9.5$, revealing a $0.5{-}1.5$\,dex underestimation from the pipeline. 
Moreover, the other lobe of sources shows a $-0.2$\,dex offset with our values, which might only be due to model assumptions.
Selecting only sources with reliable measurements $R_{\mathcal{M}_\star}>0.5$ does not improve the comparison with the pipeline estimates.
\Cref{fig:compa_phosphoros} strongly demonstrates the need to include an AGN model to derive host galaxy properties.

\end{appendix}

\end{document}